\documentclass[12pt]{template}
\usepackage[english]{babel}

\usepackage{geometry}
\usepackage{physics}
\usepackage{threeparttable}
\usepackage{hyperref}

\begin{document}
\title{Quantum Communication \\Using Semiconductor Quantum Dots}
\maketitle


\author{Daniel A. Vajner,}
\author{Lucas Rickert,}
\author{Timm Gao,}
\author{Koray Kaymazlar,}
\author{Tobias Heindel*}

\dedication{}

\begin{affiliations}
Institute of Solid State Physics,  Technische Universität Berlin,  10623 Berlin,  Germany\\
* Email Address: tobias.heindel@tu-berlin.de

\end{affiliations}

\begin{justify}


\begin{abstract}
Worldwide enormous efforts are directed towards the development of the so-called quantum internet. Turning this long sought-after dream into reality is a great challenge that will require breakthroughs in quantum communication and computing. To establish a global, quantum-secured communication infrastructure, photonic quantum technologies will doubtlessly play a major role, by providing and interfacing essential quantum resources, e.g., flying- and stationary qubits or quantum memories. Over the last decade, significant progress has been made in the engineering of on-demand quantum light sources based on semiconductor Quantum Dots, which enable the generation of close-to-ideal single- and entangled-photon states, useful for quantum cryptography tasks.

This review focuses on implementations of, and building blocks for, quantum communication using quantum-light sources based on epitaxial semiconductor quantum dots. After reviewing the main notions of quantum cryptography (section 1) and introducing the devices used for single-photon and entangled-photon generation (section 2), it provides an overview of experimental implementations of cryptographic protocols using quantum dot based quantum light sources (section 3). Furthermore, recent progress towards quantum-secured communication networks as well as building blocks thereof is summarized (section 4). The article closes with an outlook, discussing future perspectives in the field and identifying the main challenges to be solved.
\end{abstract}

\newpage
\section{Introduction} \label{section1}
The task of storing and exchanging information lies at the heart of our society. In many cases information has to be transmitted in a secure fashion. Numerous examples can be found where a secure information exchange is crucial, ranging from sensitive private information such as one’s own genome, to financial information, military equipment, or critical infrastructure. While various encryption schemes, as well as the strategies of breaking them, have been employed over time (see Singh for a historic overview \cite{Singh1999}), most current security standards rely on the computational complexity of so-called one-way functions \cite{Impagliazzo1989}. However, considering the steady increase in computational power, today's secrets might not stay secret forever. Additionally, breakthroughs in the fields of quantum computing might endanger security schemes such as the Rivest–Shamir–Adleman (RSA) scheme, which is based on prime factorization of large numbers, a problem that could be solved efficiently on future quantum computers using Shor’s algorithm \cite{Shor1997}. In addition classical encryption protocols do exist, which claim to be post-quantum secure - but none of them offers ultimate security \cite{Bernstein2017}. \\
This is where quantum key distribution (QKD) comes into play, which is a method of establishing a random bitstring, securely shared between authenticated parties. This key can then be used to encrypt messages that are indeed provably secure, protected not by computational complexity, but by the laws of quantum mechanics \cite{Shor2000,Gisin2002}. In brief, the communicating parties exchange quantum two-level systems (also known as quantum bits or qubits), prepared in one of two incompatible bases, so that any eavesdropping attempt requires a measurement on the qubits which causes detectable errors if the wrong basis was chosen by the eavesdropper. The eavesdropper hereby has to decide for only one basis, since quantum states cannot be copied \cite{Wootters1982}. By that, any eavesdropping attempt can be noticed and the particular bit string is not used as a key to encrypt the message. Also, in contrast to classical encryption schemes, a postponed measurement of the transmitted qubit is not possible. Hence, securely transmitted data will remain secure also in future. 
\\
Equally important, photonic quantum channels not only provide a means of perfectly secure communication, but they also enable other quantum technologies. “Flying” qubits, i.e. photons, are vital for distributed quantum computing schemes and to interface different parts of a future quantum computer, as famously formulated in the DiVincenzo criteria \cite{DiVincenzo2000}. 
\\
Once several nodes are securely interconnected, one speaks of a quantum network and a future vision would be to connect parties all over the world via perfectly secure quantum channels - a vision often referred to/called the quantum internet \cite{Kimble2008}. This review intends to summarize and discuss some of the progress made towards achieving that vision, with a focus on implementations of quantum communication using quantum light sources based on epitaxial semiconductor quantum dots (QDs). 
\\
Let us begin by introducing quantum key distribution in more detail. Historically, the ideas of quantum cryptography date back to the late 1960s, when Wiesner first formulated his ideas on conjugate coding, a work published more than a decade later in 1983 \cite{Wiesner}. We refer to Bennett et al. \cite{Bennett1992a} for a first-hand historical review. Motivated by the ideas of Wiesner, Bennett, and Brassard proposed the first actual key distribution protocol in 1984, later referred to as the BB84 protocol, using the quantum mechanical properties of single photons to detect eavesdropping attempts \cite{Bennett1984}. In their seminal work, the authors proposed to use the polarization of single photons to encode the bits in different, randomly chosen bases (see \textbf{Figure \ref{fig:fig1}a}). Once the photons are transmitted from Alice to Bob (typical names for the communicating parties), Bob also measures them in a randomly selected basis. After Alice and Bob are authenticated, they share the selected bases over a public channel and keep only the measurement results of the cases in which both of them had measured a photon in the same basis (a process called “key sifting”), without communicating the actual measurement results. By comparing a subset of the results, potential eavesdropping can then be detected, since by the laws of quantum mechanics every measurement of an adversary in the wrong basis perturbs the quantum state, which leads to detectable errors. After post-processing steps, which reduce errors and enhance the secrecy, the remaining bits form the secure (or secret) key. This scheme is also referred to as prepare-and-measure setting, since Alice prepares the photon states and Bob measures them. \par
A few years later, in 1991, Artur Ekert proposed the first QKD protocol using entangled photon pairs \cite{Ekert1991} (Figure \ref{fig:fig1}b). The Ekert91 protocol relies on detecting an eavesdropper by monitoring the amount of violation of a Bell-like inequality \cite{Clauser1969,Bell1964}, which quantifies the degree of remaining entanglement. In addition, one can also use the distributed entangled photons directly for measurements in the BB84 bases, compare some of the results and deduce the security from the identified error rates just as in the BB84 protocol - a protocol known as BBM92 \cite{Bennett1992}. 
\\
Note that sending single photons from Alice to Bob (prepare-and- measure) or Alice and Bob both receiving photons from a central source (entanglement based) are equivalent. One can imagine QKD with entangled photons also in the way that Alice measures a photon, which then travels backwards in time to the source and then forward in time to Bob. 
\\
Note also, that an initial misconception was that, since the source can be placed at half of the distance between Alice and Bob, the achievable distance is doubled in the Ekert protocol. This is not the case, since although each photon needs to travel only half the distance, there are two photons that need to reach the receiver, thus canceling the effect of the reduced individual distance. The arrival probability is proportional to the transmittance of the quantum channels for the respective photons $T_{1(2)}$ for a channel of length $L_{1(2)}$, which are described by an exponential absorption loss inside the optical fiber given by an absorption constant $\alpha$ (about 0.2 dB/km in ultralow-loss fibers). The probability that both photons arrive is $P \sim T_1 \cdot T_2$ , which is the same as for one photon passing through a channel of the summed lengths.
\begin{equation}
    P \sim T_1 \cdot T_2 = e^{-\alpha L_1} \cdot e^{-\alpha L_2} = e^{-\alpha (L_1+L_2)}
\end{equation}
Nowadays, numerous other QKD protocols are known beyond the simple BB84 and Ekert protocols, ranging from decoy protocols for attenuated lasers, to asymmetric versions of BB84, protocols with more states (6-state), less states (2-state) or multi-dimensional protocols, full or partly device independent protocols, two-way protocols and coherent-one-way protocols, reference-frame independent protocols, twin-field protocols, continuous variable protocols and many more. For a recent overview we refer to Pirandola et al. \cite{Pirandola2019}. 
\\
Having established a secure key by some QKD protocol, how can the two parties now exchange a message and be sure that nobody else knows its content? The only provably secure encryption method – in an information theoretical sense – known to date, is the one-time pad (OTP) \cite{shannon1949} (see Figure \ref{fig:fig1}c). This symmetric encryption method, first postulated by G. Vernam in 1926 \cite{Vernam1926}, is based on four requirements:
\begin{itemize}
    \item The key of Alice and Bob is secret
    \item It is randomly generated
    \item It has at least the same length as the message to be encrypted
    \item It is used only once (hence the name one-time-pad).
\end{itemize}
If these requirements are met, the secret key can be applied to the message using an XOR (eXclusive OR) logical operation (a modulo-2 addition). The resulting encrypted message appears to third parties as random as the key itself and therefore does not disclose any useful information. Then, the encrypted message can be sent over a public channel. The other party can now use the same key and apply it with an XOR operation to the received, encrypted bit string, by which the original message is recovered. The problem for any classical key distribution process however is to guarantee the secrecy of the key distribution. Exactly this secrecy can be guaranteed by means of QKD.
\\
Quantum cryptography is physically secure in the sense that an eavesdropping attempt can always be detected. The most basic eavesdropping scenario is the intercept-resend strategy where an adversary just measures Alice’s photons in a random basis and then sends new photons to Bob, this is indicated in the upper panel of Figure \ref{fig:fig1}d. In BB84 half the choices then lead to wrong bases and thus to a disturbance of the quantum states. When an eavesdropping attempt is detected, it can either be compensated (if the error is low enough) by distilling a shorter key that is again secure, or the key is discarded and a new key exchange is initiated, until the final key is provably secure. This requires in all cases an authenticated classical channel to communicate some information such as the used measurement bases in the BB84 protocol. To realize that authentication, some shared, secret bits must already be present to begin with, which is why QKD is technically a key expansion scheme. Note furthermore, that in most of the QKD protocols random numbers are needed in one way or the other, either to encode a random bit or to pick measurement bases randomly. Thus, true random number generators are essential for QKD schemes to be absolutely secure (see section \ref{section4.5}).
It is worth mentioning that quantum cryptography in general is the art of using quantum mechanical effects to perform cryptographic tasks, which includes much more than only exchanging a secure key (see e.g. Broadbent et al. for an overview \cite{Broadbent2016}). Nevertheless, in this review we focus on the task of QKD, since the building blocks that are necessary for successful QKD with single photons (sources, channels, memories, repeaters, random numbers etc.) will also enable many other cryptographic applications. \\
\begin{figure}[h]
  \includegraphics[width=\linewidth]{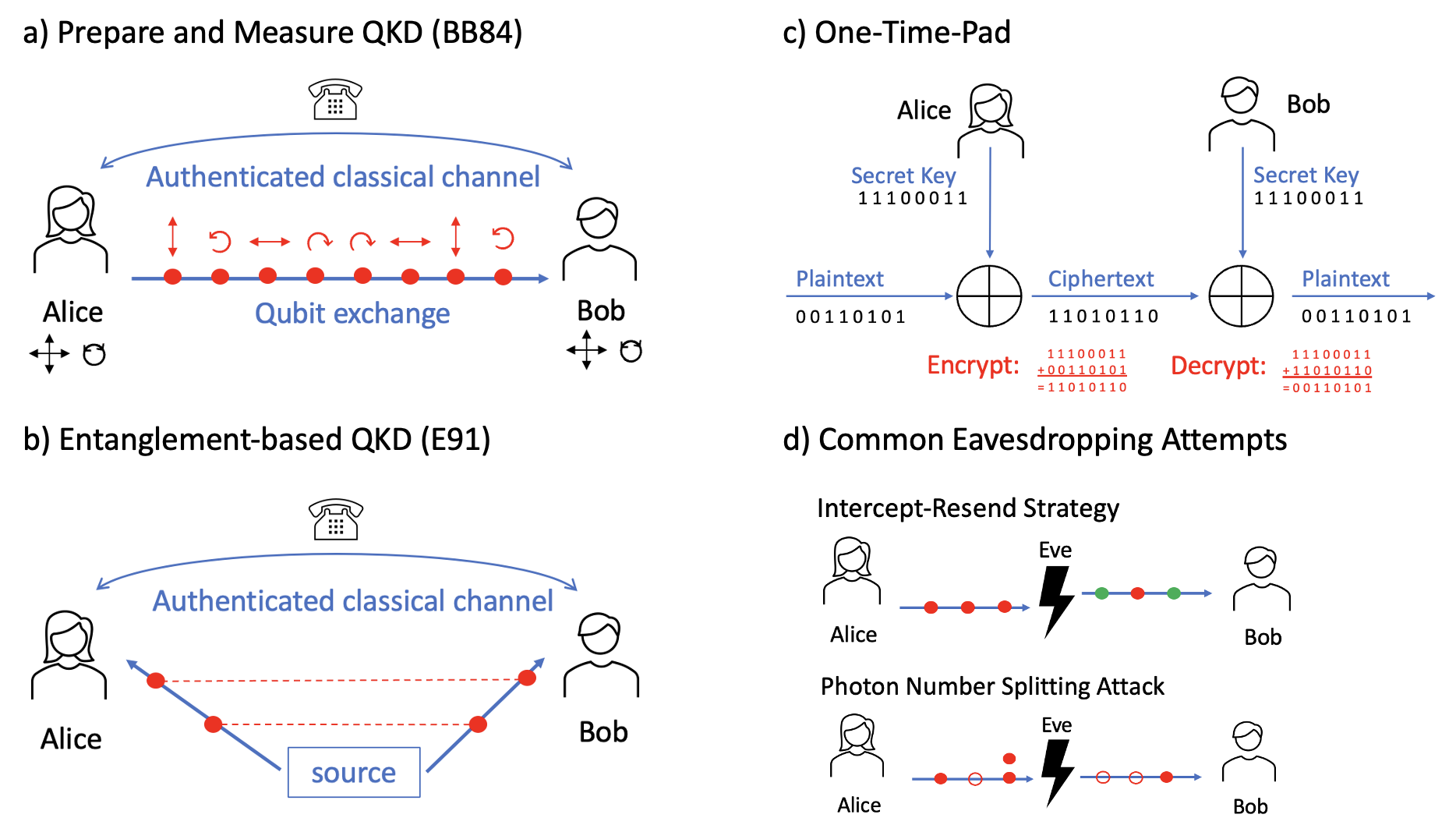}
  \caption{Principle of QKD: Prepare and Measure based (a) and entanglement based (b) protocols can be used to generate a secret key that can be used with the one-time-pad (c) for absolutely secure communications. Common eavesdropping strategies (d).}
  \label{fig:fig1}
\end{figure}
When implementing a QKD protocol, one of the first choices to take is the technology platform and the respective qubits used.  To realize a “flying” qubit, photons are the obvious choice. They can be transmitted over long distances in optical fibers or through free-space links and hardly interact with their environment which leads to unmatched coherence properties \cite{OBrien2009}. Note that in principle other types of exchangeable qubit realizations could be imagined such as electrons in wires, but compared to photons they face more loss and decoherence which strongly limits the maximum distances \cite{Hermelin2011,McNeil2011}. Due to the photon's coherence properties, speed, and the possibility to encode information in various degrees of freedom (e.g., polarization, time-bin, energy, path, phase or orbital angular momentum, see \cite{flamini2018} for an overview), they are a promising candidate for many quantum information tasks such as quantum computing, sensing, metrology and of course quantum cryptography.
For QKD different encoding schemes have been developed, which will be briefly summarized in the following \cite{Gisin2002}. 
\\
It is straightforward to encode qubits in the polarization states of photons, which is why the very first demonstration of QKD in 1992 \cite{Bennett1992a}, as well as the first field experiment \cite{Muller1993}, used polarization encoding. With polarization optics arbitrary polarization states can be prepared, both in optical fibers and in free space. Typically, the rectilinear and the circular bases are used for the two non-commuting BB84 bases, as they can be prepared actively in a random fashion with fast electro-optical modulators on the sender side. On the receiver side, a beam splitter followed by polarization optics that project in two different bases realizes a passive random basis choice. It is crucial to maintain the polarization state in this encoding scheme, which is why effects such as polarization mode dispersion in optical fibers must be compensated and the quantum channel must be stabilized to compensate polarisation drifts over time. 
\\
A coding scheme that is better suited for fiber-based communication is phase encoding. In the simplest case, a single Mach-Zehnder-Interferometer (MZI) is spanned between Alice and Bob via two optical fibers and two couplers, at which each of the parties can apply phase shifts to their half of the MZI. In this configuration, the condition for constructive or destructive interference at the receiving detector depends only  on the applied phase shifts (given the MZI pathlength is stable to a fraction of the wavelength) corresponding to the encoded states. As the required path length stability in the single MZI phase encoding scheme is difficult in practice, QKD is typically done with a double MZI phase encoding, in which two unbalanced MZI are put in series, with one being controlled by Alice and one by Bob. If the MZI imbalance is the same at both MZIs, photon paths including the short and long arm exactly once are indistinguishable, leading to interference, which again depends on the applied phase shifts at Alice and Bob \cite{Yuan2008}. As the photons are in a superposition of different time bins, this is also a way of time-bin encoding. An interesting variation are photons that travel back and forth, which improves the stability , as photons experience the same quantum channel twice in opposite directions thus compensating phase distortions, however at the cost of reduced communication distances \cite{Martinelli1992}. 
\\
Several other encoding schemes exist such as encoding in the relative phase between sidebands of a central optical frequency \cite{Sun1995}, the photon path \cite{Monteiro2015}, or using correlations between time bins and photon energy, enabling high-dimensional encoding \cite{Zhong2015}. Also, several ways exist to convert the different encoding types into each other \cite{Kupchak2017}.The probability that a photon is detected in a wrong basis in a given encoding scheme is called the quantum bit error ratio (QBER). 
\\
Ideally, all these encoding schemes require practical, high-quality sources of single photons (SPSs) or entangled photon pairs (EPSs). Here, high quality means that all emitted photons are in the single photon Fock states with low multi-photon emission probability (purity), the majority of excitation pulses lead to collected photons (brightness) and photons emitted from the same source are similar (indistinguishability). Being practical implies that they are easy to operate, durable, spectrally tunable, affordable, and scalable. While many non-classical light sources have been developed to date (see e.g. \cite{Lounis2005,shields2010,Eisaman2011} for an overview), ideal SPSs or EPSs are challenging to realize. Semiconductor QDs, however, are considered one of the most promising candidates to solve this challenge \cite{Aharonovich2016}. \\
\begin{figure}
  \includegraphics[width=\linewidth]{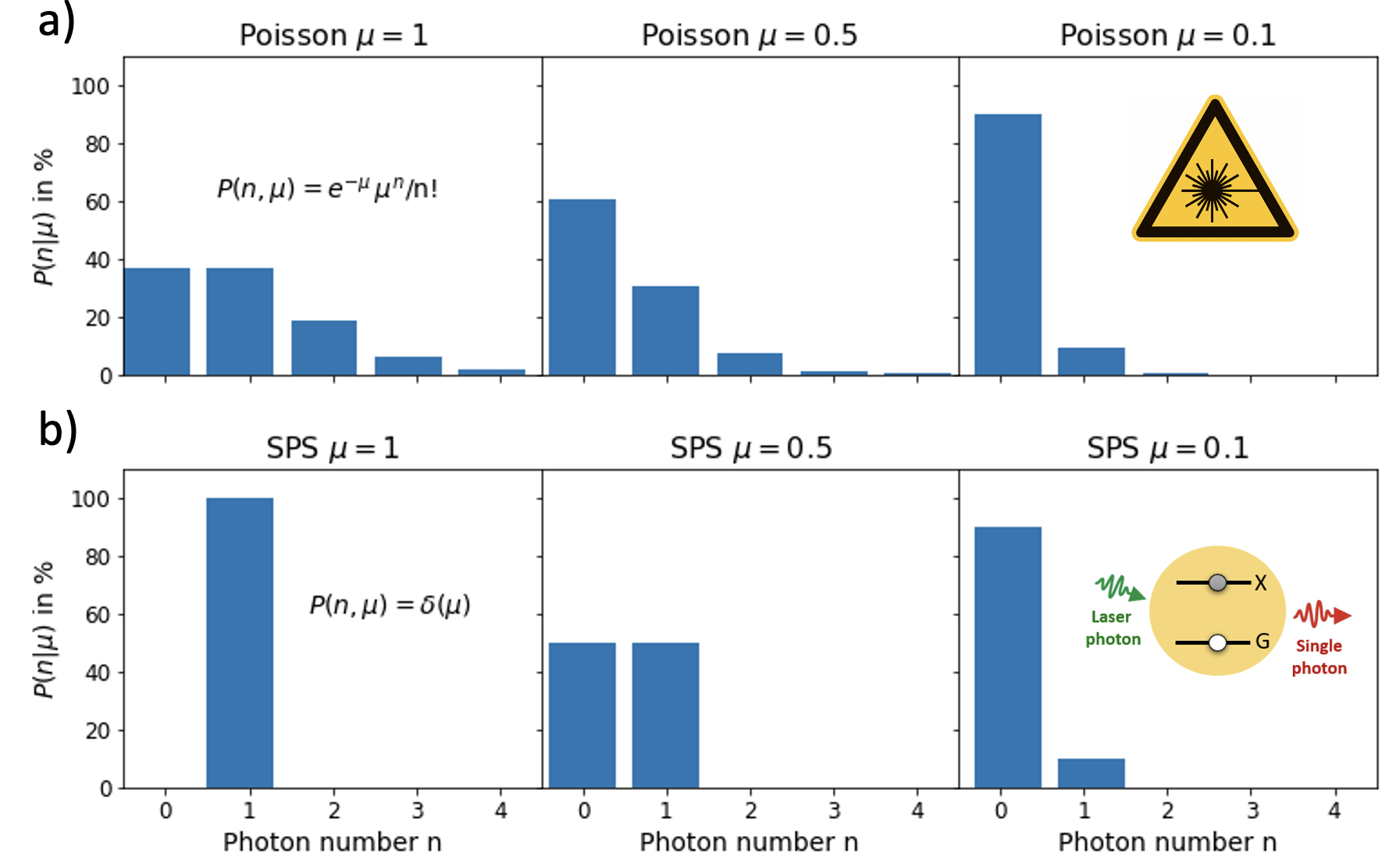}
  \caption{Comparison of Photon Statistics for laser sources (a) and single photon sources (b) for different mean photon numbers $\mu$.}
  \label{fig:fig2}
\end{figure}
Due to the lack of close-to-ideal quantum light sources, the majority of QKD implementations, and all commercial realizations, made use of phase-randomized attenuated laser pulses, also known as weak coherent pulses (WCP) \cite{Chen2009,Zhao2006,Jennewein2000a,Poppe2004,Boaron2018a}. These can be realized more easily, but can only approximate a single photon state, since the photon number still obeys the Poisson statistics. The probability to find $n$ photons in a state with mean photon number $\mu$ is $P(n,\mu) = e^{-\mu} \mu^n /n!$ Therefore, the conditional probability to find more than one photon, given the pulse is not empty, is \cite{Gisin2002}
\begin{equation}
P(n>1, \mu \mid n>0, \mu)=\frac{P(n>1, \mu)}{P(n>0, \mu)}=\frac{1-P(0, \mu)-P(1, \mu)}{1-P(0, \mu)}=\frac{1-(1+\mu) e^{-\mu}}{1-e^{-\mu}} \approx \frac{\mu}{2}
\end{equation}
This means that the probability that a non-empty pulse contains more than 1 photon can be made arbitrarily small, however at the cost of an extremely low number of photons in the quantum channel: $P(n=0, \mu) \approx 1-\mu$. QKD experiments with WCPs typically use $\mu = 0.1$, a mean photon number at which about $5\%$ of the non-empty pulses still contain more than 1 photon. This leads to possible attacks such as photon number splitting attacks (Figure \ref{fig:fig1}d) or the more realistic beam splitting attacks \cite{Lutkenhaus2002}. 
\\
In these attacks an adversary replaces the channel with one of lower loss, measures the photon number non-perturbatively in each pulse, blocks pulses with 1 photon and steals all but 1 photon from the pulses with more than 1 photon. In this way the eavesdropper possesses a copy of all photons that Bob receives and can measure them according to the basis information disclosed over the public channel. Since an ideal SPS would emit at maximum 1 photon per pulse, the photon number splitting attack would not be possible. In (\textbf{Figure \ref{fig:fig2}}) the photon number distribution of WCP (Poissonian) and single photons are compared for different mean photon numbers. While for a typical average number of 0.1 photons per pulse both sources have a similar number of single photon events, the attenuated laser still has a considerable amount of multi-photon contributions. One might argue that a solution is to reduce the mean photon number in WCP setups strongly and, in turn, increase the repetition rate. In this case, however, the photon detectors receive many empty pulses significantly increasing the influence of dark counts at the detectors. 
\\
Using so-called decoy state protocols \cite{Wang2005}, these attacks on WCP implementations can be mitigated, by estimating the amount of multi-photon pulses in-situ with the help of decoy states. But significant shrinking of the sifted key is required since the decoy pulses cannot be used to generate a key. Provided a single-photon source (in contrast attenuated laser pulses) with high efficiency, i.e. with an average number of photons per pulse $\mu_\text{SPS}$ comparable to average intensities typically used for signal states in decoy protocols (e.g. $\mu_\text{WCP} = 0.4$), the reduced protocol overhead gives a clear benefit for the SPS. For an overview of the state-of-the-art in attenuated laser QKD covering decoy and other types of protocols we refer the reader to \cite{Xu2020}.
\\
Even more general, Poisson statistics predict a maximum probability to emit 1 photon in a WCP source of $37\%$, which means that a QD SPS with an average photon number in the quantum channel above that value, outperforms the WCP source when driven at the same rate. The same argument holds true for entangled photon sources. By sending laser pulses to nonlinear crystals, spontaneous parametric down conversion (SPDC) can be used to generate entangled photon pairs \cite{Kwiat1995,Couteau2018}. This process however is not only inefficient (conversion rates of $10^{-6}$ are typical) but also does not provide entangled photon pairs with single-photon statistics. The number of photon pairs follows a thermal distribution on timescales below the coherence time and a Poisson distribution for larger times \cite{Tapster1998,Avenhaus2008}. As a consequence, the maximum number of created photon pairs in SPDC sources is limited, since the fidelity to the maximally entangled states reduces with increasing multi-pair contributions. As a consequence, the higher probability to emit more than one entangled photon pair in down-conversion processes ultimately limits the key rate, no matter how fast the clock rate of the system is, as was recently shown by Hosak et al. \cite{Hosak2021}. QD SPSs on the other hand can provide entangled photon pairs with single photon pair statistics via the biexciton-exciton emission cascade (see section \ref{section2.1}) which have already exceeded the fundamental limits of SPDC sources \cite{Chen2018,Liu2019,Wang2019}. 
\\
Even more important, when considering realistic finite key lengths, the number of secure bits reduces significantly due to finite-size corrections, which arise from statistical errors due to the reduced size of the parameter estimation set. These corrections, that are necessary to still claim a secure key exchange for finite-size keys, have a much greater effect on attenuated laser pulses than on SPSs \cite{Scarani2009,cai2009}, which makes SPS superior even at lower average photon numbers. Such finite-size effects are especially severe in scenarios where the maximum key length is limited, for example in satellite links or mobile systems \cite{Nauerth2013,Liao2017}.
\\
Moreover, in many advanced protocols such as measurement-device-independent (MDI), fully device independent (DI) QKD (see section \ref{section3.4}) or in essential building blocks of quantum networks such as quantum repeaters (see section \ref{section4.2}), successful Bell state measurements are essential. These however suffer strongly from multi-photon input states which occur more often in WCP systems due to the Poisson statistics. These measurements also require a high two-photon interference visibility which can in principle reach up to $100\%$ for quantum light sources, while it is fundamentally limited to $50\%$ in classical WCPs \cite{Mandel1983}. Also in this case, SPSs would therefore provide a clear advantage.
\\
Another aspect is that nowadays, even the state-of-the-art single-photon detectors can only operate at detection rates below 100 MHz \cite{SingleQuantum}. The repetition rates of modern QKD systems however have already reached the GHz regime, hence many QKD system clock-rates are already limited by the photon detector \cite{Gordon2005,Dixon2009}. In such a case and following the arguments presented above, using SPSs with relatively low brightness can already increase the secret key rate. Semiconductor QDs are a promising candidate for such a SPS and therefore at the core of this review.
\\
While this review article will focus mainly on the experimental advances in the field, also theoretical work has advanced significantly since the proposal of the BB84 protocol. First rigorous security proofs were soon extended to more realistic scenarios, also considering imperfections of Alice and Bob, unavoidable in physical implementations. A common way of calculating asymptotic secure key rates for imperfect photon sources was introduced in the so-called GLLP paper by D. Gottesmann, H.-K. Lo, N. Lütkenhaus, and J. Preskill \cite{Gottesman2004}. Typically, secure key rates are calculated in the asymptotic regime, i.e. assuming that an infinite amount of exchanged qubits is available so that protocol parameters can be estimated without errors from an infinite set. In that case the secure key rate $K_{\text{sec}}$ is simply the clock rate of the QKD experiment $R_0$ multiplied with the mean photon number that is coupled into the channel $\mu$, the channel transmissivity $t$, the detection setup transmission $\eta_{\text{Bob}}$, the detector efficiency $\eta_\text{det}$, the sifting factor $q$ and the secure bit fraction $r$ as
\begin{equation} \label{eq:GLLP}
    K_\text{sec} = R_0 \cdot \mu \cdot q \cdot  t \cdot \eta_\text{Bob} \cdot \eta_\text{Det} \cdot r
\end{equation}
The secure bit fraction $r$, following the GLLP security proof, is the amount of uncertainty an adversary has over the key, quantified by the binary Shannon entropy $h_2(e) = -e \log_2(e) - (1-e) \log_2(1-e)$ as a function of the QBER $e$, reduced by the information leakage due to the error correcting code with an efficiency of $f_\text{EC}$, and corrected for the amount of multi-photon emission events that may enable photon number splitting attacks. This leads to a secure bit fraction of
\begin{equation} \label{eq:GLLP_correction}
    r=A\left[1-h_{2}\left(\frac{e}{A}\right)\right]-f_\text{EC} \cdot h_{2}(e)
\end{equation}
Here, $A$ is the correction factor used to incorporate multi-photon emission events as $A = P_{1}/P_\text{click} $, which describes the ratio of detector clicks resulting from single-photon events with probability $P_1$, to all detector clicks $P_\text{click}$. The quantum bit error ratio $e$ is the ratio of erroneous detection events normalized by the sum of all detection events, which can be estimated from a detection system intrinsic error $e_\text{det}$ (probability that a photon encoded in one basis is detected in the other one) and the number of dark counts $P_\text{dc}$ as
\begin{equation} \label{eq:GLLP_QBER}
    e = e_\text{false clicks} + e_\text{dark counts} = \frac{P_\text{click } \cdot e_\text{det} + \frac{1}{2} \cdot P_\text{dc}}{P_\text{click}+P_\text{dc}} .
\end{equation}
Note that also more refined channel models exist which take into account detector imperfections such as dead-time and after-pulsing and allow channel multiplexing \cite{eraerds2010quantum}. The GLLP equations (\ref{eq:GLLP}-\ref{eq:GLLP_QBER}) discussed above will be used at the end of section \ref{section3.2} to put the QKD experiments reported to date into perspective. An important development in the field of QKD rates over the last decade was to consider even more realistic settings beyond the asymptotic limit, in which only a finite number of qubits can be exchanged. This is an important practical scenario, e.g. for communication links to and between moving platforms such as air-crafts or satellites. Several important adaptations have to be made to the estimation of the secure key rate, which are well-described in the works of Cai, Scarani and Renner \cite{Scarani2008,Renner2008,Scarani2009,cai2009}. Importantly, incorporating finite-size effects is not just a theoretical consideration to certify security of the scheme more precisely. On the contrary, not incorporating these effects entails actual security risks, as shown by Chaiwongkhot et al. by hacking a commercial QKD system exploiting finite-size effects \cite{Chaiwongkhot2017}. Noteworthy, secure key rates can also be calculated numerically, by maximizing Eve's information over all attacks that are allowed by the laws of physics \cite{Coles2016}, which resulted in key rates that agreed with the analytically known ones \cite{Winick2018,George2021}.

\section{Quantum Dot Based Quantum Light Sources}
While the notion of "single photons" has been around for a 100 years from the early 20th century on \cite{Einstein1905,LEWIS1926}, the concepts and practical realizations of light sources emitting single photons one by one took their fair time to be developed. Theoretically, a driven two-level system is an ideal single-photon emitter. While driving transitions in single atoms fulfills that goal \cite{kimble1977,ripka2018}, this route is not always practical. This was one reason for researchers to consider low-dimensional solid-state systems, which mimic atoms but provide advantages for device integration. Important steps in this endeavor where made in the 1990s, where research groups succeeded in encapsulating small islands of semiconductor material with a lower bandgap by another semiconductor with a larger bandgap, effectively forming a quasi zero-dimensional structure known as QD \cite{Leonard1993,Grundmann1995}. Technically this is possible by exploiting self-organized processes, e.g. in the Stranski–Krastanov growth-mode \cite{Stranski1939}, during hetero epitaxy of one material onto another \cite{Lay1978}. After their discovery, QDs quickly gained interest with applications ranging from laser physics  \cite{Arakawa1982} to quantum information \cite{Imamoglu1999}. Many excellent review articles and books on epitaxial semiconductor QDs and QD-based quantum light sources have been published over the years, e.g. \cite{Michler2009a,Shields2007,Buckley2012,Senellart2017,Trivedi2020,Rodt2020}. This section therefore intends to briefly summarize the most important properties of QDs and engineered QD-based quantum light sources used for implementations of quantum communication to date.

\subsection{Semiconductor Quantum Dots} \label{section2.1}
In QDs the three-dimensional confinement of charge carriers in a small volume of typically a few nanometer leads to quantized energy levels in the valence and conduction band, of which the lowest ones approximate a two-level system (cf. \textbf{Figure \ref{fig:fig3} a)-c)}). Filling the QD potential with one electron-hole pair forming a bound exciton (X) using optical or electrical pumping, the energy levels can be occupied. After a characteristic period of time, i.e. the spontaneous emission lifetime (typical $~1\,$s), the exciton recombines radiatively under emission of a single photon, leaving the empty QD in its ground state. As long as the exciton has not recombined, the electronic state is occupied, thus providing the basic mechanism required for a SPS emitting photons subsequently, as first demonstrated by Michler et al. in 2000 \cite{Michler2000}. Note, that QDs require typically cryogenic cooling down at temperatures arround $4-70\,$K for optimal performance. There are however material combinations, such as CdSe and Nitride materials, as well as organic structures, which lead to higher exciton binding energies, resulting in higher possible operating temperatures up to room temperature \cite{Michler2000a,Kako2006,Arians2008,Deshpande2013,Cui2019}. In addition, also cryotechnologies advanced significantly in recent years, nowadays enabling compact and benchtop cryogenic quantum light sources as recently demonstrated by Schlehahn et al. \cite{Schlehahn2018} (see section \ref{section4.1}).\\
Due to the semiconductor environment, QDs can host multiple charge carrier in different combinations of excited states, resulting in various distinct emission lines (cf. Figure \ref{fig:fig3}d and e). Spectrally selecting either one of these emission lines results in single-photon emission which can be experimentally verified via second-order photon autocorrelation experiments in a HBT setup (see Figure \ref{fig:fig3} f and g). Advantages of QDs for quantum light generation are their excellent single photon properties, recently demonstrated to be superior to any other known type of SPS \cite{Schweickert2018}, in combination with the possibility to realize engineered devices with designed emission wavelength (see also section \ref{section2.2}). This is highly beneficial for applications in quantum information, as the generated quantum light states can be spectrally matched to the Telecom O- and C-band (at wavelengths around 1300 nm and 1500 nm) for long distance communication in optical fibers, or atomic transitions in Alkali vapor cells for quantum memory applications (see section \ref{section4.3}). As the growth of high-quality QDs at Telecom wavelengths is still a technological challenge, an alternative approach is the wavelength conversion of single photons emitted from a QD at lower wavelengths to the Telecom range exploiting nonlinear optical effects \cite{Rakher2010}. The state-of-the-art in this field enables conversion efficiencies of $35\%$ \cite{Morrison2021}. Another important development in the field of QD-device engineering refers to the availability of several deterministic fabrication technologies, where the QDs are either grown directly at a pre-determined location (site-controlled growth) or QDs are pre-selected post-growth (according to the optical properties) and subsequently integrated in photonic devices (see Refs. \cite{Rodt2020,Liu2021} for recent review articles). These deterministic technologies turned out to become a game changer enabling high device yields for applications in photonic quantum technologies.
\\
\begin{figure}
  \includegraphics[width= \linewidth]{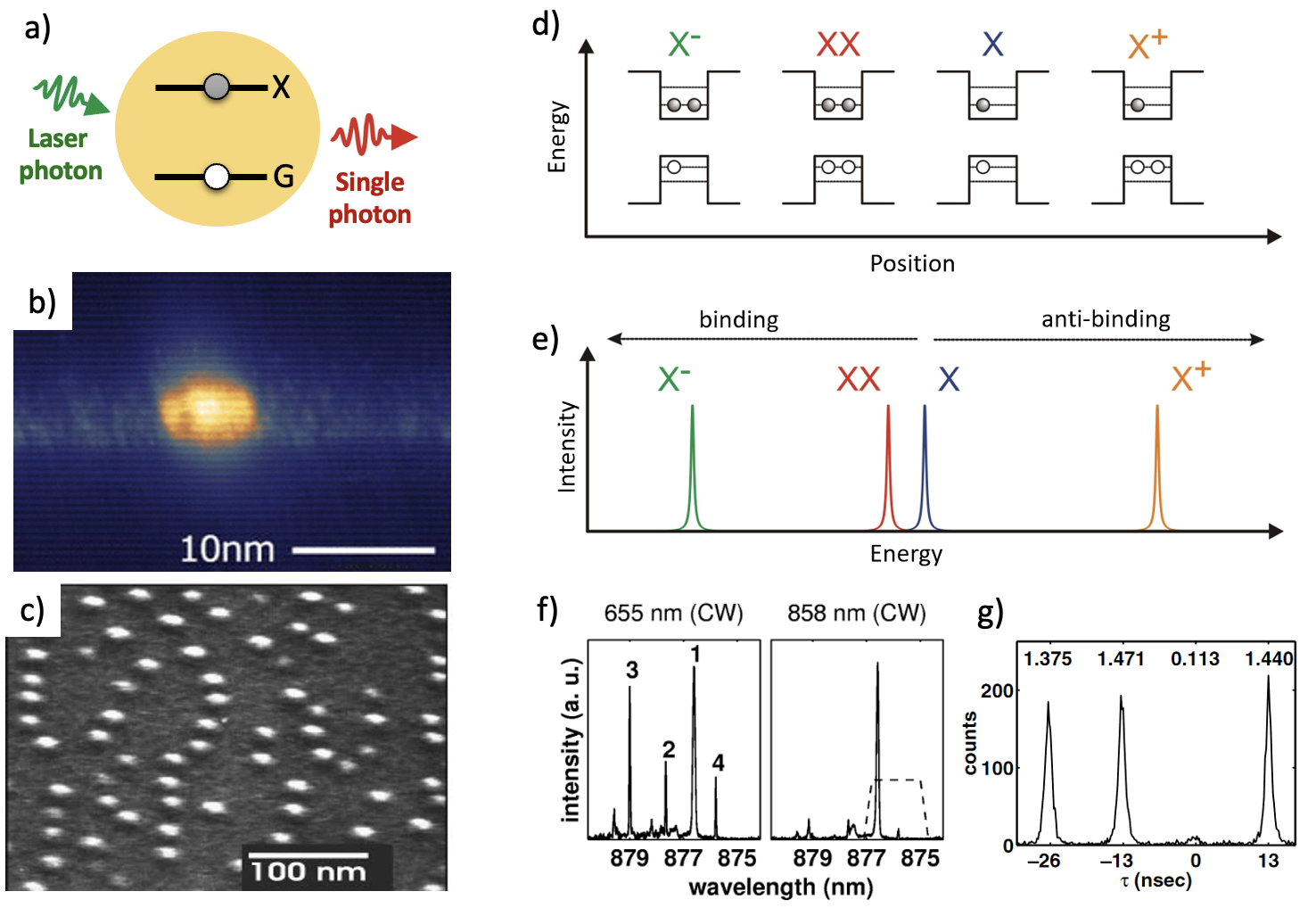}
  \caption{Two-level-scheme of a single photon emitter (a), Scanning Tunneling Microscope image of a single InAs/GaAs QD (b), scanning electron microscope image of epitaxially grown InGaAs QD islands (c), overview of different excited states in a QD (d) and their respective emission lines (e), different emission lines when exciting a QD off resonantly (f left) and only one emission line in quasi-resonant excitation (f right), low correlation at zero temporal delay in a HBT experiment indicates single photon emission (g). (b) reprinted from \href{https://aip.scitation.org/doi/full/10.1063/1.4770371}{\textit{Keizer et al. 2012}} \cite{keizer2012} with the permission of AIP publishing, (c) reprinted by permission from \href{https://www.nature.com/articles/nature02969}{\textit{\textit{Reithmaier et al. 2004}}} \cite{reithmaier2004} Copyright 2004 Springer Nature, (f,g) reprinted with permission from \href{https://journals.aps.org/prl/abstract/10.1103/PhysRevLett.86.1502}{\textit{Santori et al. 2001}} \cite{santori2001} Copyright 2001 by the American Physical Society.}
  \label{fig:fig3}
\end{figure}
Importantly, again thanks to the semiconductor environment, different schemes are possible for the excitation of excitonic states in QDs. The highest purity of single-photons can be achieved in resonant excitation, where the excitation laser matches the QD emission wavelength \cite{Muller2007}. This scheme, however, requires a suppression of the scattered laser light of typically 6 orders of magnitude using e.g. cross-polarized excitation and collection paths \cite{Kuhlmann2013}, excitation from the side of the sample \cite{Ates2009a}, via coupling to photonic waveguides \cite{Makhonin2014} or by using dichroic excitation profiles \cite{He2019}. Alternatively, one can excite quasi-resonantly (also known as p-shell excitation) where the closest energy level above the emission state is resonantly excited to limit the excitation of excess charge carriers and non-radiative relaxation induced dephasing \cite{holewa2017}. The third possibility is non-resonant excitation with photon energies exceeding the band gap of the matrix material, which results in an excited excitonic reservoir, from which single excitons relax to the QD states and subsequently recombine radiatively emitting single photons. Non-resonant excitation is from a technical viewpoint the easiest excitation scheme, which however limits the single-photon quantum optical properties due to large dephasing, fluctuating charge traps and timing jitter \cite{Vural2020}.
\\
The physics of QDs, however, are by far not limited to single-photon emission. When exciting two (or more) excitons, cascaded emissions are possible, where the individual photons have slightly different energies (\textbf{Figure \ref{fig:fig4}}c) due to the Coulomb interaction of the confined excitons and can thus be separated \cite{Hu1990,Kulakovskii1999,Moreau2001,Rodt2003,Seguin2006,Sarkar2006}. Under certain conditions, photon pairs emitted by the biexciton-exciton radiative cascade reveal polarization-entanglement, as proposed theoretically in Ref. \cite{Benson2000} followed by several experimental demonstrations, e.g. \cite{Benson2000,Akopian2006,Young2006} (Figure \ref{fig:fig4} d-g). Noteworthy, also other peculiar configurations of the energy level scheme can be realized in this cascade, enabling the generation of polarization-entanglement via time-reordering \cite{Avron2008} or so-called twin photon states \cite{Heindel2017,Moroni2019}.
\\
Note that in order to obtain entangled photons from a QD, the so-called fine-structure-splitting (FSS), which quantifies the difference in energy of the photons emitted with different polarizations and which originates from intrinsic anisotropies in the QD, has to vanish (Figure \ref{fig:fig4} a, b). As it was shown, this FSS can be made zero via careful high quality growth, external strain or by applying magnetic fields \cite{Stevenson2006}. The possibility to generate polarization entanglement - being one crucial ingredient for many schemes of quantum communication - is a major advantage of using engineered QD devices. We refer the interested reader to Refs. \cite{huber2018semiconductor} and \cite{Schimpf2021} for a recent review and perspectives article on this topic, respectively.
\\
\begin{figure}
  \includegraphics[width=\linewidth]{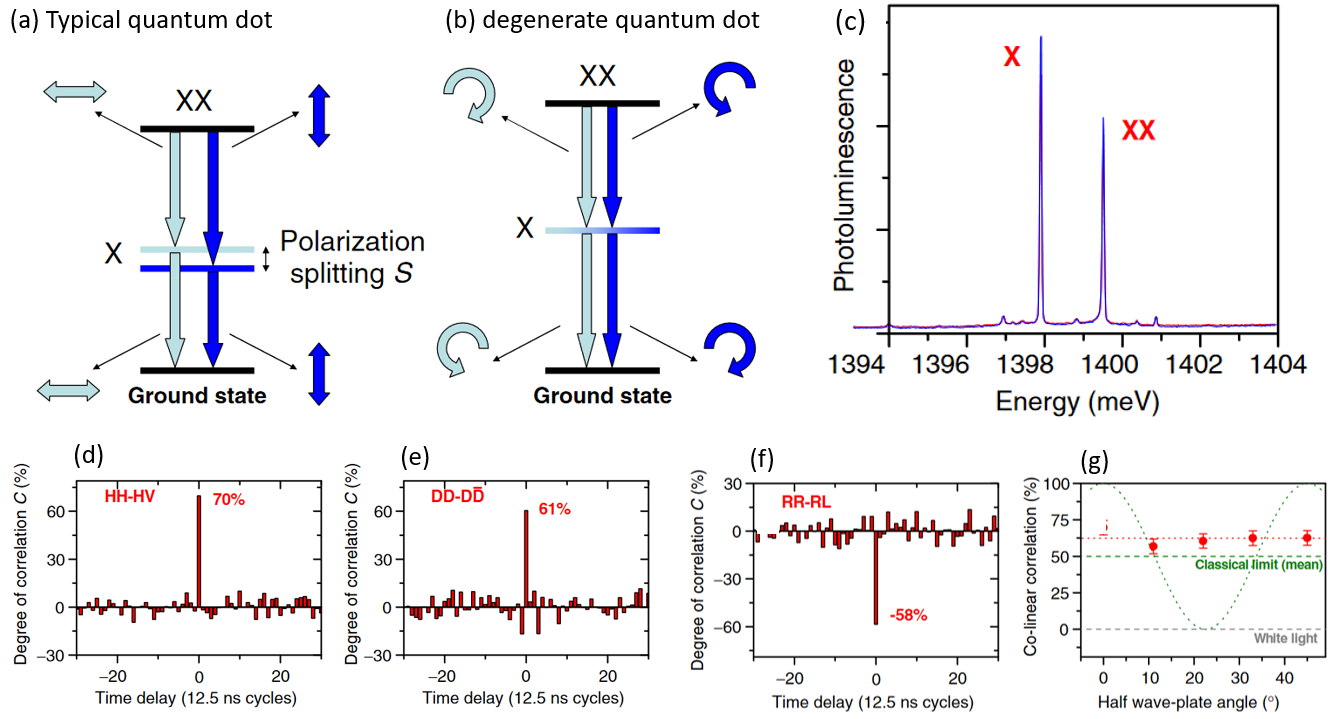}
  \caption{Generation of polarization entangled photon pairs from single QDs: By reducing the polarization splitting between the different decay paths of the biexciton-exciton emission cascade (from a to b), the ‘which-path’ information is erased, resulting in entangled photons. The emitted photons have different energies (c) which allows separating them. Looking at the polarization correlations in different bases confirms the entanglement (d-g). Figures reproduced from \href{https://iopscience.iop.org/article/10.1088/1367-2630/8/2/029}{\textit{Young et al. 2006}} \cite{Young2006} under Creative Commons BY license.}
  \label{fig:fig4}
\end{figure}
How QDs compare to other SPSs such as defects, carbon nanotubes, atomic-vacancy centers in crystals, or two-dimensional materials was reviewed by Aharonovich et al. \cite{Aharonovich2016}, with the result that QDs have the best overall performance due to the short lifetimes and the high purity and indistinguishability that are possible. There are also solid-state emitters such as crystal defects that emit directly in the Telecom range, even at room temperature \cite{Zhou2018}, but the photon purity is worse than what is possible in QD sources. Overall, semiconductor QDs are by now the closest there is to ideal quantum light sources. 

\subsection{Engineered Quantum Light Sources for Quantum Communication} \label{section2.2}
As mentioned in the previous section, QDs provide all assets to achieve fast emission of single, indistinguishable and entangled photons at high rates. To further boost their performances in terms of out-coupling efficiency and emission rate, solid-state QDs are commonly incorporated into photonic structures \cite{Barnes2002}. This section will provide an overview of the photonic structures used for QD-based SPSs employed in QKD experiments to be discussed in detail in section \ref{section3}.
\\
\begin{figure}
  \includegraphics[width= \linewidth]{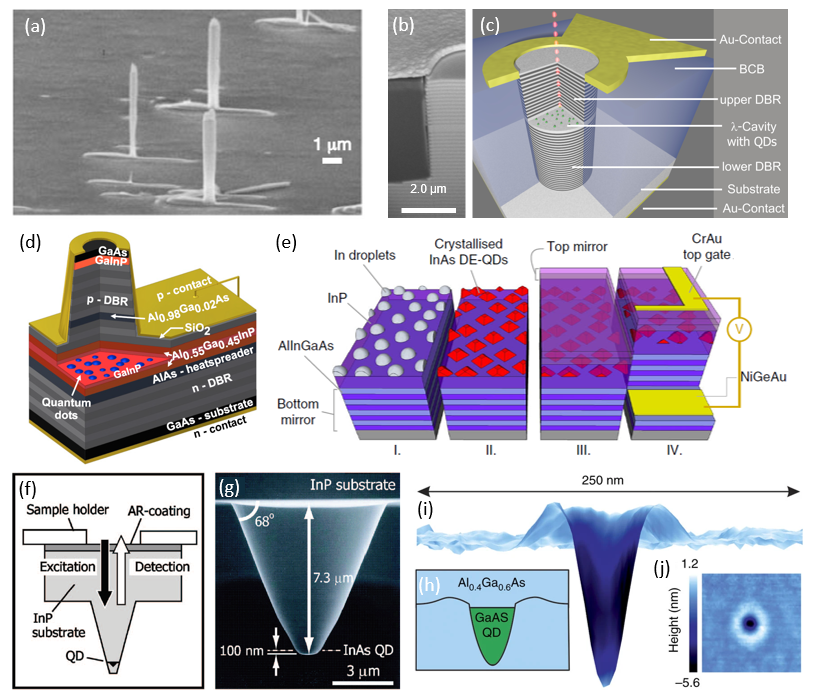}
  \caption{(a) SEM image of early QD micropillar cavities for optical excitation. (b,c) Cross-sectional SEM image and illustration of an single-photon light emitting diode (LED) based on electrically contacted p-i-n doped micropillar cavities with self-organized InAs/GaAs-QDs. (d) Schematic of an single-photon LED with embedded InP-based QDs. (e) Schematic fabrication process and design layout of an entanlged light emitting diode. (f,g) Schematic representation and SEM image of a QD embedded in an optical horn structure, whose emission is excited and collected through the backside of the substrate-wafer using an AR-coating. (h) Schematic representation of a GaAs quantum dot grown by the droplet etching method in an AlGaAs layer. (i) 3D cross-sectional atomic force microscopy measurement of a nanohole inside a AlGaAs layer after droplet etching. (j) Top view of the atomic force microscopy measurement in (i). (a) reprinted by permission from Springer Nature: \href{http://www.nature.com/articles/nature01086}{\textit{ Santori et al.}} \cite{Santori2002} Copyright 2002, (b,c) adapted from \href{http://aip.scitation.org/doi/10.1063/1.3284514}{\textit{Heindel et al. 2010}} \cite{Heindel2010} with the permission of AIP Publishing. (d) reprinted from \href{http://aip.scitation.org/doi/10.1063/1.3497016}{\textit{Reischle et al. 2010}} \cite{Reischle2010} with permission from AIP Publishing,  (e) reprinted from \href{http://dx.doi.org/10.1038/s41467-018-03251-7}{\textit{Müller et al. 2018}} \cite{Muller2018} under Creative Commons CC BY license. (f,g) reprinted from \href{http://aip.scitation.org/doi/10.1063/1.2723177}{\textit{Takemoto et al. 2007}} \cite{Takemoto2007} Copyright 2007 by American Institute of Physics. (h,i,j) reprinted from \href{https://www.nature.com/articles/ncomms15506}{\textit{Huber et al. 2017}} \cite{huber2017highly} under Creative Commons CC BY license.}
  \label{fig:fig5}
\end{figure}
Micropillar (or micropost) cavities, since their first application for the enhancement of QD-emission at the turn of the century \cite{Gerard1998,Solomon2001}, have been widely used and employed also for QKD experiments using optically-excited QD-based single-photon sources in the following chapter \cite{Waks2002a,Intallura2009}. A scanning electron microscope (SEM) image of micropillar cavities is shown in \textbf{Figure \ref{fig:fig5}a}. The etched pillars consist of a QD layer sandwiched between two distributed Bragg reflector (DBR) mirror sections. The top DBR mirror is usually designed to have a lower reflectivity than the bottom DBR mirror, to promote emission into the upper hemisphere towards a collecting optics. Purcell enhancements of a factor 5-6 of the emission rate and extraction efficiencies above $60\%$ can be typically achieved with micropillar structures \cite{Santori2002,Gazzano2013,Somaschi2016,Ding2016}.
\\
Micropillar structures were also used to realize electrically triggered QD-based single-photon sources emitting in the near-infrared (900 nm) \cite{Heindel2010} (cf. Figure \ref{fig:fig5} b,c) or visible (650 nm) \cite{Reischle2010} (cf. Figure \ref{fig:fig5}d) spectral range. These highly engineered devices have in turn be employed for the first QKD experiments with electrically injected QD-devices \cite{Heindel2012}. In both device approaches a layer of self-organized QDs is embedded in an undoped (intrinsic) section of a micropillar structure sandwiched between a top, p-doped and bottom, n-doped DBR mirror forming a p-i-n diode structure electrically contacted via gold bars \cite{Boeckler2008}. The operation wavelength can thereby be determined by the choice of the QD-material, being InAs/GaAs and InP/GaAs in case of the device emitting in the near-infrared and visible spectral range, respectively, highlighting the flexibility QDs offer for application in quantum technologies. Electrically driven micropillar-based single-photon sources were reported to reach overall efficiencies (including electrical loses) exceeding $60\%$ \cite{Schlehahn2016a} and Purcell enhancement of close to 5 \cite{Heindel2010}, values which are similar to non-electrical counterparts but, in contrast, excitation pulse-repetition rates in the GHz-range are straighforward to achieve \cite{Hargart2013,Schlehahn2016a}.
\\
In another approach, QDs were embedded in diode structures to electrically generate polarization entangled photon pairs via the biexciton-exciton radiative cascade (cf. section \ref{section2.1}) \cite{Salter2010,Muller2018}. This so-called entangled light emitting diode (ELED) has later been employed for the first entanglement-based QKD-experiments using QD-devices \cite{Dzurnak2015}. The fabrication scheme to realize InP-based ELEDs for entangled photon emission in the telecom C-Band is shown in Figure \ref{fig:fig5}e.
\\
A different type of photonic structure that was used for QKD experiments in the telecom C-Band \cite{Takemoto2010} is an so-called optical horn  (cf. Figure \ref{fig:fig5}f). It consists of a QD embedded in a fabricated cone on a substrate with the horn acting as a reflector to direct photons upwards through the substrate with an antireflection coating, towards the collecting optics. A SEM image of a fabricated optical horn structure is shown in Figure \ref{fig:fig5}g. The horn structure does not show Purcell enhancement, but photon collection efficiencies of close to $11\%$ were achieved \cite{Takemoto2007}.
\\
The last type of QD-device that shall be mentioned here was used for a recent implementation of entanglement-QKD experiments \cite{Basset2021,Schimpf2021a} and utilized optically-pumped symmetric GaAs QDs grown by the droplet-etching method, which show short radiative lifetimes and small fine-structure splittings enabling large entanglement fidelities \cite{huber2017highly}. A schematic representation of such a symmetric GaAs QD situated in a hole in an AlGaAs-matrix is shown in Figure \ref{fig:fig5}h. A three-dimensional atomic force microscopy cross-section image is shown in Figure \ref{fig:fig5}i, with the top view of the measurement in Figure \ref{fig:fig5}j. Here, the QDs were additionally combined with solid immersion lenses to increase their out-coupling efficiency to about $8\%$.
\\
Other photonic structures to enhance the performance of solid-state QDs include nanowires \cite{Claudon2010}, photonic crystal cavities \cite{Madsen2014,Kim2016}, circular Bragg gratings \cite{Davanco2011,Liu2019,Wang2019}, open cavity systems \cite{Tomm2020}, on-chip waveguide based structures \cite{Uppu2020}, and monolithic microlenses \cite{Gschrey2015}. The latter proved to be useful for the development of practical plug\&play single-photon sources \cite{Schlehahn2018,Musial2020} as well as tools for the performance optimization of single-photon QKD \cite{Kupko2020}, both evaluated very recently in a QKD-tesbed operating at O-band wavelengths \cite{Kupko2021} (see section \ref{section4} for details).

\section{Realizations of Quantum Key Distribution using Quantum Dots}\label{section3}
In  the previous section, we introduced QD-based quantum light sources as promising candidates for applications in quantum information. In this section, we review the implementations of QKD based on respective QD-devices reported to date. Noteworthy,  also other types of quantum emitters have been used successfully for QKD experiments with sub-Poissonian light states.  Among the two very first demonstrations of the BB84 protocol with single photons, one of them used Nitrogen vacancy (NV) centers in diamond to create the single photons \cite{Beveratos2002}. Also later, NV and Silicon vacancy (SiV) -centers were used to implement QKD protocols \cite{Alleaume2004,Leifgen2014}, but their secure key rates were smaller since excited states in NV centers have higher radiative lifetimes. In this review we restrict ourselves however to QD-based implementations. 
\\
Concerning the QKD scheme, there are two major groups of protocols for which QD sources are used. As discussed in section \ref{section1}, QDs can either be used for prepare-and-measure type QKD (like in the BB84 protocol) or the QD source can be placed in between Alice and Bob creating polarization entangled photon pairs distributed to Alice and Bob (like in the Ekert protocol). Let us begin by discussing BB84-like implementations of sub-Poissonian QKD.

\subsection{Quantum Key Distribution using Single Photons} \label{section3.1}
The very first implementation of single-photon QKD by Waks et al. dates back to 2002 \cite{Waks2002a}. The authors implemented the BB84 protocol with the bits being polarization-encoded in single-photon states from a triggered QD source \cite{Santori2002}. Here, InAs QDs were encapsulated in micro-pillar cavities made from distributed Bragg reflectors (such as the one shown in Figure \ref{fig:fig5}a). In their setup (\textbf{Figure \ref{fig:fig6}a}) the QD sample was kept in a liquid Helium cryostat and was optically excited with a pulsed Ti:Sapphire laser in a reflection configuration. The average single photon rate was measured by sending a part of the photon stream to a detector, while the polarization of the rest was modulated using an electro optical modulator (EOM), after selecting a fixed polarization in the polarizing beam splitter (PBS). The non-resonant optical excitation with a rate of 76 MHz led to an average rate of $\mu = 0.007$ photons/s injected into the quantum channel. The photons were then sent to Bob via a free-space link with variable attenuation, where the polarization state was measured. The beam splitter (BS) realized a random basis choice and a time interval analyzer (TIA) was used for synchronization.
\\
By this, they measured bit rates and QBER for different attenuations, from which they could calculate the asymptotic key rate values shown as the red dots (Figure 6b), with the green line being the calculated values. In their work, they also introduced an upper bound on the probability for two-photon emission of SPSs of $P_{2, S P S} \leq \frac{1}{2} \mu^{2} g^{(2)}(0)$, where $g^{(2)}(0)$ is the auto-correlation value at zero delay measured in a Hanbury-Brown andTwiss (HBT) experiment \cite{Brown1956}, which is a measure for the probability of two photons being emitted into the same pulse. By calculating a simple, asymptotic secure key-rate \cite{Waks2002} with a measured QBER of $2.5\%$ they obtained a maximum communication rate without channel losses of 25 kbit/s. By introducing an additional loss, they found a maximum tolerable channel loss up to which communication is possible, i.e. positive key rate, of 28 dB.
\\
Repeating the same experiment with attenuated laser pulses,without applying decoy states at that time, with an average photon number small enough to have the same multi-photon probability as the QD source, an experimental and calculated key rate was obtained. While for small losses the laser gave a higher communication rate, the QD single photons outperformed the laser at larger losses, since compensating potential photon number splitting attacks used up more bits for the attenuated laser pulses \cite{Lutkenhaus2000}. Ultimately the QD could tolerate about 4 dB higher losses than the laser without decoy states. Using their secure bits, Waks et al. encoded an image of Stanford University Memorial’s Church and decode it again using the transmitted secure key (Figure \ref{fig:fig6}c). \\
Note however, that nowadays decoy-state protocols allow for the in-situ estimation of the multi-photon contribution to mitigate photon-number-splitting attacks and hence much higher average photon numbers in the laser pulses \cite{Wang2005}, which is why the asymptotic rate of Waks et al. would not beat a decoy-state implementation with WCPs. On the other hand, the upper-bound on the probability of multi-photon events used by Waks et al. for their QD-source is not tight, and that, following recent discussions, the measured  value is a too pessimistic measure for the purity of the photon source \cite{Grunwald2019}. Hence, the true probability for a multi-photon emission event of a QD-SPS is likely to be even lower. This further increases the advantage sub-Poissonian SPSs can have over attenuated laser pulses in implementations of QKD.
\\
\begin{figure}[h]
  \includegraphics[width=0.5 \linewidth]{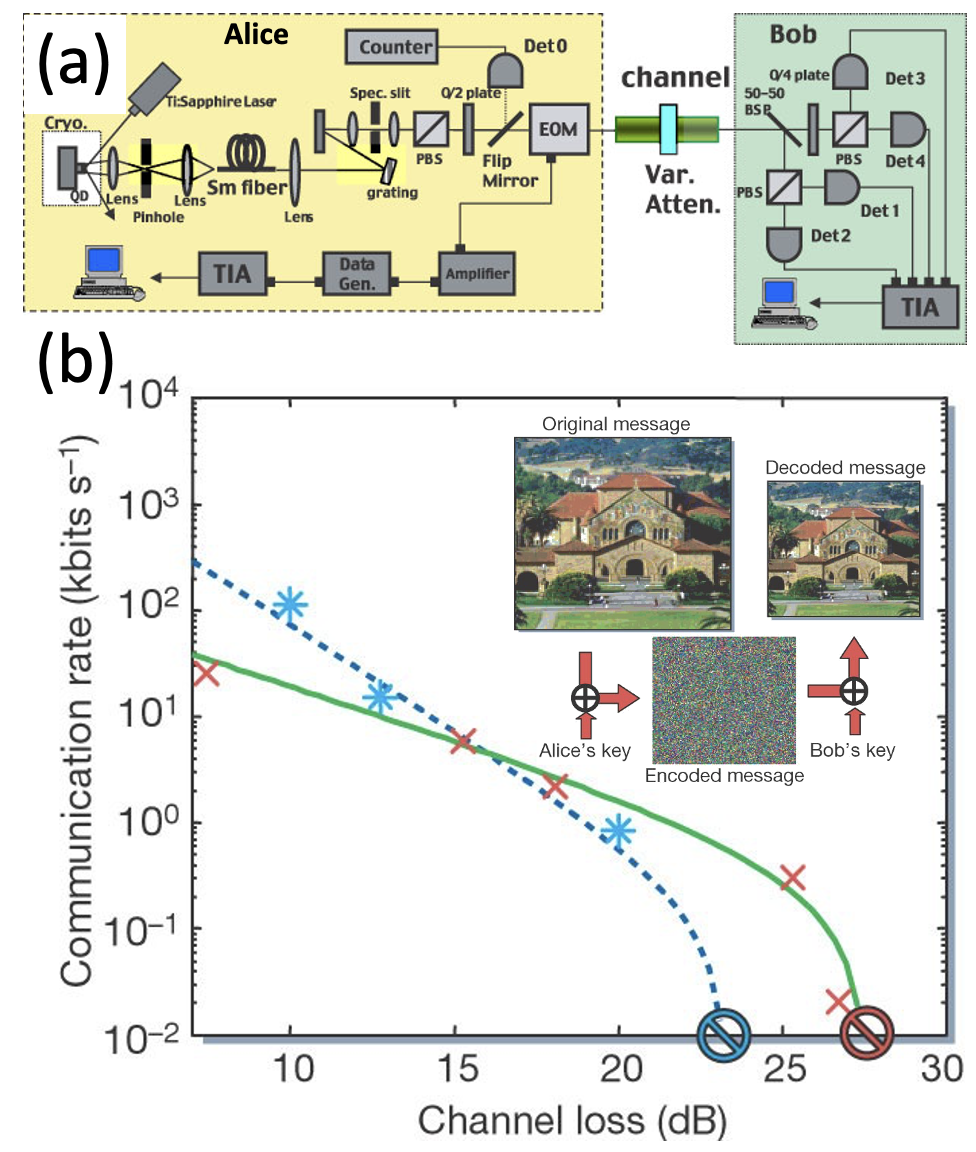}
  \caption{Sketch of the setup of the first QD single photon QKD experiment by Waks et al. (a), the measured and calculated asymptotic secure key rates for the QD single photons (red crosses) and attenuated laser pulses (blue stars) (b) and the image that was securely transmitted (c). Reprinted from \href{https://www.nature.com/articles/420762a}{\textit{Waks et al. 2002}} \cite{Waks2002a} with permission of Springer Nature: Copyright 2002 Springer Nature.}
  \label{fig:fig6}
\end{figure}
An appealing scheme to improve secure key rates in BB84-QKD for a given QD-source was presented bei Aichele et al. in 2004 \cite{Aichele2004}. Here, the cascaded emission of biexciton and exciton state was used to generate two single-photons at different energies from one excitation pulse, which was used to effectively double the achievable key rate. Using a Michelson-interferometer to introduce a temporal delay between both photons, each photon’s polarization could be modulated individually (time-multiplexing) on Alice’s side before coupling them to the same free-space quantum channel. At Bob, a second Michelson interferometer was used to separate the photons again and measure their polarization separately. Note that the separating and combining of the two photons is done in the time domain instead of separating them by energy to make the setup less vulnerable to emission energy fluctuations. Using this approach, Aichele et al. demonstrated a rate of secure bits per pulse of $5 \cdot 10^{-4}$, which results in a communication rate of 38 kbit/s at the given laser repetition rate (76 MHz).
\\
While these first implementations used free-space optical (FSO) links as quantum channel, which is ideal for achieving low losses at large distances in air-ground \cite{Nauerth2013} or space-ground \cite{Bedington2017,Yin2020,Sidhu2021} link scenarios, the use of optical fibers as quantum channels has many practical advantages for ground-based communication scenarios. Besides the fact that no direct line of sight is required in this case, which are susceptible to weather conditions and atmospheric turbulence, the technology can be made compatible with the world-wide fiber-based communication infrastructure. Several QD-based QKD experiments have been reported employing single photons coupled to an optical fiber acting as a quantum channel. Collins et al. were the first to report on a fiber-based QKD experiment using single photons generated by a QD emitting at a wavelength of 900 nm, sent through 2 km of optical fiber \cite{Collins2010}.
\\
Since optical fibers provide lowest transmission losses at wavelengths in the second and third Telecom window, i.e. O- and C-band, it is beneficial to use QDs operating at these wavelengths, such as the ones first fabricated by Ward et al.\cite{Ward2005}. Incorporating such QDs into micropillar cavities, Intallura et al. demonstrated single-photon QKD at 1300 nm in 2010, optically exciting the QD-device above bandgap \cite{Intallura2009}. The quantum channel was represented by a standard SMF-28 optical fiber of 35 km length. As mentioned in section 1, it can be difficult to maintain a certain polarization state over a long fiber transmission. For this reason Intallura et al. used a phase encoding scheme, which also employed a multiplexed reference laser to match the path differences in Alice’ and Bob’s MZIs (\textbf{Figure \ref{fig:fig7}a}). For the QKD demonstration, the entire system ran at a clock rate of 1 MHz, limited by the single photon detectors response and dead time. Using the GLLP \cite{Gottesman2004}, which is asymptotic, but incorporates multi-photon events (see also section 3.3), a measured QBER of $5.9\%$ and an error correction efficiency of 1.17, the authors calculated a maximum secure key-rate of about 160 bit/s and managed to achieve positive key rates at a distance of 35 km, overcoming the maximum distance achieved by WCPs (without decoy states) in their setup (Figure \ref{fig:fig7}b). \\
\begin{figure}[h]
  \includegraphics[width=0.5 \linewidth]{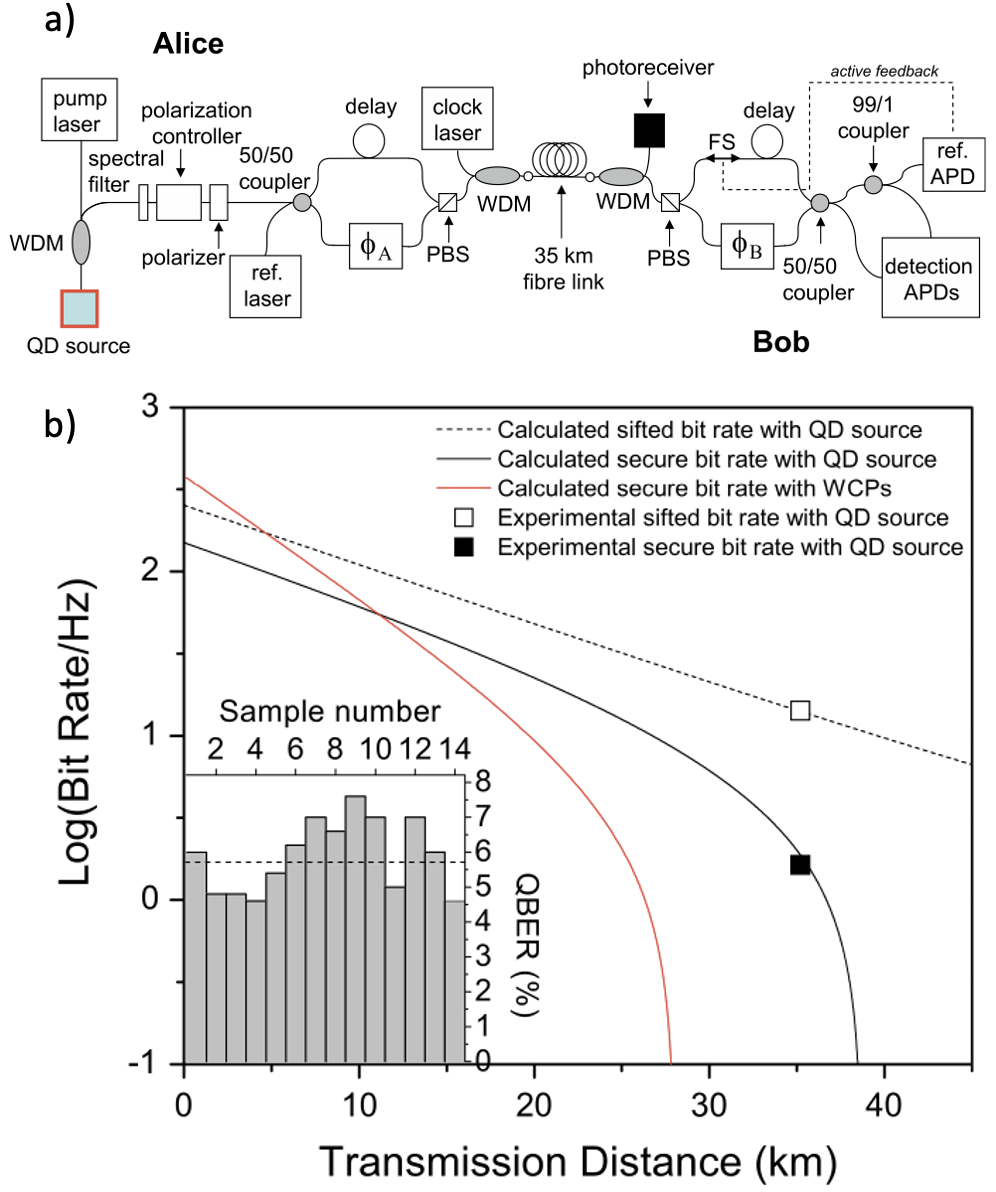}
  \caption{First QKD demonstration at Telecom wavelengths in 2009 by Intallura et al. Using a phase-encoding scheme (a), they achieved QKD over a 35 km distance (b) via optical fiber transmission. ©IOP Publishing. Figure reproduced with permission from \href{https://iopscience.iop.org/article/10.1088/1464-4258/11/5/054005}{\textit{Intallura et al. 2009}}  \cite{Intallura2009}. All rights reserved.}
  \label{fig:fig7}
\end{figure}
Only one year later, Takemoto et al. presented the first implementation of single-photon QKD at C-band wavelength (1560 nm), benefitting from even lower losses in optical fibers \cite{Takemoto2010}. Using a QD  incorporated into horn-like structure (depicted in Figure \ref{fig:fig5}f) \cite{Takemoto2007} and a QKD setup for phase encoding (see \textbf{Figure \ref{fig:fig8}}a for the setup), the authors achieved a maximum secure communication distance of 50 km, calculated with the asymptotic GLLP rate equation, setting a new record for single-photon QKD at that time.
\\
A few years later, the same group improved their QKD-implementation further, by using better detectors and QDs of higher quality. At close to maximum distance, e.i. strong channel losses, the number of detected signal photons is on the order of the dark counts of the detectors, ultimately limiting the achievable communication distance. Using single-photon detectors based on superconducting nanowire \cite{Hadfield2005}, the dark count contribution could be significantly reduced and hence the maximum achievable distance increased. Moreover, the authors also improved the single-photon purity of their QD source, achieving an auto-correlation value at zero time delay of only $g^{(2)}(0) = 0.005$, so that the necessary corrections for multi-photon emission events, which would allow photon number splitting, became much smaller. With these improvements Takemoto et al. achieved with 120 km the so far longest distance in fiber-based BB84 QKD using a sub-Poissonian SPS in 2015 \cite{Takemoto2015} (cf. Figure \ref{fig:fig8}b).
\\
\begin{figure}[h]
  \includegraphics[width=0.5 \linewidth]{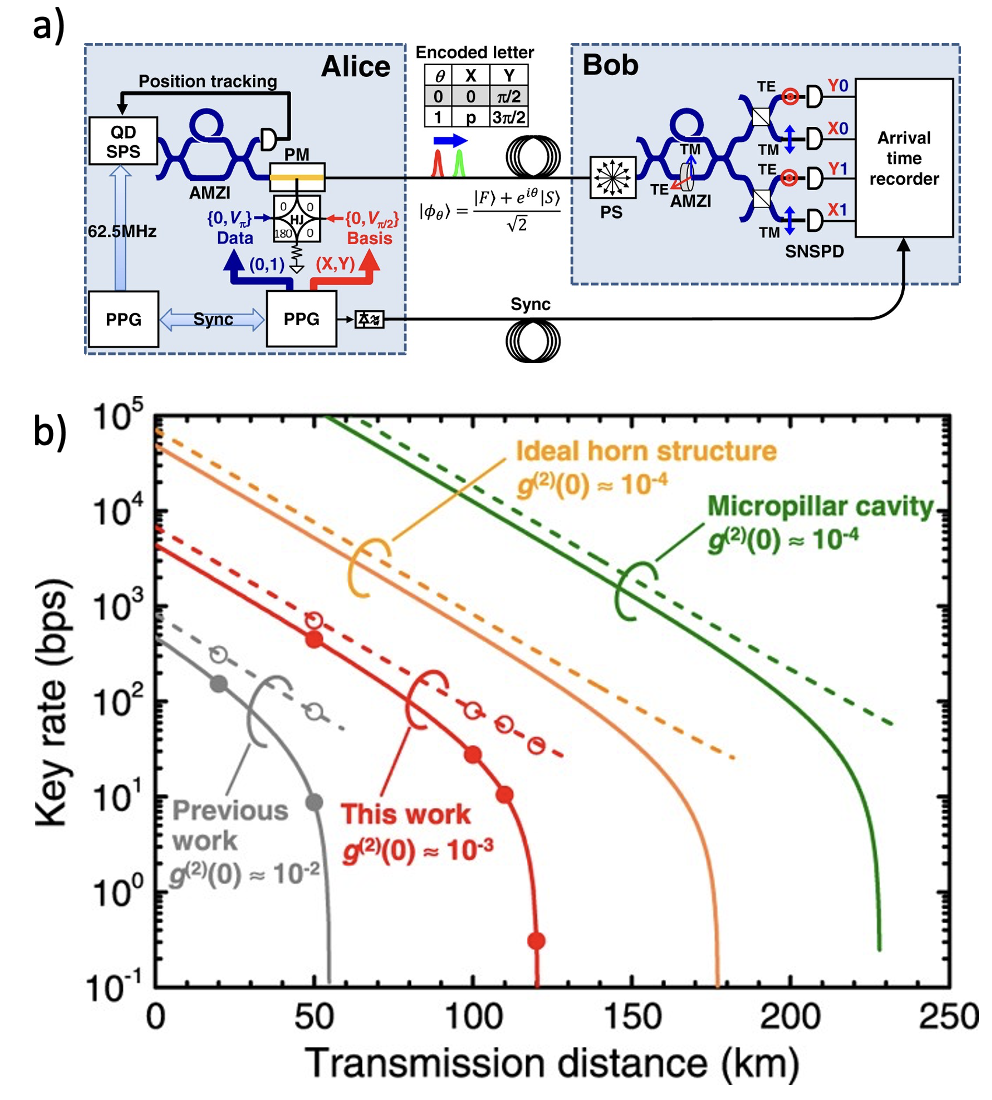}
  \caption{Implementation of BB84 QKD protocol using single photons at 1550 nm created by InAs QDs in horn-like structures. Using a phase-encoding scheme (a), 50 km QKD was achieved (grey line in b) \cite{Takemoto2010} and later in an improved version of the experiment, that used Superconducting Nanowire Single Photon Detectors, more than 120 km were possible (red line in b), while improved structures promise even longer distances (other lines in b) \cite{Takemoto2015}. Figures reproduced from \href{http://www.nature.com/articles/srep14383}{\textit{Takemoto et al. 2015}} \cite{Takemoto2015} under Creative Commons CC BY license.}
  \label{fig:fig8}
\end{figure}
In all QKD experiments presented so far, the QD-devices were excited optically using pulsed laser systems. A major advantage of semiconductor-based quantum light sources, however, is the possibility to realize complex engineered devices including diode structures for the injection of electrical charge carriers, which in turn allows for electrical triggering of QD emission. This is highly beneficial for applications, not only because higher degrees of device integration become possible (bulky laser systems become obsolete), but also the clock rate of implementations of quantum cryptographic protocols are easily adjustable (see also section 4.1). While the first electrically injected QD-based SPS was reported already in 2002 by Yuan et al. \cite{Yuan2002}, it took one decade until the first QKD experiments could be realized. In the work by Heindel et al. in 2012 \cite{Heindel2012}, three research groups joined forces to demonstrate lab-scale BB84-QKD experiments with two different types of single-photon light emitting diodes emitting in the near infrared (NIR) and visible (VIS) spectral range, at 897 nm and 653 nm, respectively (see \textbf{Figure \ref{fig:fig9}}). Using engineered QD-devices based on different material systems and growth techniques, their work highlighted the flexibility semiconductor-based quantum light sources offer for implementations of quantum information. While the NIR-SPS was based on an electrically contacted micropillar cavity exploiting the Purcell effect to enhance the photon extraction efficiency \cite{Heindel2010}, the QDs were integrated in a quasi planar DBR cavity structure in case of the VIS-SPS \cite{Reischle2010} (cf. Figure \ref{fig:fig5}b). Using the NIR-SPS, the authors achieved sifted key rates of 27.2 kbit/s (35.4 kbit/s) at a QBER of $3.9\%$ ($3.8\%$) and a g(2)(0) value of 0.35 (0.49) at moderate (high) excitation under pulsed current injection at a clock-rate of 182.6 MHz. The VIS-SPS was triggered at 200 MHz, delivering sifted keys at a rate of 95.0 kbit/s at a QBER of $4.1\%$ and a g(2)(0) value of 0.49. While both the achieved suppression of multi-photon events as well as the key rates left room for future improvements, these first proof of principle QKD experiments using electrically operated semiconductor single-photon sources can be considered as a major step forward in photonic quantum technologies. Shortly after the lab-scale QKD experiments reported in 2012, the authors integrated the NIR-emitting SPS in a, at that time, compact quantum transmitter setup to be employed for QKD field experiments in downtown Munich (see Figure \ref{fig:fig9}d). As reported by Rau et al. \cite{Rau2014}, the QKD experiments comprised a 500 m FSO link between two buildings of the Ludwigs-Maximilians-Universität Munich, with the transmitter and receiver units synchronized via GPS-disciplined oscillators. Using their single-photon light emitting diode modulated at a clock-rate of 125 MHz, the authors achieved sifted key rates of 7.4 kbit/s (11.6 kbit/s) at a quantum bit error ratio of $7.2\%$ ($6.3\%$) and a g(2)(0) value of 0.39 (0.46) at low (moderate) excitation.
\\
\begin{figure}
\centering
  \includegraphics[width= 0.75 \linewidth]{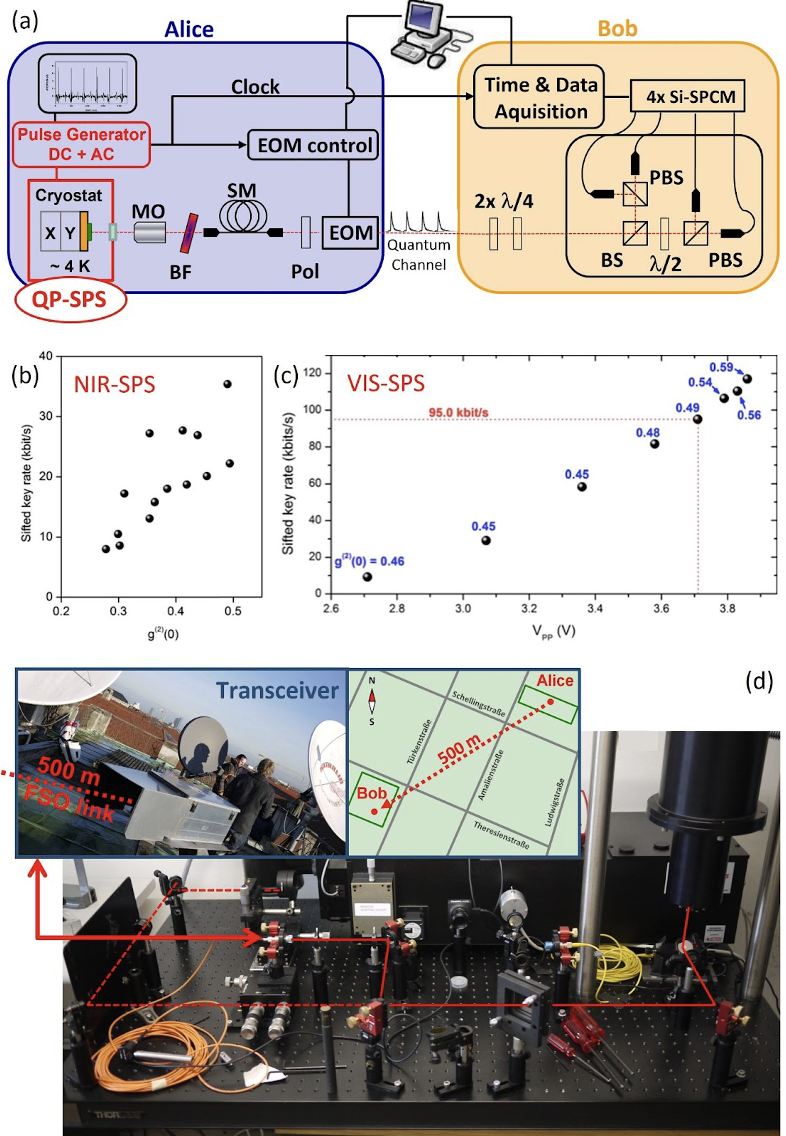}
  \caption{First QKD experiments with two electrically triggered QD single photons emitting at 900 nm (InAs) and 650 nm (InP). The setup in (a) was used to measure sifted key rates as well as photon purity under different excitation conditions for both QD structures (b,c) \cite{Heindel2012}. (d) In a second experiment, the single-photon emitting diode emitting at 900 nm was employed for field-experiments in down-town Munich using a 500 m free-space optical (FSO) link connecting two building of the Ludwigs-Maximilians-University Munich  \cite{Rau2014}, (a-c) reproduced with permission from \href{https://iopscience.iop.org/article/10.1088/1367-2630/14/8/083001}{\textit{Heindel et al. 2012}} \cite{Heindel2012} © IOP Publishing and Deutsche Physikalische Gesellschaft. Reproduced by permission of IOP Publishing.}
  \label{fig:fig9}
\end{figure}
In order to become compatible with attenuated laser systems using decoy-state protocols, the efficiency and clock-rate of electrically contacted QD SPSs still has to be significantly increased. Promising steps in this direction were reported by Schlehahn et al. with achieved photon extraction efficiency of single-photon light emitting diodes of up to $61\%$ (into the first lens) and trigger rates in the GHz range \cite{Schlehahn2016a}. A promising route to improve also the $\mu$ inside the quantum channel is a tighter integration of the SPSs, e.g. by the  direct coupling to optical fibers allowing for practical plug-and-play quantum light sources (see section \ref{section4.1}). Very recently, Kupko et al. evaluated for the first time the performance of a state-of-the-art plug-and-play SPS operating at O-band wavelengths for BB84-QKD \cite{Kupko2021}.

\subsection{Quantum Key Distribution using Entangled Photon Pairs} \label{section3.2}
The implementations discussed in the previous section were all based on the BB84 protocol in a prepare-and-measure configuration. In this section we will review QKD experiments using entangled photon states reported to date. By each measuring the two photons in a random basis and then keeping only the results in which they had used the same basis, they obtain perfectly correlated bit strings and by quantifying the remaining degree of entanglement for instance via violations of the Bell-type CHSH inequality \cite{Clauser1969}, they can uncover eavesdropping attempts as described in the Ekert protocol \cite{Ekert1991} explained in section 1. Entanglement monogamy guarantees that if Alice’s and Bob’s photons are maximally entangled, then the system cannot be entangled with any other system, hence an adversary’s state is separable from the state of Alice and Bob, thus the adversary cannot have any information. From the degree of deviation from a maximum violation, the amount of necessary privacy amplification can be deduced.
\\
Two important questions must be answered in entanglement-based implementations. How is the entanglement created and how is it distributed? QDs provide an excellent source of entangled photon pairs, via the biexciton-exciton emission cascade, as explained in section 2. Since the photons obey single-photon statistics, higher generation rates of entangled photons are possible than with SPDC sources \cite{Chen2018,Wang2019,Liu2019}. In case of SPDC sources, the fidelity decreases significantly above pair emission efficiencies of 0.1, due to the emission of multiple pairs, which reduces the purity of the emitted states. The distribution of the entangled photons works like the distribution of single photons. They are either distributed via free-space links or via optical fibers which should maintain their polarization states.
\\
The first proof-of-concept demonstration of QD-based entanglement QKD was reported by Dzurnak et al. \cite{Dzurnak2015} in 2015. The authors performed an in-lab experiment using entangled photons generated via an electrically driven QD-device, referred to as an entangled-light emitting diode (ELED) as introduced first by Salter et al. \cite{Salter2010} (similar to Figure \ref{fig:fig5}c). The photon pairs were distributed in optical fibers to two detectors with random basis choices (\textbf{Figure \ref{fig:fig10}a}). Since the entangled photons emitted from a QD via the biexciton-exciton cascade have slightly different energies, they could be spectrally separated to send them to different receivers (Figure \ref{fig:fig10}b). By this, they were able to transfer a total of 2000 secure bits. To prove that photons emitted during the same excitation pulse were entangled, the violation of the CHSH equation was tested, as indicated by the S-parameter exceeding 2 for vanishing time delays in Figure \ref{fig:fig10}c. The authors obtained 1 MHz of photon counts on each detector, which, due to tight temporal filters, resulted in 10 sifted bits of key per minute. \\
\begin{figure}
  \includegraphics[width= \linewidth]{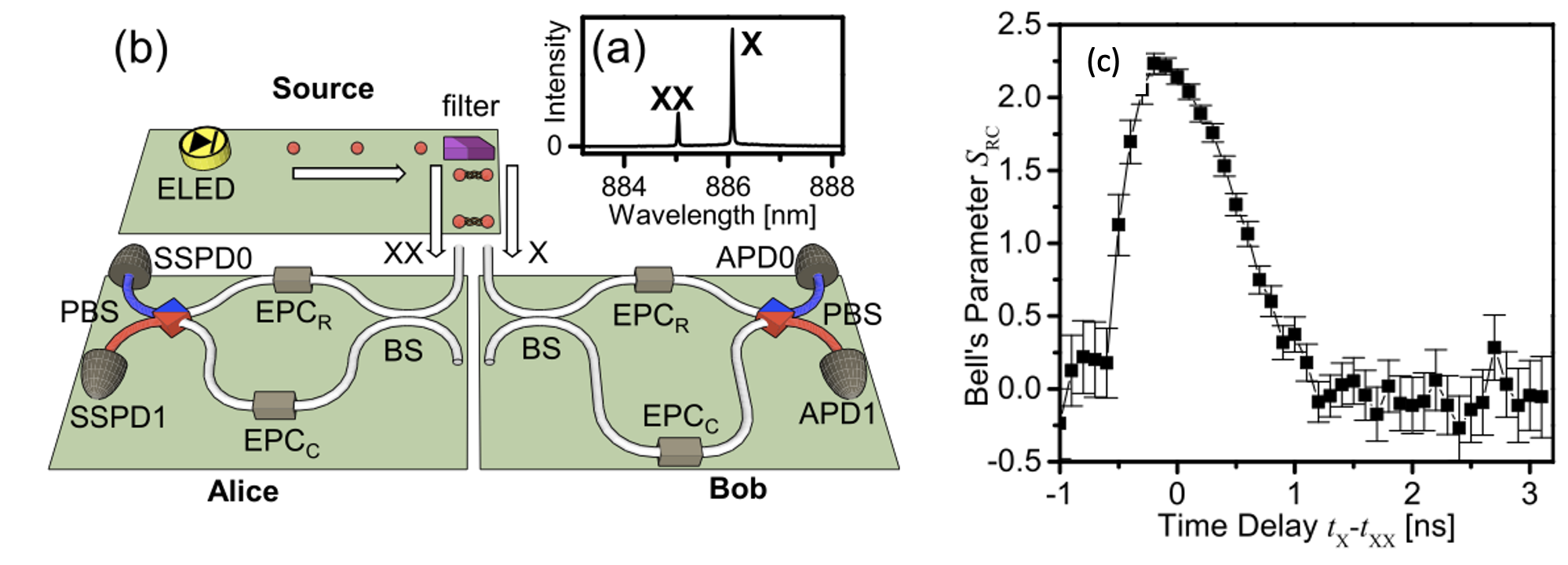}
  \caption{First realization of entanglement based QKD with entangled photons from a single, electrically excited QD presented by Dzurnak et al. Since the photons have different wavelengths (b) they could be spectrally separated and distributed to the receivers (a) where the preserved entanglement was measured via the CHSH equation violation, as shown by the S-parameter $>$ 2 (c). Reprinted from \href{http://aip.scitation.org/doi/10.1063/1.4938502}{\textit{Dzurnak et al. 2015}} \cite{Dzurnak2015}, with the permission of AIP Publishing.}
  \label{fig:fig10}
\end{figure}
Recently, two other successful implementations of entanglement based QKD were reported back-to-back, that both used the same type of optically excited QD source emitting at 785 nm (embedded between a bottom DBR and a top solid immersion lense). Importantly, and in contrast to all previous QKD-implementations, both groups used a coherent excitation scheme, i.e. they optically excited the biexciton state of the quantum emitter via two-photon resonant laser pulses. While this is not required per se for the implementation of QKD summarized in the following, it is nevertheless an important step towards quantum repeaters and other advanced schemes of quantum information, relying on high photon indistinguishabilities. 
\\
In the implementation reported by Schimpf et al. the photons were transmitted via a 350 m optical fiber, resulting in an asymptotic secure bit rate of 86 bits/s \cite{Schimpf2021a}. Basset et al. realized both a fiber link and a free-space link, allowing for a direct comparison of both channel types operated with the same source \cite{Basset2021} (\textbf{Figure \ref{fig:fig11}a}). The authors found that for their communication distance of 250 m, the fiber link enabled more stable conditions for entanglement distribution, resulting in the observation of a larger Bell parameter showing less fluctuations (Figure \ref{fig:fig11}b). This led to a higher communication rate of about 500 secure bits/s, compared to about 100 secure bits/s in the free-space link (Figure \ref{fig:fig11}c), the main reason being the difficulty of correcting instabilities and drift in the free-space optics, as manifested in the higher QBER in the free-space channel (Figure \ref{fig:fig11}d). The authors counteracted distortions of the polarization state propagating in the fiber,  by actively monitoring and compensating for the change in polarization during the experiment.
\\
Interestingly, the two groups used slightly different protocols for entanglement QKD. Basset et al. \cite{Basset2021} implemented an asymmetric version of the original Ekert protocol. For this, the authors used a subset of the transmitted bits to evaluate the violation of the CHSH and quantified the amount of entanglement left after the transmission of the photons. Hereby, they determined the amount of eavesdropping that could have occurred and the amount of privacy amplification that was necessary. To do so, they measured the photons in the basis set, which is known to maximally violate the CHSH inequality, but only on Alice’s side, while Bob measured them in the conventional BB84 bases. This asymmetric approach reduced the number of necessary detectors.
\\
Schimpf et al. \cite{Schimpf2021a}, on the other hand, implemented an entanglement-based version of the BB84 protocol known as BBM92 \cite{Bennett1992}. Here, Alice and Bob measure their respective halves of the entangled state in two conjugate bases and the amount of necessary privacy amplification is determined solely from evaluating the deviations of a subset of results measured and compared by Alice and Bob (as in the BB84). Thus, the amount of entanglement is not actively monitored and no Bell-like inequality violation is measured.
\\
\begin{figure}
  \includegraphics[width= \linewidth]{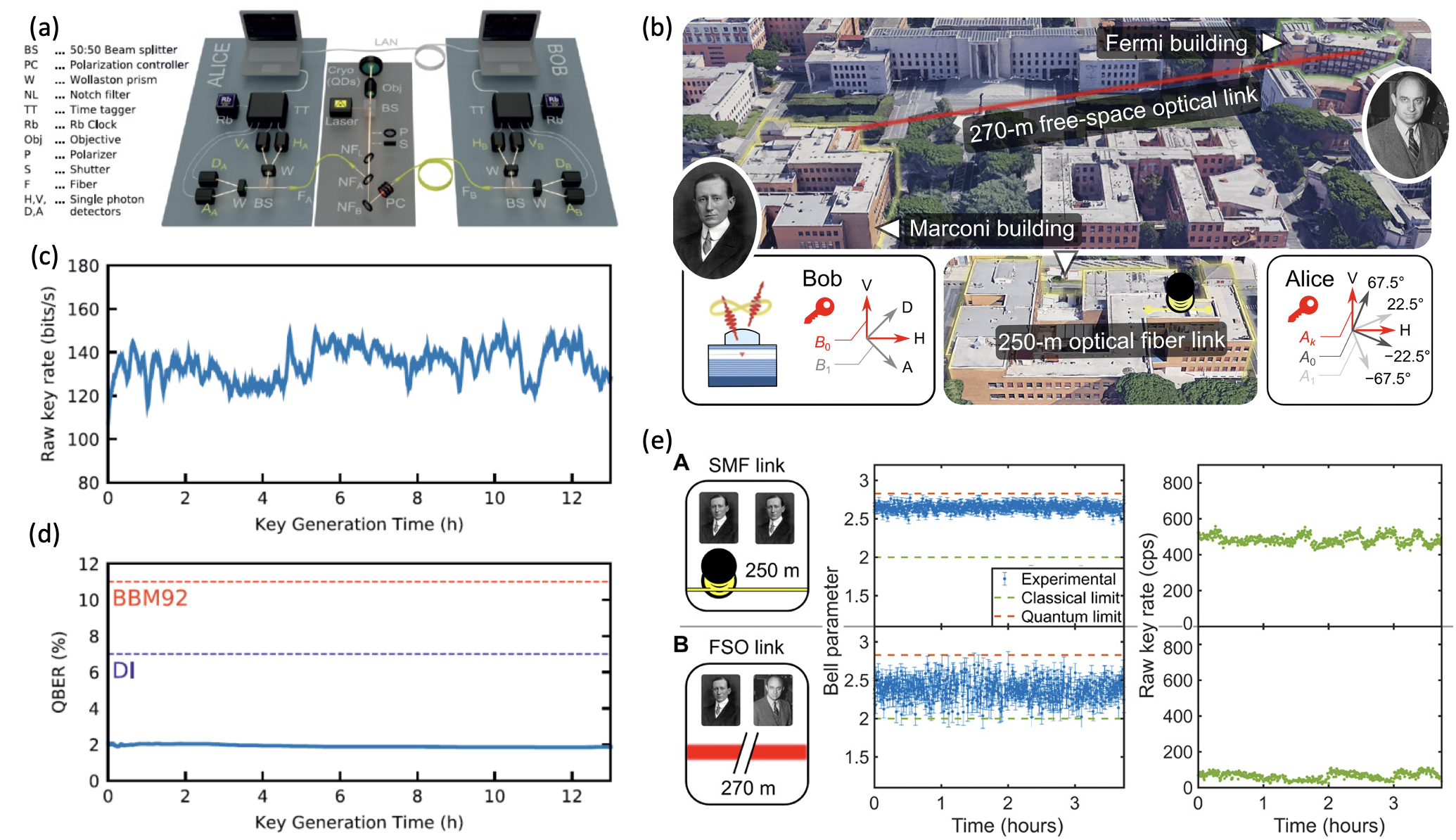}
  \caption{Entanglement-based QKD experiments: Schimpf et al. realized the BBM92 QKD protocol \cite{Schimpf2021a}. Their setup (a) allowed them to achieve secure key transmission with the rate shown in (c) and a QBER (d) far below the maximum allowed value for secure transmission. Basset et al. used an asymmetric Ekert protocol \cite{Basset2021} to perform QKD between two buildings (b) and the photons were transmitted both through an optical fiber (top panel in e) and a free-space (bottom panel in e) and the degree of entanglement (second column in e) and the key rate (third column in e) were compared. (a,c,d) reproduced from \href{https://advances.sciencemag.org/lookup/doi/10.1126/sciadv.abe8905}{\textit{Schimpf et al. 2021}} \cite{Schimpf2021a} under Creative Commons Attribution License 4.0, (b,e) reprinted from \href{https://advances.sciencemag.org/lookup/doi/10.1126/sciadv.abe6379}{\textit{Basset et al. 2021}} \cite{Basset2021} © The Authors, some rights reserved; exclusive licensee AAAS. Distributed under a \href{http://creativecommons.org/licenses/by-nc/4.0/}{CC BY-NC 4.0 license.}}
  \label{fig:fig11}
\end{figure}
Interestingly, QKD has so far not been implemented with entangled photons generated by QDs at Telecom wavelengths. The generation of entangled photons at a wavelength of 1550 nm from a QD SPS has already been shown in experiments by Olbrich et al. \cite{Olbrich2017}. While in their experiment the quantum emitter was optically excited with a continuous-wave laser, Shooter et al. recently realized a pulsed electrically excited QD-source generating entangled photon pairs at GHz clock rate in the Telecom C-band \cite{Shooter2020}. The entangled photon pairs showed a fidelity to the maximally entangled state of $89\%$  and were distributed over a fiber link of 4.6 km length at record rates, representing an important step towards high-performance QKD systems exploiting sub-Poissonian entanglement sources. Additionally, using electrically excited InAs-QDs, the same group reported already entanglement generation via the biexciton-exciton cascade up to a temperature of 93 K \cite{Muller2018} - a temperature not requiring liquid Helium anymore and allowing for the integration into compact Stirling cryocoolers \cite{Schlehahn2015} (see also section 4.1). \\
In another experiment Xiang et al. achieved $91\%$ fidelity of polarization entangled photons from electrically excited QDs, that had been transmitted over an 18 km long metropolitan optical fiber \cite{Xiang2019}. Sufficiently maintaining the photon polarization over the fiber transmission for over a week was made possible, by multiplexing a polarization reference signal through the fiber and actively stabilizing it. Note that the current record for the longest distance of entanglement distribution via fiber-links is almost 100 km and was achieved by sending polarization entangled photons through a submarine optical fiber between the islands Malta and Sicily \cite{Wengerowsky2019}. Although the entangled photons were created via a nonlinear downconversion process here and are thus not ideal for QKD applications due to the Poisson statistics, it is remarkable how the entanglement was preserved over such a long distance within the optical fiber in a ‘real-world-scenario’. The success of the entanglement preservation was indicated by the clear violation of the CHSH inequality, with a value for the S-parameter of 2.5 after the entanglement distribution. The experiment achieving the to-date longest distance of free-space entanglement distribution sent photons from a SPDC-based source on the quantum satellite ‘Micius’ to earth \cite{Yin2017}. Here, the photons propagated over a distance of 500-1000 km and the polarization entanglement was preserved with a fidelity of $86\%$.
\\
The status quo of QD-based QKD experiments is summarized in \textbf{Table \ref{tab:table1}}, listing all implementations reported to date. For a comparison of different implementations, the reader should keep in mind the following comments. While some experiments were proof-of-concepts with a fixed communication distance and/or lab-based experiments, others were performed at varying distance or attenuation to simulate different channel losses. In these cases, we give a range of parameters. Moreover, some references quote an average value for the QBER in their system, while this value in general depends on the channel loss. Furthermore, some of the quoted parameters rely either on assumptions made in the key rate calculation (which imperfections to include and which attacks to consider), while others depend highly on the equipment used (such as number of detector dark counts), making a direct comparison difficult. Finally, key rates that we calculated according to the asymptotic GLLP equation (cf. equation \ref{eq:GLLP}, section \ref{section1}) from parameters given in the respective publication are indicated by footnotes.
\\
\newgeometry{left=1cm,right=1cm}
\begin{table}
\centering
\caption{Summary of QKD implementations employing QD-based quantum light sources (abbreviations: polarization (Pol), single-photon source (SPS), entangled photon-pair source (EPS), free space optical (FSO), fiber-coupled (FC))}
\label{tab:table1}
\begin{threeparttable}
\begin{tabular}{ccccccccccl}
\hline
Protocol &
  Coding &
  \begin{tabular}[c]{@{}c@{}}Source/ \\ Pump\end{tabular} &
  \begin{tabular}[c]{@{}c@{}} Clock-rate \\ {[}MHz{]} \end{tabular} &
  $\lambda$ {[}nm{]} &
  \begin{tabular}[c]{@{}c@{}}Max Sifted \\ Key Rate\end{tabular} &
  \begin{tabular}[c]{@{}c@{}}Max Secure \\ Key Rate\end{tabular} &
  \begin{tabular}[c]{@{}c@{}}QBER \\ {[}\%{]}\end{tabular} &
  \begin{tabular}[c]{@{}c@{}}FSO/ \\ FC \end{tabular} &
  \begin{tabular}[c]{@{}c@{}}Max\\ distance\end{tabular} &
  Ref. \\ \hline
BB84    & Pol   & SPS / optic.    & 76    & 880  & -             & 25 kbps      & 2.5        & FSO & In-Lab & \cite{Waks2002a}  \\
BB84    & Pol   & SPS / optic.    & 0.01  & 635  & 15 bps        & 5 bps        & 6.8        & FSO & In-Lab & \cite{Aichele2004} \\
BB84    & Phase & SPS / optic.    & 1     & 1300 & 10 bps        & 1 bps        & 5.9        & FC  & 35 km  & \cite{Intallura2009} \\
BB84    & Pol   & SPS / optic.    & 40    & 895  & - & 8-600 bps & 1.2-21.9 & FC  & 2 km   & \cite{Collins2010}\\
BB84    & Phase & SPS / optic.    & 20    & 1580 & 15-386 bps    & 3-9 bps      & 3.4-6      & FC  & 50 km  & \cite{Takemoto2010} \\
BB84    & Pol   & SPS / elect. & 182.6 & 898  & 8-35 kbps     & -            & 3.8-6.7    & FSO & In-Lab & \cite{Heindel2012} \\
BB84\tnote{ a)}    & Pol   & SPS / elect. & 200   & 653  & 9-117 kbps    & -            & 4.1-6.0    & FSO & In-Lab & \cite{Heindel2012}  \\
BB84    & Pol   & SPS / elect.& 125   & 910  & 5-17 kbps     & -            & 6-9        & FSO & 500 m  & \cite{Rau2014} \\
BB84    & Phase & SPS / optic.    & 62.5  & 1500 & 34 bps        & 0.307 bps    & 2-9        & FC  & 120 km & \cite{Takemoto2015} \\
E91     & Pol   & EPS / elect. & 50\tnote{ b)} & 885  & 0.2 bps & 0.1 bps & -          & FC  & In-Lab & \cite{Dzurnak2015} \\
E91\tnote{ c)} & Pol   & EPS / optic.   & 320   & 785  & 243 bps       & 69\tnote{ d)}      & 3.4        & FC  & 250 m  & \cite{Basset2021}\\
E91     & Pol   & EPS Optic. & 320   & 785  & 30 bps        & 9 bps        & 4.0        & FSO & 270 m  & \cite{Basset2021} \\
BBM92   & Pol   & EPS / optic.    & 80    & 785  & 135 bps       & 86 bps       & 1.9        & FC  & 350 m  & \cite{Schimpf2021a} \\ \hline
\end{tabular}
a) Same publication as above but QKD experiment performed by a different group; b) Time-multiplexed detector effectively reduced clock rate to below 1 MHz; c) Modified asymmetric Ekert91 protocol; d) Calculated from stated parameters using the asymptotic GLLP equation;
\end{threeparttable}
\end{table}
\restoregeometry
In order to put the QKD implementations reviewed above into perspective, we now address the question which performance level, in terms of secure key rate and communication distance, can be expected with current state-of-the art QD-based quantum light sources. For this purpose, we use equations (\ref{eq:GLLP}-\ref{eq:GLLP_QBER}) to extrapolate the achievable asymptotic key rate from parameters reported in the literature as described in the following and assuming an operation wavelength of 1550 nm. Note, while the values states in the following were not yet realized at C-band wavelengths, the advances in the development of QD-based Telecom-wavelength quantum light sources make us optimistic that this can be achieved in the not too distant future.
\\
As stated earlier, the amount of multi-photon events is upper-bounded by $P_\text{M} \leq \frac{1}{2} \mu^{2} g^{(2)}(0)$ \cite{Waks2002}. Assuming a mean photon number of $\mu=0.3$ and the record antibunching value of $g^{(2)}(0) = 7.5 \cdot 10^{-5}$ \cite{Schweickert2018}, an upper bound for the achievable multi-photon contribution can be calculated. Note that even larger values of $\mu$ were recently demonstrated \cite{Tomm2020}. In this regime, however, the increase of multi-photon contributions might effectively reduce the key rate at large loss on Bob's side. We further assume a setup transmission of $\eta_\text{Bob} = 0.3$ \cite{gobby2004}, alignment errors of $e_{\text{det}} = 1\%$ \cite{rusca2018finite}, and superconducting nanowire detectors with a few dark counts per second, resulting in $P_\text{dc} = 10^{-8}$ dark counts per pulse with an detection efficiency of $\eta_\text{det} = 0.9$ \cite{SingleQuantum}. Using ultralow-loss fibers, losses in the quantum channel of $\alpha = 0.17\, $dB/km are possible at 1550 nm \cite{rusca2018finite}. We further assume a value of $f_\text{EC} = 1.1$ for the efficiency of the error correction protocol \cite{elkouss2009} (with $f_\text{EC} = 1$ being the ideal Shannon limit), $q=0.5$ for the sifting factor of the symmetric BB84 protocol, and a clock rate (for excitation and encoding) of $R_0=100\,$MHz \cite{Boaron2018a}.
\\
Using these parameters, we deduce an asymptotic secure key rate exceeding 3$\,$Mbit$\,\text{s}^{-1}$ at short distance, a key rate larger than 1$\,$kbit$\,\text{s}^{-1}$ after a distance of 200 km in optical fiber, and a maximum achievable communication distance beyond 250$\,$km. 
These realistic extrapolations highlight the prospects for the substantial advances possible in the field of sub-Poissonian QKD, which are in turn also beneficial for schemes of quantum communication beyond direct point-to-point links. Also note, that this performance can be further improved significantly, e.g. by using higher clock-rates \cite{Schlehahn2016}, asymmetric bases choices allowing for larger sifting factors, or temporal filtering \cite{Kupko2020} as will be detailed in section \ref{section4.6}. Keep in mind, however, that such high asymptotic key rates can only be achieved when sufficiently large block sizes of qubits are transmitted (about $10^{15}$), as otherwise finite-size effects drastically reduce the secure key rate \cite{Scarani2009,cai2009}.
\\
\subsection{Advanced QKD protocols – Towards Device Independence} \label{section3.4}
In recent years, also advanced types of protocols have been proposed with a special emphasis on device independence. Knowing that the bulk of quantum hacking attacks target either the qubit source or the detector module, it would be a great advancement if the communication protocol did not have to rely on the integrity of any of these. While fully device-independent QKD (DI-QKD) is still only a theoretical proposal at this point \cite{Acin2007,Masanes2011,Zapatero2019,Vazirani2019,Schwonnek2021}, intermediate steps such as source-independent \cite{Koashi2003} or measurement-device-independent (MDI-QKD) might be feasible \cite{Lo2012}. MDI-QKD guarantees independence of the detection setup which could in principle be controlled by an adversary. It is basically a time-reversed version of the Ekert protocol, requiring Alice and Bob each sending single photons to a central detector, which projects them into an entangled two-photon state via a Bell measurement. By learning the outcome of the Bell measurement and by communicating the preparation bases, Alice and Bob can establish a secret key. Knowing only the outcome of the Bell measurement does not reveal anything about the key which is why the detector does not need to be trusted. 
\\
This protocol has so far only been implemented using attenuated laser pulses \cite{Rubenok2013,Liu2013,DaSilva2013,Cao2020,Semenenko2020,Wei2020} and also, as a proof-of-concept, with stored and released photons from down-conversion sources \cite{Kaneda2017}. Implementing it with single photons is more difficult but will ultimately pay out. To allow for a successful and efficient Bell measurement, the photons must be indistinguishable. To obtain sufficiently indistinguishable single photons from remote sources, QDs are a promising candidate. Although high indistinguishability in practice is a challenge (due to local dephasing effects and spectral diffusion \cite{Vural2020}) several demonstrations have already already surpassed the classical limit of $50\%$ two photon interference visibility of laser light (see section \ref{section4.2}) \cite{Zhai2021}. The main advantage of QD single photons over the attenuated laser implementations is that, due to the limited visibility and the Poisson photon statistics, a Bell measurement using coherent light is less precise than using true single photons \cite{Lee2021}. 
\\
The achievable secure key rate has already been analytically estimated for MDI-QKD based on attenuated laser pulses in the asymptotic \cite{Lo2012}, as well as in the finite-length regime \cite{Curty2014}, but not yet for true single photons, incorporating their better indistinguishability. MDI-QKD protocols are also a possible way of realizing many-parties metropolitan QKD networks with all members surrounding a central detection node in a star-like topology \cite{Tang2016}. An experimental demonstration of this type of quantum network with sub-Poissonian quantum light source would be a major step towards the quantum internet.

\section{Recent Progress on Building Blocks for Quantum Networks}\label{section4}
Having the ultimate aim of a world-wide quantum internet in mind, the establishment of a QKD-secured communication link, as discussed in the previous section, is only the first step. In this section, we discuss building blocks necessary for the extension toward networks (see \textbf{Figure \ref{fig:fig12}}). Noteworthy, many of these building blocks are equally important for other types of quantum technologies such as distributed quantum computing, optical quantum computing, quantum sensing / metrology and many more. This is why there are already extensive reviews on the building blocks for future quantum technologies \cite{OBrien2009,Barnett2017,Uppu2021}. Therefore, in this section, we highlight recent advances in developing building blocks with a special focus on QD based communication networks. This includes practical QD SPSs (4.1), quantum memories compatible with QD single photons (4.3), teleportation of QD single photons (4.4) and quantum repeater schemes suitable for entangled photons from QDs (4.2). For completeness, we highlight recent advances in random number generation (4.5) and quantum network optimization tools (4.6).
\\
\begin{figure}
  \includegraphics[width= \linewidth]{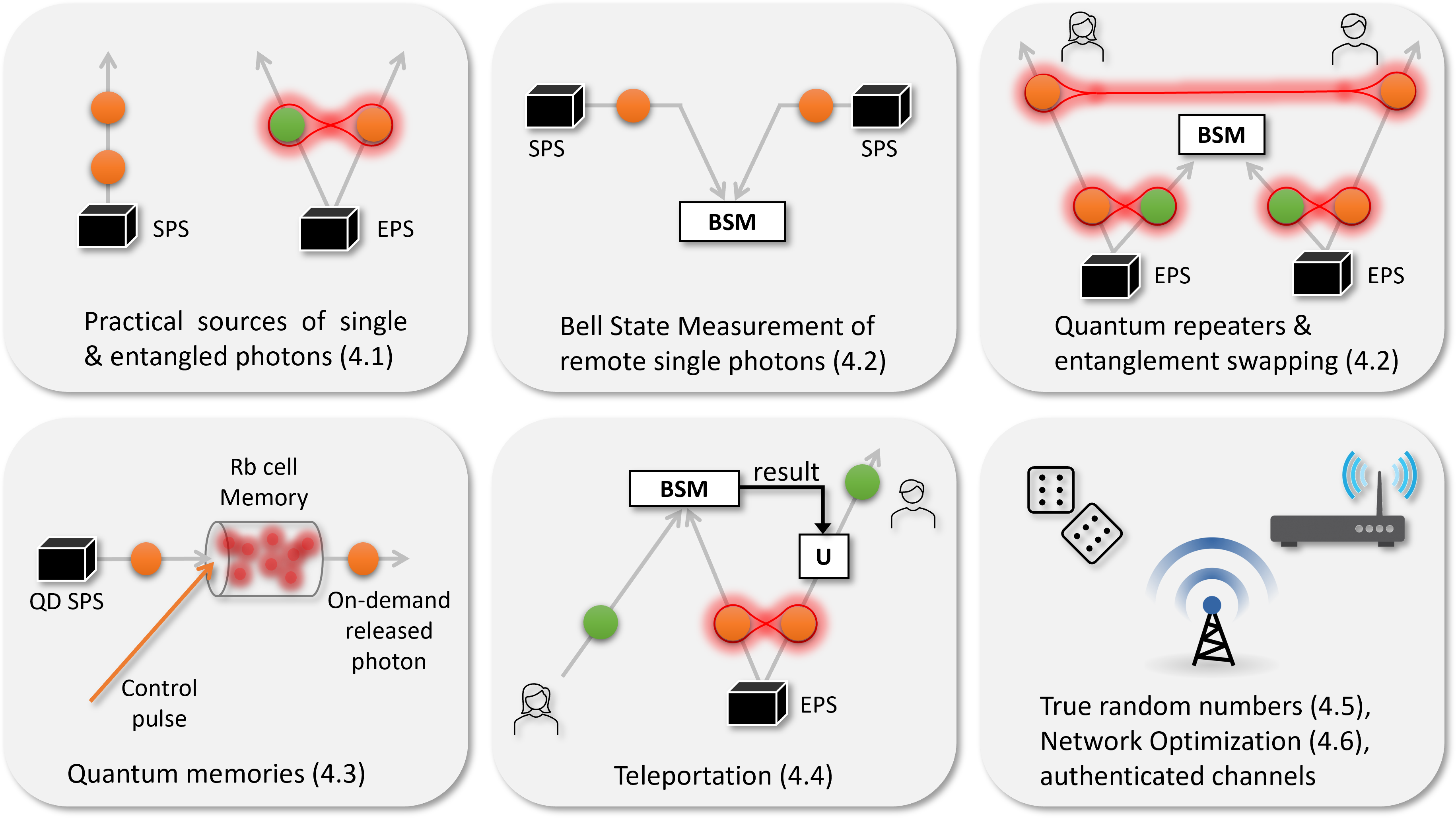}
  \caption{Building blocks for future quantum networks. The numbers in brackets point at the section in which recent advances concerning that building block are discussed. In this review we focus on building blocks that are built on (or compatible with) QD light sources.}
  \label{fig:fig12}
\end{figure}
\subsection{Towards practical quantum light sources} \label{section4.1}
A practical QD-based quantum light source is ideally “plug-and-play”. This means that it is operable independent of special laboratory infrastructure (e.g. without the need for  liquid coolants and large laser systems) and provides the generated quantum light “ready-to-use” via an optical fiber output. Furthermore, it needs to be robust and durable in use, as compact as possible, and ideally only requires standard mains voltage as power supply. Not least, a practical source needs to be benchmarked regarding long-term stability.
\\
A key requirement to integrate QD-SPSs into practical modules as described above is a technology for the precise alignment and permanent coupling of the QD-device to an optical fiber. Such a direct connection has been the interest of research for more than a decade, starting with the probabilistic coupling of QDs using bundles of hundreds of single mode fibers \cite{Xu2007}, then exploring deterministic coupling possibilities using open cavity systems \cite{Muller2009} and deterministic selection of nanowire SPSs \cite{Cadeddu2016}. 
\\
Recently, fiber-scanning techniques in combination with epoxy resists or optical (UV-) adhesives have shown to be suitable for permanent fiber-coupling of micropillar cavities \cite{Haupt2010,Snijders2018,Rickert2021}, microlens and micromesa SPSs \cite{Schlehahn2018,Zonacz2019,Musial2020}, as well as for nanowires \cite{Northeast2021}. A schematic of an electrically controllable, fiber-coupled micropillar-cavity with a detection and excitation fiber and capable of resonant excitation schemes published by Snijders and coworkers in 2018 \cite{Snijders2018} is shown in \textbf{Figure \ref{fig:fig13}a}. While permanently coupled devices reported to date reached overall efficiencies in the range of only few percent \cite{Snijders2018}, microcavities based on hybrid circular Bragg gratings (hCBGs) have recently been evaluated as promising a strategy for achieving large fiber coupling efficiencies up to unity. In their study, Rickert et al. presented numerically optimized designs for devices operating at O-band wavelengths \cite{Rickert2019}, indicating that overall efficiencies exceeding $80\%$ are possible using off-the shelf single-mode fibers. A schematic depiction of a single-mode fiber-coupled hCBG-cavity is shown together with simulation data in Figure \ref{fig:fig13}b.  This device approach appears particularly promising, considering the potential for state-of-the art SPS performances \cite{Wang2019}.
\\
The QD-based SPSs used for QKD experiments so far, entirely relied on laboratory infrastructure including bulky cryogenics in particular. A way to realize these low temperatures in a more practical fashion is to use off-the-shelf Stirling-cryocoolers \cite{sunpower}. Such cryocoolers, operable with 230 V standard net supply voltage, allow operation of a suitable quantum emitter at temperatures below 30 K, an approach first introduced by Schlehahn et al. in 2015 using free-space optics \cite{Schlehahn2015}. Subsequently, the integration of  fiber-coupled QD-based SPSs in state-of-the-art Stirling cryocoolers proved to be a promising route for realizing plug-and-play QD-based quantum light sources  as demonstrated for a fiber-pigtailed QD-device emitting around 925 nm in 2018 \cite{Schlehahn2018}. The cryocoolers were compact enough to integrate them into a standard 19” rack module (Figure \ref{fig:fig13}c) and the achievable base-temperature of 40 K could be reached easily within 30 min. In the work by Musiał et al. the source module additionally housed a fiber-based pulsed laser as well as a fiber-based spectral filter, providing single-photon pulses in the telecom O-Band at the SMF-28 fiber output \cite{Musial2020}.
\\
Another aspect for practical single photon sources, it is worth also considering alternative ways of source excitation to achieve practicability. While in principle an electrically-contacted SPS allows efficient, fast and practical excitation, optical excitation required for resonant excitation of the source can also be desirable. In a pioneering work in 2013, Stock et al. proposed to use electrically driven microlaser sources in close vicinity to excite QDs in micropillar cavities, and showed Purcell enhanced emission for a QD in a micropillar exited in such a way \cite{stock2013}. In 2017  Munnelly et al. used this on-chip excitation concept to show single-photon emission with an emission rate above 100 MHz and the possibility of wavelength tuning via the quantum-confined Stark-effect \cite{Munnelly2017}. In a work in 2017 following a similar concept, Lee and coworkers realised an on-chip excited quantum dot light emitting diode (LED) and used the quantum-confined Stark effect to also tune the fine-structure of the emitting QD \cite{Lee2017}. Work of such an on-chip driven QD entangled LED deployed in an urban fiber network was published recently \cite{Xiang2020}. Although so far not demonstrated for (quasi-) resonant excitation, the concept of using an on-chip pumped QD SPS  shows the potential for a very compact optical excitation compatible with the discussed Stirling technology.
\\
In summary, the integration of directly fiber-coupled QD-based SPSs into compact cryocoolers  offers a promising approach for high-performance plug-and-play quantum light sources. While Stirling-type refrigerators are the most compact solution to date, the achievable base temperatures are presently limited to about 27 K. Applications which rely on the excellent coherence properties of QDs, e.g. for the generation of highly indistinguishable photons, small-footprint Gifford-McMahon (GM) cryocoolers in combination with compact compressors are an alternative. Beyond the promising proof-of-concept experiments with fiber-coupled QD-SPSs in Stirling cryocoolers discussed above, a next important step is to show that this concept can also exploit the full potential QDs offer in terms of the efficiency and single-photon purity.
\begin{figure}
  \includegraphics[width=0.5 \linewidth]{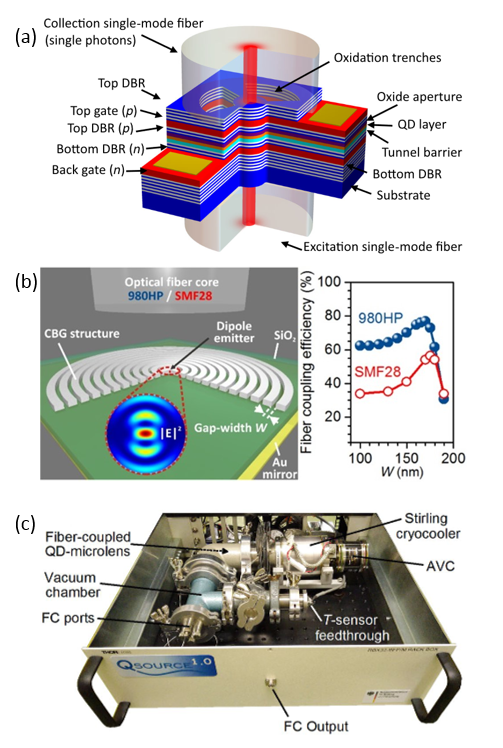}
  \caption{(a) Schematic of a fiber-coupled micropillar cavity with electrical control, and two fiber connections for excitation and collection. (b) (left) Schematic representation of a hybrid circular Bragg grating (hCBG) cavity coupled to a optical single mode fiber. (right) Simulated fiber coupling efficiencies of a hCBG cavity with two different single mode fibers as function of the grating gap width W (see Ref. \cite{Rickert2019} for details). (c) QD-based SPS source module based on a Stirling cryocooler with active vibration cancellation (AVC) employing a fiber-coupled QD-based microlens SPS (from Ref. [Schlehahn2018]\cite{Schlehahn2018}). (a) Reprinted with permission from \href{https://link.aps.org/doi/10.1103/PhysRevApplied.9.031002}{\textit{Snijders et al. 2018}} \cite{Snijders2018} Copyright 2018 by the American Physical Society, (b) Copyright 2021 by the authors of this work. (c) reprinted from \href{http://www.nature.com/articles/s41598-017-19049-4}{\textit{Schlehahn et al. 2018}} \cite{Schlehahn2018} under Creative Commons CC BY license.}
  \label{fig:fig13}
\end{figure}
\subsection{Towards Quantum Repeaters} \label{section4.2}
The maximum distance over which a QKD scheme can exchange a provably secure key is limited due to photon loss, both in free-space and in fiber links. To cover larger distances, one has to rely on trusted intermediate nodes, which reduces the overall security. However, the maximum distance, as well as the rate can be extended without trusted nodes, by using so-called quantum repeaters. These allow to transfer an encoded qubit, without travelling the entire distance, with the help of entangled photon pairs and entanglement swapping. In this subsection we will briefly introduce the quantum repeater protocol and entanglement swapping, before we highlight recent advances in the key ingredient, two-photon interference from remote QD-based SPSs.
\\
The original repeater protocol was proposed by Dür, Briegel, Cirac and Zoller in 1998 (now known as BDCZ protocol), as a way of reducing the errors in quantum channels, which depend exponentially on the transmission loss \cite{Briegel1998,Dur1999}. This protocol enables the distribution of a maximally entangled photon pair, such as the well-known EPR state, in reference to the Einstein-Podolski-Rosen paradox \cite{Einstein1935}, over arbitrary distances. A version which uses atomic ensembles both as the memory and the photon source is known as the DLCZ protocol \cite{Duan2001}. The entangled photon pair can then be used to realize QKD protocols based on entangled photon pairs or to teleport any quantum state from one end to the other. To distribute the entanglement, the authors proposed to use multiple EPR states. The entangled photons are each split up and sent to opposite directions, thus covering a small part of the quantum channel length. At intermediate nodes, the photons are stored and then, via a joint Bell state measurement (BSM) between two photons of different EPR pairs, the entanglement is swapped, entangling the two remaining photons. By repeating this many times in a nested fashion, arbitrary distances can in principle be covered (\textbf{Figure \ref{fig:fig14}a}).
\\
Note, that as all photons together travelled the total distance in case of a successful run, an improvement of the photon loss sensitivity is only achieved, if each intermediate node has access to a quantum memory. In this case, redundant EPR pairs can be sent, until the memory is filled, and the BSM can be made, thus preventing any loss between two of such nodes. In other words, the quantum channel is cut into shorter segments and over each segment entanglement purification can create maximally entangled, distributed photon pairs in a memory, before entanglement swapping connects all these segments.
\\
This protocol can be implemented with different quantum emitter platforms \cite{Loock2020}, but to be implemented with entangled photon pairs from QDs, a quantum memory compatible with QD single photons is required, which will be discussed in \textbf{section 4.3}. It also requires the ability to successfully project two single photons into a joint two-photon Bell state to realize the entanglement swapping. This was demonstrated for the first time by Pan et al. using a SPDC source \cite{Pan1998}. The authors created two pairs of entangled photon states, projected one photon of each pair in a joint two-photon Bell state via two-photon-interference (TPI) and finally proved entanglement of the remaining two photons that never interacted before. 
\\
A quantum repeater node implementing the BDCZ protocol was first experimentally realized by Yuan et al. \cite{Yuan2008} using two atomic ensembles as quantum memories. By projecting photons, entangled with the states of their respective memory, in a joint Bell state, the entanglement was swapped, entangling the two atomic ensembles, which was confirmed by measuring the entanglement between the photons emitted from the memories. 
\\
In addition to memory-based quantum repeaters, all-photonic, measurement-based repeater schemes have been proposed which do not need a quantum memory to operate \cite{Zwerger2012,Azuma2015,Zwerger2016}. The necessary resource states for such a repeater protocol are photonic cluster states, which were recently used in a proof-of-principle quantum repeater experiment \cite{Li2019}. Photonic cluster states are also relevant for photonic quantum computing \cite{raussendorf2001} and have already been generated using QD SPSs \cite{Schwartz2016,Istrati2020}.
\\
\begin{figure}
  \includegraphics[width= \linewidth]{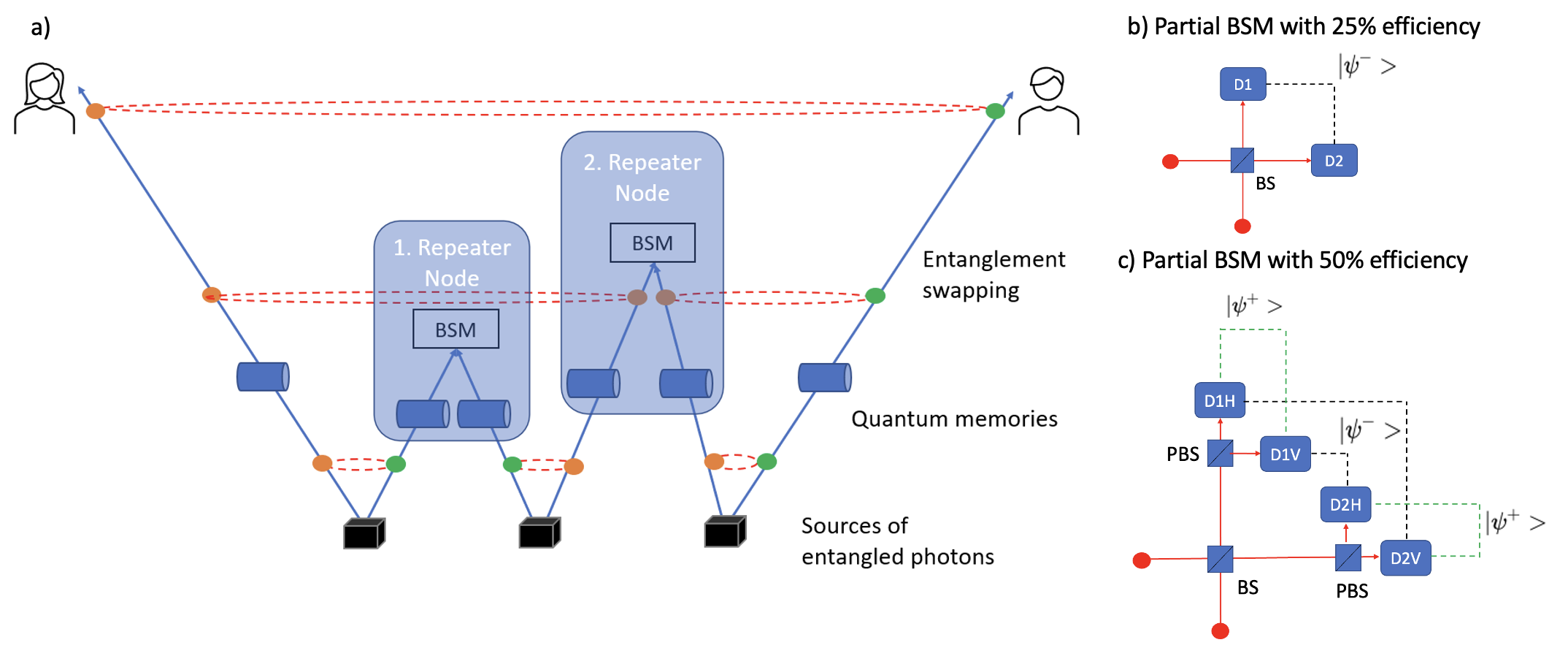}
  \caption{The concept of a quantum repeater to distribute entanglement over long distances (a). The two types of BSMs using linear optics with (b) only 1 BS to identify 1/4 Bell states and (c) with PBS to identify 2/4 Bell states from coincidence detections.}
  \label{fig:fig14}
\end{figure}
The crucial ingredient for all repeater protocols and many other schemes in quantum information  is to perform a Bell State measurement, which is typically implemented via two-photon-interference experiments at a beam splitter. Before reviewing recent advances in QD-based TPI, we remark a few general things concerning joint BSMs. The Bell basis is the natural basis of the 4-dimensional Hilbert space in which a joint two-qubit state is described. It consists of four maximally entangled states, the Bell states, that are non-separable superpositions of the tensor products of the respective one-particle Hilbert space bases. For the case of polarization as the only relevant quantum number (the two photons are ideally indistinguishable in all other quantum numbers) and using the rectilinear {H,V}-basis, they can be written as $|\Psi^{\pm}> = \frac{1}{\sqrt{2}} \left( \ket{HV} \pm \ket{VH} \right)$ and $|\Phi^{\pm}> = \frac{1}{\sqrt{2}} \left(\ket{HH} \pm \ket{VV} \right)$.
\\
A fundamental limit to BSMs is that, using linear optics, only two states out of the four Bell states can be discriminated \cite{Braunstein1995a}, thus without photon-photon interaction, only a partial BSM with a success rate of $50\%$ is possible. Braunstein et al. were the first to point out that two-photon-interference at a BS can be seen as a BSM, since the Bell states are eigen-states of the unitary that describes the BS. If the photons go to opposite outputs of the BS, one can be sure that a projection into the $\ket{\Psi^{-}}$ Bell state took place (Figure \ref{fig:fig14}b). The reason is that only the $\ket{\Psi^{-}}$ Bell state is antisymmetric under particle exchange and thus displays fermionic anti-bunching statistics, causing the particles to leave from opposite output ports of the BS. But this only leads to a successful BSM in $25\%$ of the cases. Furthermore, the photons need to be indistinguishable in all other quantum numbers, so that only their projected polarization state determines the coincidences. Otherwise, if the photons are too distinguishable, also a projection into the other Bell states would show up as an erroneous $\ket{\Psi^{-}}$ coincidence event between the two outputs. But if the photons are indeed indistinguishable and not projected into the $\ket{\Psi^{-}}$  state, they will go to the same output, as observed in the Hong-Ou-Mandel (HOM) effect, where it was first shown that completely indistinguishable single photons always leave the BS together \cite{Hong1987}.
\\
One can increase the BSM efficiency to $50\%$ for polarization encoded photons, by adding a PBS at each output of the interference BS (Figure \ref{fig:fig14}c) to further distinguish the polarization of the states that leave the BS from the same output. Photons which arrive at the same output side, with the same polarization, will go to the same detector and thus do not lead to coincidences (the $\ket{\Phi}$  states), while the $\ket{\Psi^{+}}$ state photons leave from the same output with different polarizations and thus lead to coincidences. Hence, one can now identify both $\ket{\Psi}$ states, because they are anti-correlated in polarization. Thus, $\frac{1}{2}$ of the possible Bell basis can be detected at maximum.
\\
Note that if the incoming photons are not sufficiently indistinguishable, erroneous coincidences in the PBS basis can be detected, while states prepared in any other basis (that are randomly projected at the PBS) lead to false coincidences, as was shown by Basset et al. by comparing the achieved teleportation fidelities under the $25\%$ and $50\%$ BSM (without and with PBSs), as well as with high and low indistinguishability \cite{Basset2020a}. Since SPSs allow higher indistinguishability, they cause less erroneous coincidence events than photons from SPDC sources. Note furthermore that, also due to more multi-photon events, a Bell measurement is less precise with attenuated laser pulses, than with true single photons \cite{Lee2020}. The reason is that false coincidences due to two photons in one BS input cannot always be distinguished from true coincidences from one photon in each BS input. Therefore, it is of critical importance for many applications to achieve TPI with indistinguishable single photons from remote quantum emitters.
\\
The indistinguishability is quantified by the coalescence probability (for pure states the maximum wave packet overlap of the two photons $P_{Coal} = \abs{\bra{\phi_1}\ket{\phi_2}}$ , which is estimated by measuring the TPI visibility V in a Hong-Ou-Mandel interference experiment \cite{Hong1987}. Hereby, one typically measures the suppression of coincidences between two BS outputs due to indistinguishable paths at vanishing temporal delay, and normalizes it by comparison to the case of maximum distinguishability (selecting a cross-polarized configuration, detuned wavelengths or detuned arrival times). While phase-randomized Poissonian light is limited to a visibility of 0.5 \cite{Mandel1983}, single photons can reach higher visibilities up to unity, for more efficient BSMs.
\\
In order to measure a high TPI visibility between photons from remote emitters, they must be as indistinguishable as possible, i.e. they must agree in all quantum numbers. So far, TPI between photons from remote emitters has already been demonstrated with many different types of photon sources, such as parametric down conversion sources \cite{DeRiedmatten2003,Llewellyn2020}, trapped atoms \cite{Beugnon2006}, atoms in combination with QDs \cite{Vural2018}, trapped ions \cite{Maunz2007}, silicon and nitrogen vacancy centers in diamond \cite{Sipahigil2014,Bernien2012,Humphreys2018}, and also molecules \cite{Lettow2010}. In the following, we will discuss recent progress on TPI experiments with photons emitted by remote QD-based quantum light sources.
\\
In the case of QDs, due to their self-organized nature and the semiconductor environment, the spectral properties of the quantum emitters are of particular importance. While photons that are emitted subsequently by the same QD have shown almost ideal TPI visibilities  $>99\%$ \cite{Somaschi2016}, such high values have not yet been achieved for TPI between photons from remote QDs. When photons are emitted by QDs located at different positions and interacting with their unique environment, the wavelength of their emitted photons are likely to be different. Coarse spectral matching of quantum emitters can thus be achieved using pre-selection of suitable QDs, if deterministic fabrication technologies are used even in engineered devices \cite{Rodt2020,Liu2021}. Subsequently, a spectral fine-tuning is typically required, e.g. via temperature control (\textbf{Figure \ref{fig:fig15}}f) \cite{Giesz2015,Thoma2016}, strain tuning (Figure \ref{fig:fig15} c-e) \cite{Flagg2010,Beetz2013,Reindl2017,Moczaa-Dusanowska2020,zhai2020a}, or electrical tuning via the quantum-confined Stark effect in diode-like structures (Figure \ref{fig:fig15} a, b) \cite{Patel2010}. These methods are summarized in Figure \ref{fig:fig15} and the obtained visibilities of the respective TPI experiments are collected in Table 2.
\\
\begin{figure}[h]
  \includegraphics[width= \linewidth]{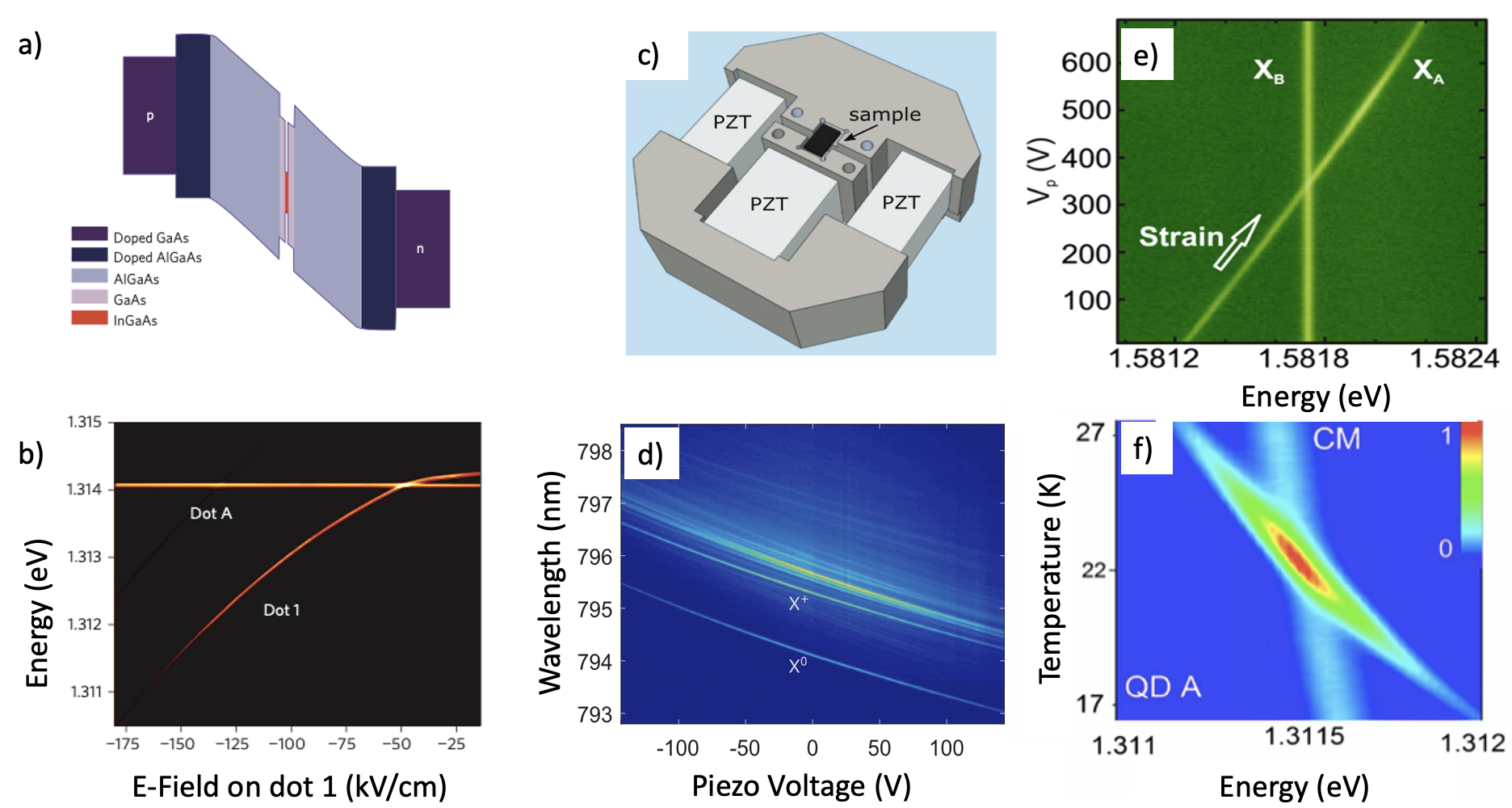}
  \caption{Different ways to tune the emission energies of remote QDs in TPI experiments. Patel et al. used electrical tuning of QDs in a diode-like structure (a) which allowed them to match the energies (b) \cite{Patel2010}. Zhai et al. employed piezo actuators to apply strain to the QD (c) which enabled tuning of the emission energies (d) \cite{zhai2020}, as did Reindl et al. to match the QD energies (e) \cite{Reindl2017}. Giesz et al. used temperature to tune the QD energy into the cavity resonance (f) \cite{Giesz2015}. (a,b) reprinted by permission from Springer Nature: \href{http://dx.doi.org/10.1038/nphoton.2010.161}{\textit{Patel et al. 2010}} \cite{Patel2010} Copyright 2010, (c,d) reprinted from \href{https://aip.scitation.org/doi/full/10.1063/5.0017995}{\textit{Zhai et al. 2020}} \cite{zhai2020}, with the permission of AIP Publishing, (e) reprinted from \href{https://pubs.acs.org/doi/10.1021/acs.nanolett.7b00777}{\textit{Reindl et al. 2017}} \cite{Reindl2017} under Creative Commons  license CC-BY, (f) reprinted Figure with permission from \href{https://link.aps.org/doi/10.1103/PhysRevB.92.161302}{\textit{Giesz et al. 2015}} \cite{Giesz2015} Copyright 2015 by the American Physical Society.}
  \label{fig:fig15}
\end{figure}
However, even if the nominal emission energies of two QDs are matched perfectly, the achievable TPI visibility can be limited by several effects. Phonon induced dephasing, fluctuation of electron spins and fluctuation of surrounding charges can lead to emission energy fluctuations (also known as spectral diffusion) as well as emission line broadening \cite{Vural2020}. Recently, an analytical model including dephasing as well as spectral diffusion has been developed, which can be used to predict the maximum expected TPI visibilities from experimentally accessible parameters of each individual QD \cite{Kambs2018} - previous models numerically investigated achievable visibilities for non-ideal SPSs \cite{Fischer2016}. To reduce the inhomogeneous line broadening mechanisms of QDs, different countermeasures can be taken, such as working at ultra-low temperatures to freeze out the majority of phonon contributions to the dephasing, resonant excitation schemes combined with small amounts of off-resonant light, as well as external control via electric and magnetic fields to saturate charge traps and align the nuclear spins inside the QD to reduce fluctuations. All the necessary requirements can be summarized as follows: For maximum indistinguishability, one needs Fourier-limited SPSs, which have no inhomogeneous line broadening. Such QD-SPSs were  reported by Wang et al. \cite{Wang2016} and Kuhlmann et al. \cite{Kuhlmann2013}, but they are far from being the norm.
\\
Additionally, to reduce relative spectral drifts between the two QD emission energies, an active feedback can be used \cite{Schmidt2020}. Such a stabilization was applied to remote TPI experiments by Zopf et al., who applied piezo strain to compensate for any spectral drift to stabilize the emission energies of the QDs \cite{Zopf2018}. In their setup both QDs were tuned via piezos glued to the QD. A fraction of the emitted photons went through a Faraday filter made from a Rubidium vapor cell, to identify small shifts in emission wavelengths and then a feedback loop corrected them. By doing so, they achieved a remote QD TPI visibility of $41\%$, as predicted by the Kambs et al. model \cite{Kambs2018} for their QD parameters. The active stabilization resulted in higher average TPI visibility compared to the unstabilized case (\textbf{Figure \ref{fig:fig16}a}).
\\
Another technique, which was employed by Weber et al., is to use frequency conversion of both QDs (Figure \ref{fig:fig16}b) to do the TPI at 1550 nm, in the Telecom window \cite{Weber2019}. Although the two QDs themselves were spectrally distinguishable, laser-tuning in the upconversion process allowed the authors to match the photon’s wavelengths (Figure \ref{fig:fig16}c), resulting in an observed TPI visibility of $29\%$. By now, TPI visibilities exceeding the classical limit of $50\%$ were achieved in several experiments using QDs. Reindl et al. in 2017 for instance used a phonon-assisted two-photon excitation scheme in combination with piezo strain tuning to match the emission energies \cite{Reindl2017} resulting in a TPI visibility of $51\%$.
\\
Very recently, a new record has been set by Zhai et al. \cite{Zhai2021}, reaching visibilities of up to $93\%$ between QDs located in different cryostates. The high visibility was achieved without Purcell enhancement, tight spectral filtering, post-selection or any active stabilization, simply by using high quality, low noise, electrically tunable QD samples, leaving even room for further improvement \cite{zhai2020a,zhai2020}. In another recent experiment You et al. observed interference of remote QD single photons that were converted to 1583 nm via quantum frequency conversion and separated by a 300 km optical fiber \cite{You2021}, setting a remarkable record for the distance achieved between interfering QD-sources.
\\
\begin{figure}
  \includegraphics[width= \linewidth]{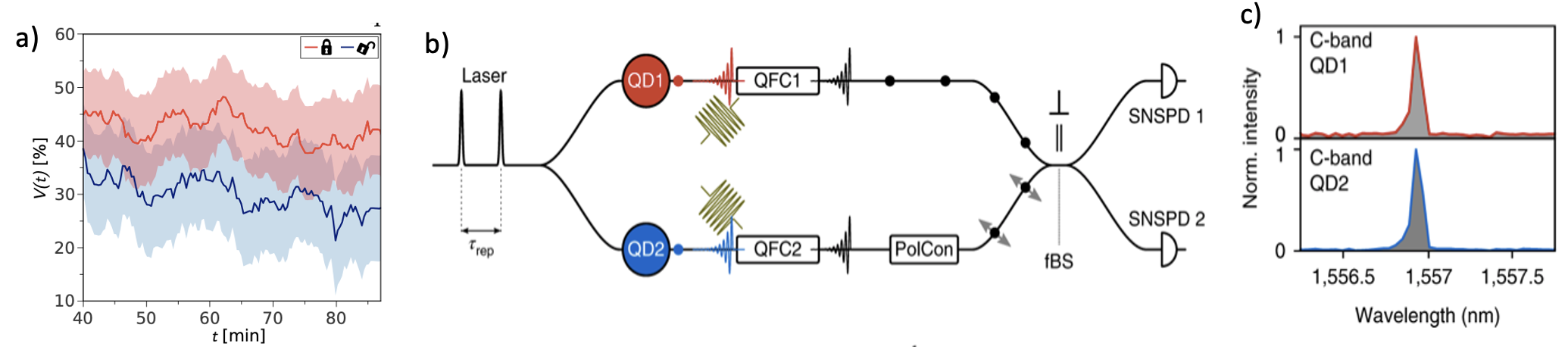}
  \caption{Active stabilization of QDs in remote TPI (red curve in a) improves visibility compared to no stabilization (blue curve in a) \cite{Zopf2018}. Matching of remote QD wavelengths (c) via Quantum Frequency Conversion (b) \cite{Weber2019}. (a) reprinted with permission from \href{https://doi.org/10.1103/PhysRevB.98.161302}{\textit{Zopf et al. 2018}} \cite{Zopf2018} Copyright 2018 by the American Physical Society, (b,c) reprinted with permission from Springer Nature: Nature Nanotechnology \href{http://dx.doi.org/10.1038/s41565-018-0279-8}{\textit{Weber et al. 2019}} \cite{Weber2019} Copyright 2019.}
  \label{fig:fig16}
\end{figure}
Noteworthy, indistinguishable photons are not only important for entanglement swapping and quantum repeaters, but also to erase the which-path-information in schemes that generate entanglement between remote solid-state qubits \cite{Cabrillo1999}. This entanglement scheme was realized experimentally with remote QDs using hole- \cite{Delteil2015} as well as electron-spins \cite{Stockill2017}. An overview of achieved TPI visibilities with remote QD single photons is given in \textbf{Table \ref{tab:table2}}.
\\
\begin{table}
\centering
\caption{Chronological overview of achieved visibilities in remote QD TPI experiments}
\label{tab:table2}
\begin{threeparttable}
\begin{tabular}{cccc}
\hline
Wavelength in nm & Tuning mechanism & TPI visibility [$\%$] & Reference \\ \hline
 940 &  Electrical   &  33 $\pm$ 1\tnote{(a)} & Patel et al. 2010 \cite{Patel2010}    \\
920 &  Piezo strain &  18 $\pm$ 1  &  Flagg et al. 2010 \cite{Flagg2010}    \\
930 & Temperature  & 39 $\pm$ 2    & Gold et al. 2014 \cite{gold2014}    \\
945 &  Temperature &  40 $\pm$ 4  & Giesz et al. 2015 \cite{Giesz2015}  \\
 955 &  Electrical   &  91 $\pm$ 6    & Delteil et al. 2015 \cite{Delteil2015} \\
933 & Temperature  &  29 $\pm$ 6    & Thoma et al. 2016 \cite{Thoma2016}   \\
1250 & Laser-induced Evaporation & 33 $\pm$ 1\tnote{(a)} & Kim et al. 2016 \cite{Kim2016a}\\
750 & Piezo strain &  51 $\pm$ 5    & Reindl et al. 2017 \cite{Reindl2017}   \\
 968 & Electrical  & 93 $\pm$ 1     & Stockill et al. 2017 \cite{Stockill2017} \\
795 & Piezo strain &  41 $\pm$ 5\tnote{(b)} & Zopf et al. 2018 \cite{Zopf2018} \\
1550 & Frequency Conversion & 29 $\pm$ 3  & Weber et al. 2019 \cite{Weber2019}  \\
780   & Electrical  & 93 $\pm$ 1 & Zhai et al. 2021 \cite{Zhai2021}  \\
1583 & Frequency Conversion & 67 $\pm$ 2\tnote{(c)}  & You et al. 2021 \cite{You2021} \\ \hline
\end{tabular}
 a) used temporal post-selection (CW experiment); b) with active feedback; c) (93 $\pm$ 4)$\,\%$ with temporal filtering
\end{threeparttable}
\end{table}
A proof-of-concept entanglement swapping (\textbf{Figure \ref{fig:fig17}a}) experiment with polarization entangled photons from QDs was reported by Zopf et al. \cite{Zopf2019} as well as by Basset et al. \cite{Basset2019}. Here, the entangled photon pairs did not yet originate from two different QDs, since the remote TPI visibility was not sufficient. Instead, two entangled photon pairs, that are subsequently emitted from the same QD, were each split up spectrally, with one photon of each pair being sent to a receiver module, where a partial BSM is performed on the joined state of one photon from each entangled photon pair (Figure \ref{fig:fig17}b). Note that it is important to pick photons with the same energy for them to be indistinguishable, here they use the XX-photon from both entangled photon pairs. 
\\
In at best $25\%$ of the TPI events, a coincidence indicates a projection into the $\ket{\Psi^{+}}$ Bell state (note that despite the photons leaving from different output ports of the BS here, a phase shift of $\pi$ rotates them into the $\ket{\Psi^{+}}$ state). For these cases, the two remaining photons violated the CHSH inequality, in other words were entangled, which means that entanglement swapping took place. State tomography revealed that after the swapping, the two X photons were in the entangled $\ket{\Psi^{+}}$ state (Figure \ref{fig:fig17}d), but without the BSM, their density matrix was maximally mixed (Figure \ref{fig:fig17}c). Using the determined density matrices, Zopf et al. obtained a fidelity of the joint state of the remaining photons, to the Bell state, of $81\%$ \cite{Zopf2019}. 
\\
In addition to being the first proof-of-principle experiment of entanglement swapping with entangled photons from QDs, the emission wavelength was 780 nm which is close to the D2 optical-transition line in Rubidium, making it compatible to Quantum memories (see \textbf{section 4.3}). Note also, that entanglement swapping of QD entangled photon pairs creates two entangled photons at the same energy, which is different from the typical XX-X emission cascade in QDs.
\\
Basset et al. did a similar experiment but triggered their coincidences on projections into the $\ket{\Psi^{-}}$ Bell state, using coincidences from two different outputs of the BS \cite{Basset2019}. As for Zopf et al., the obtained fidelity of the entangled state after the swapping is mainly limited by the indistinguishability of the used photon pairs and their initial degree of entanglement. To quantify that, the authors introduced a model which incorporates the initial entanglement fidelity, mainly determined by the amount of FSS, and the HOM visibility, which reproduces their measured results (Figure \ref{fig:fig17}e). 
\\
Using the parameters of state-of-the-art QD entangled photon sources \cite{Huber2018,Liu2019}, they predict a maximum achievable fidelity of the swapped entanglement of $83\%$ with current technology. Even higher values can be expected with further improvements in the foreseeable future, opening up the route for entanglement swapping using remote QD-sources and ultimately quantum repeaters based on sub-Poissonian quantum light sources.
\\
\begin{figure}[h]
  \includegraphics[width= \linewidth]{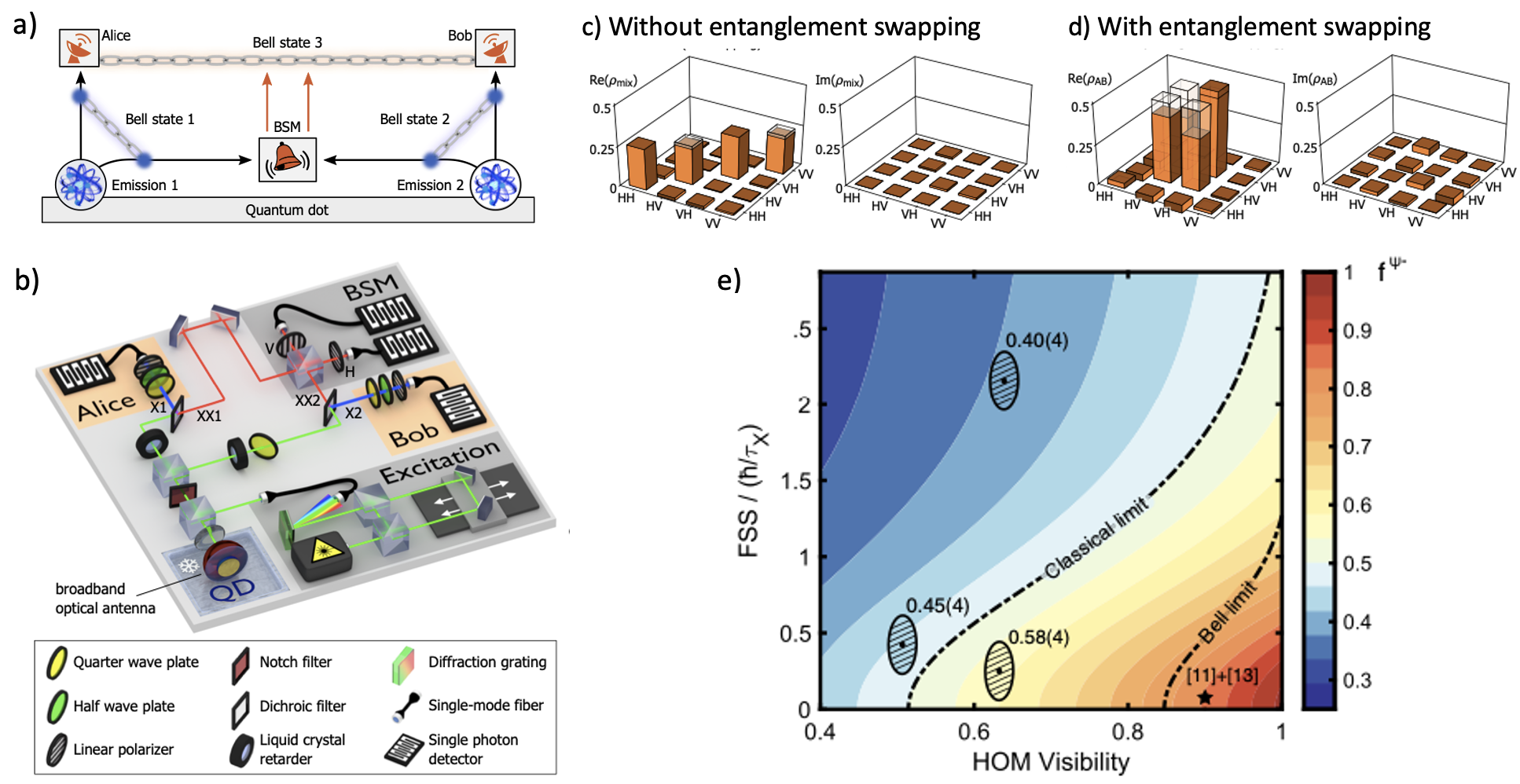}
  \caption{Realization of entanglement swapping (a) with entangled photons emitted by the same QD (b). State tomography revealed that the two X photons were afterwards in the entangled psi-plus state (d), but without the BSM, their density matrix is maximally mixed (c) \cite{Zopf2019}. Achievable fidelity of the state after entanglement swapping as a function of TPI visibility and FSS (f) \cite{Basset2019}. (a-d) Reprinted with permission from \href{https://doi.org/10.1103/PhysRevLett.123.160502}{\textit{Zopf et al. 2019}} \cite{Zopf2019} Copyright 2019 by the American Physical Society, (e) reprinted from \href{https://doi.org/10.1103/PhysRevLett.123.160501}{\textit{Basset et al. 2019}} \cite{Basset2019} under Creative Commons License CC BY 4.0.}
  \label{fig:fig17}
\end{figure}
To summarize, while making remote single photons from different QDs indistinguishable enough to project them reliably into a joint BSM is a challenge that requires high quality samples, resonant excitation schemes and very good experimental control, it has already been demonstrated several times with TPI visibilities surpassing the classical limit of $50\%$ for WCPs. This promises further advances in QKD schemes and quantum repeaters relying on remote sources.

\subsection{Quantum Memory} \label{section4.3}
Another crucial building block for many applications in quantum information is the quantum memory. An ideal quantum memory stores a quantum state with zero decoherence for an infinite amount of time and allows on-demand retrieval of the same quantum state for further use. It does so at a high bandwidth for fast operation and without introducing additional photon noise. In a quantum computation scenario, a quantum memory is necessary to delay computational steps, temporarily store quantum states, and hereby allow for more complex algorithms or new computation schemes such as linear optical quantum computing (LOQC) \cite{Kok2005}. Other applications of quantum memories include quantum metrology \cite{Giovannetti2011}, quantum machine learning \cite{Biamonte2017}, single-photon detectors \cite{Imamoglu2002} and more (see \cite{Bussieres2013} for an in-depth review).
\\
In communication scenarios we consider in this review, quantum memories are required for the implementation of quantum repeater protocols (cf. previous subsection), which allow to distribute entanglement and enable QKD over, in principle, arbitrary distances \cite{Briegel1998}. Note, that although there exist all-photonic repeater schemes, which do not rely on quantum memories, they are complex in other ways, e.g., demanding multiple multiplexed quantum channels or cluster states \cite{Li2019,Borregaard2020}. Quantum communication also profits indirectly from a quantum memory, e.g. from memory-assisted two-photon interference and facilitated teleportation, as discussed in \cite{Ma2019}.
\\
Different types of memories have been proposed and demonstrated in the past, such as solid-state systems \cite{DeRiedmatten2008}, trapped atoms \cite{Bao2012} and alkali vapor cells \cite{Eisaman2005}. For an extensive overview of memory protocols and platforms we refer to \cite{Heshami2016}. Here, we want to focus on memories that are suitable for QD-SPSs, as well as practical enough for scalable, optical quantum networks.
\\
As a natural way of creating nodes for an optical quantum network is to combine QD-SPSs with compatible quantum memories, one has to ensure that the memories can keep up with the quantum emitters in terms of efficiency and bandwidth. QDs allow high emission rates, thus the memory should be able to operate at a similar rate. Since not every pulse emitted by a real QD contains a photon, memories also require low noise backgrounds so that it is possible to distinguish a retrieved photon from the noise floor. In the following, we will review recent advances in quantum memories compatible with QDs (see also Neuwirth et al. \cite{Neuwirth2021} for an in-depth review).
\\
Alkali vapor cells are promising candidates to realize QD-compatible memories (\textbf{Figure \ref{fig:fig18}a}). Here, the group velocity of light pulses propagating through the atomic ensemble can be reduced, i.e. light pulses are slowed down, which allows for photon storage \cite{hau1999}. Using warm alkali vapor cells allows one to omit the complex cooling infrastructure necessary for many solid-state memories and the laser cooling necessary for memories based on cold atoms. They are also practical for larger networks, because they can offer sufficiently long coherence times \cite{Borregaard2016} and profit from a large set of existing memory protocols \cite{Lvovsky2009}. Moreover, solid-state memories and ultracold atoms have much less noise, smaller bandwidths, and longer storage times, making them more desirable for long-term memories, but are, due to the smaller bandwidth, not compatible with QD single photons in quantum networks.
\\
To store photons in the atomic ensemble inside an alkali vapor, one uses transitions between three energy levels (a so-called $\lambda$-system), with two spectrally-close lower states and one excited state at a higher energy, which enables an effect called electromagnetically induced transparency (EIT) \cite{Fleischhauer2000,Ma2017} (Figure \ref{fig:fig18}b). While the to-be-stored quantum state is resonant with one transition, a control pulse enables the excitation by driving the other transition. On the atomic level, all the atoms are prepared in a joint ground state, before the excitation by the signal photon, in combination with a control pulse, leads to the creation of an atom-spin-wave, in which the coherence is stored. With a second control pulse, the spin-wave is transformed back into an optical excitation and the photon can be retrieved again (Figure \ref{fig:fig18}c).
\\
Typically, signal- and control- pulse are detuned from the resonant atomic transition, to reduce noise due to immediate fluorescence. One way of detuning them and still driving the transitions, is by using Raman scattering, where the anti-stokes-shift helps matching the transition energy, as shown by Reim et al. \cite{Reim2011}. While detuning reduces the noise floor, higher detuning reduces the memory efficiency and requires higher control pulse intensities and consequently a better suppression of the control pulse in the memory output, thus a compromise between low noise and efficient excitation must be found.
\\
Initial theoretical comparisons of different schemes of using Rubidium vapor cells combined with QD single photons were performed by Rakher et al. \cite{Rakher2013} and highlighted the necessary steps towards storing QD single photons in a high bandwidth memory. Such a memory was demonstrated with an end-to-end efficiency of $3.4\%$ by Wolters et al., using a warm Rubidium vapor cell to store photons (emitted by a laser) for a time of up to 50 ns exploiting EIT and achieving a bandwidth of 0.66 GHz, compatible with typical QD-SPSs \cite{Wolters2017}.
\\
The quantum memory demonstrations presented up to here only proved that a photon could be stored and retrieved with finite efficiency, but not that the retrieved photon was actually the same, i.e. indistinguishable from the initial photon. This was shown first by Hosseini et al. who  achieved $98\%$ process fidelity by performing quantum state tomography on the retrieved photons \cite{Hosseini2011}. In this context, another insightful experiment towards quantum memories was performed by Vural et al., in which the authors performed TPI measurements between two photons of which only one went through the alkali vapor cell, proving that the two photons were still indistinguishable after interacting with the atomic ensemble \cite{Vural2018}. Moreover, the state-of-the-art for hot vapor cell quantum memories in terms of efficiency was recently demonstrated by Guo et al. who achieved an efficiency of >$82\%$, using the Raman-detuned quantum memory scheme and attenuated laser pulses as the signal photons \cite{Guo2019}.
\\
However, standard EIT in vapor cells does not allow storing qubits in which the information is encoded in the polarization, since the EIT memory only preserves the phase, but not the polarization. This was solved by England et al. achieving $98\%$ fidelity of polarized photons using a dual-rail memory \cite{England2012}, as well as by Namazi et al. demonstrating storage of single polarized photons and their retrieval with a fidelity to the original polarization state exceeding $90\%$ \cite{Namazi2017}, which they also employed in QKD experiments \cite{namazi2017free}. For this purpose, both groups used polarization-dependent displacement optics (Glan-laser polarizer) to transform the polarization-encoding into a path-encoding, defining two spatially separate paths through the memory, and then converting the qubits back to polarization-states after leaving the memory (Figure \ref{fig:fig18}d). Hereby, each polarization component is individually stored by taking a different path through the memory medium. Since the entangled photons generated from the XX-X radiative cascade in QDs are polarization entangled, the demonstration of a quantum memory for polarization qubits was a major step towards memory-based quantum networks using QD quantum light sources. Recently, also the storage and retrieval of a pair of polarization entangled photons inside a quantum memory was demonstrated \cite{Ding2018}.
\\
\begin{figure}
  \includegraphics[width= \linewidth]{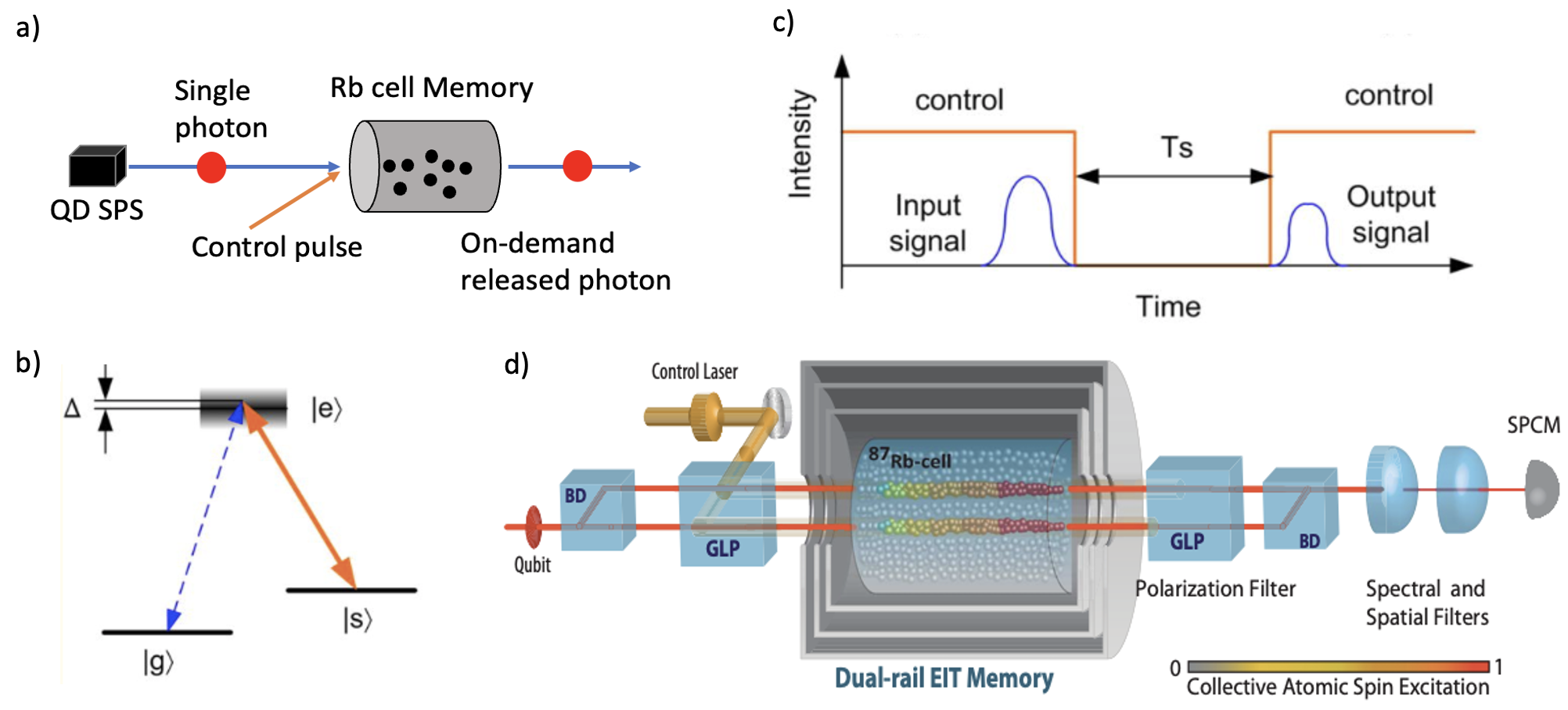}
  \caption{Storing a QD single photon in a Rb vapor cell (a) works via Electromagnetically Induced Transparency in a $\lambda$-system (b), so that a sequence of control pulses can define the light storage in the memory (c). To store polarization qubits, the polarization encoding can be translated into path encoding via beam displacer (BD) and Glan-laser polarizer (GLP in d) and transformed back after the memory \cite{Namazi2017}. (d) reprinted Figure with permission from \href{https://link.aps.org/doi/10.1103/PhysRevApplied.8.034023}{\textit{Namazi et al. 2017}} \cite{Namazi2017} Copyright 2017 by the American Physical Society.}
  \label{fig:fig18}
\end{figure}
Even with existing quantum memories that are capable of being combined with QDs in terms of bandwidth, there are still a couple of issues to be addressed. The first one is the question of how one can, in practice, precisely match the QD emission energies to the transition energies in the atomic vapor ensemble. Additionally, the memory used in a future quantum network must also be able to deal with imperfect QDs, which are for example subject to spectral diffusion and dephasing, both broadening the emission and reducing the single-photon indistinguishability. These effects were theoretically treated by Rakher et al. finding that both fluctuations in phase and wavelength of the QD photons significantly reduce the memory efficiency \cite{Rakher2013}.
\\
The challenge of spectrally matching artificial atoms with their natural counterparts, i.e. matching photons emitted by QDs to transitions of alkali atoms, was first addressed by Akopian et al. in 2011, by slowing down single photons from QDs in an atomic ensemble \cite{Akopian2011}. To do so, the authors shifted the QD emission energies to the memory transition energy via magnetic Zeeman tuning. In addition, they also showed that, after filtering the QD signal by a spectral window smaller than the linewidth of the atomic ensemble, the remaining photons, with fluctuating emission energies due to spectral diffusion, were all slowed down by the same amount (experienced the same storage times in the memory). Other ways of matching the QD emission energies with the atomic vapor transitions are piezo-strain tuning as done by Jahn et al. \cite{Jahn2015}, temperature variation \cite{Bremer2020} and over a large range covering all the relevant alkali vapor transition energies by Zhai et al. \cite{zhai2020}. Additionally, tuning can be done by using a “dressed state” resonance fluorescence \cite{Vamivakas2009}, which does not require any electric or magnetic field tuning as shown by Ulrich et al. \cite{Ulrich2014}. Finally, we would like to stress that although single photons from QDs have been successfully stored and retrieved with high fidelity from quantum memories, these types of quantum memories are not yet sufficient, since they do not allow on-demand retrieval. They can only release the photon by a control pulse which is defined by the pulsed control laser, but the storage time cannot be adapted in a feedforward manner yet. Here, further progress will be necessary to obtain a true quantum memory.
\\
Apart from storing a photon in an atomic ensemble, one can also store it in a solid-state system. Such a system could, for example, be a rare earth-ion system using an atomic frequency comb \cite{Usmani2010}. However, such systems typically have a strong dependence on the polarization, which is why storing polarization qubits was difficult initially. But, by using two rare-earth crystals of Nd$^{3+}$:YVO$_4$, Zhou et al. also demonstrated the storage of polarization qubits \cite{Zhou2012}. Later, the same group showed how a sequence of single photons emitted by a QD can be stored in their memory for 40 ns \cite{Tang2016}. Additionally, they found that even for imperfect QD sources (with significant multi-photon contributions), only one photon is stored, thus increasing the purity of the qubit through the quantum memory.
\\
Interestingly, also the ensemble of nuclear spins inside a QD can be considered a solid-state quantum memory itself, as demonstrated by Kroutvar et al. using electron spins \cite{Kroutvar2004} and also the so-called dark-exciton state in a QD can serve as a memory \cite{McFarlane2009}. The advantage of nuclear spin memories is that they are shielded much better from the environment, due to the small magnetic moments of the nuclei. Thus, long-term storage of quantum bits (a quantum hard-drive) is more likely to be possible in such solid-state memories, limited only by dipole-dipole interactions among the nuclear spins \cite{Taylor2003}, which can be suppressed as demonstrated by Kurucz et al. \cite{Kurucz2009}. Other examples of nuclear spin memories also include nuclear spins of Carbon atoms coupled to NV centers in diamonds \cite{Shim2013} or dopants in Silicon \cite{Morton2008}. Although solid-state memories show excellent properties for photon storage, their reduced coupling to the environment can also be a disadvantage, making the read and write processes more difficult.
\\
An alternative way of transferring the qubit state from a flying qubit, such as a QD single photon, onto a solid-state memory is by teleporting the state, using an entangled photon pair. In this scheme one photon of the entangled state would be stored in a local solid-state quantum memory, while the other is sent to the to-be-stored flying qubit. By projecting the two in a joint Bell state, the flying qubit gets teleported into the quantum memory, as was demonstrated by Bussières et al. using entangled photons at Telecom wavelengths, created via parametric down conversion, and using a rare-earth memory \cite{Bussieres2014}. They achieved a fidelity above $80\%$ over a distance of 12 km.
\\
In summary, significant progress has been made during the last decade in optical quantum memories, especially increasing compatibility with QD-SPSs. Challenging tasks that remain are to further enhance the efficiency and reduce the noise (increase the signal to noise ratio to allow for non-ideal QDs) as well as the implementation of on-demand retrieval mechanisms. On the QD side research must focus on creating tunable, fourier-limited single photons which are subject to less line broadening and can thus be better integrated with quantum memories.
\subsection{Teleportation} \label{section4.4}
Teleportation consists of sending a previously unknown quantum state from Alice to Bob without physically sending the qubit itself, but by ‘sacrificing’ an entangled state shared between Alice and Bob, and using classical communication (\textbf{Figure \ref{fig:fig19}a}). Hereby, the state to be teleported is projected into the Bell basis in a joint measurement together with one half of the entangled state at Alice. The result of that Bell measurement then determines the unitary that must be applied to the other half of the entangled state by Bob to retrieve the teleported state at Bob’s side \cite{Bennett1993}.
\\
Teleportation is a vital building block in future quantum networks as it provides a way of realizing non-local quantum computations and can be used to transfer qubits into a solid-state quantum memory \cite{Bussieres2014}. Teleportation can also be directly used for QKD (given a shared entangled state) by teleporting an encoded state from Alice to Bob, which is known as a quantum relay. Quantum relays using entangled photons from QDs have been demonstrated by Varnava et al. over a distance of 1 km \cite{Varnava2016} and later by Huwer et al. using photons at Telecom wavelengths \cite{Huwer2017}. In the following we discuss advances in quantum teleportation enabled by QD-based quantum light sources.
\\
As discussed earlier, the efficiency of BSMs is reduced by multi-photon events unavoidable in implementations using WCP or SPDC sources.  Therefore, also teleportation can benefit from the use of entangled photons created by deterministic QD-SPSs. The first QD-based proof-of-concept teleportation experiment was reported by Nilsson et al. using an QD entangled-light emitting diode \cite{Nilsson2013}. The authors used the entangled photon pair emitted in one pulse to teleport a photon from the subsequent photon pair (Figure \ref{fig:fig19}b) and obtained a maximum teleportation fidelity above the classical threshold of 2/3. Although teleported photon and entangled photon were created by the same device, the authors emphasized that the teleported photon can in principle also stem from an external source. In fact, their QD emission energy was tunable, to match the wavelength of an incoming external photon, to facilitate TPI and enable teleportation. But the operation wavelength at 890 nm was not yet compatible with standard fiber-optical communication networks.
\\
Teleporting external photons with a QD entangled-photon source emitting in the Telecom C-band was recently demonstrated by the same group using an attenuated laser pulse \cite{Anderson2020} (Figure \ref{fig:fig19}c). In an initial characterization measurement, they found a HOM visibility of up to $70\%$ for the TPI between the attenuated laser pulse and single photons emitted from their non-resonantly excited QD, a value significantly above the maximum value for TPI between two attenuated laser pulses of $50\%$. The fidelity observed for teleporting a laser photon was exceeding $85\%$ (Figure \ref{fig:fig19}d).
\\
\begin{figure}
  \includegraphics[width= \linewidth]{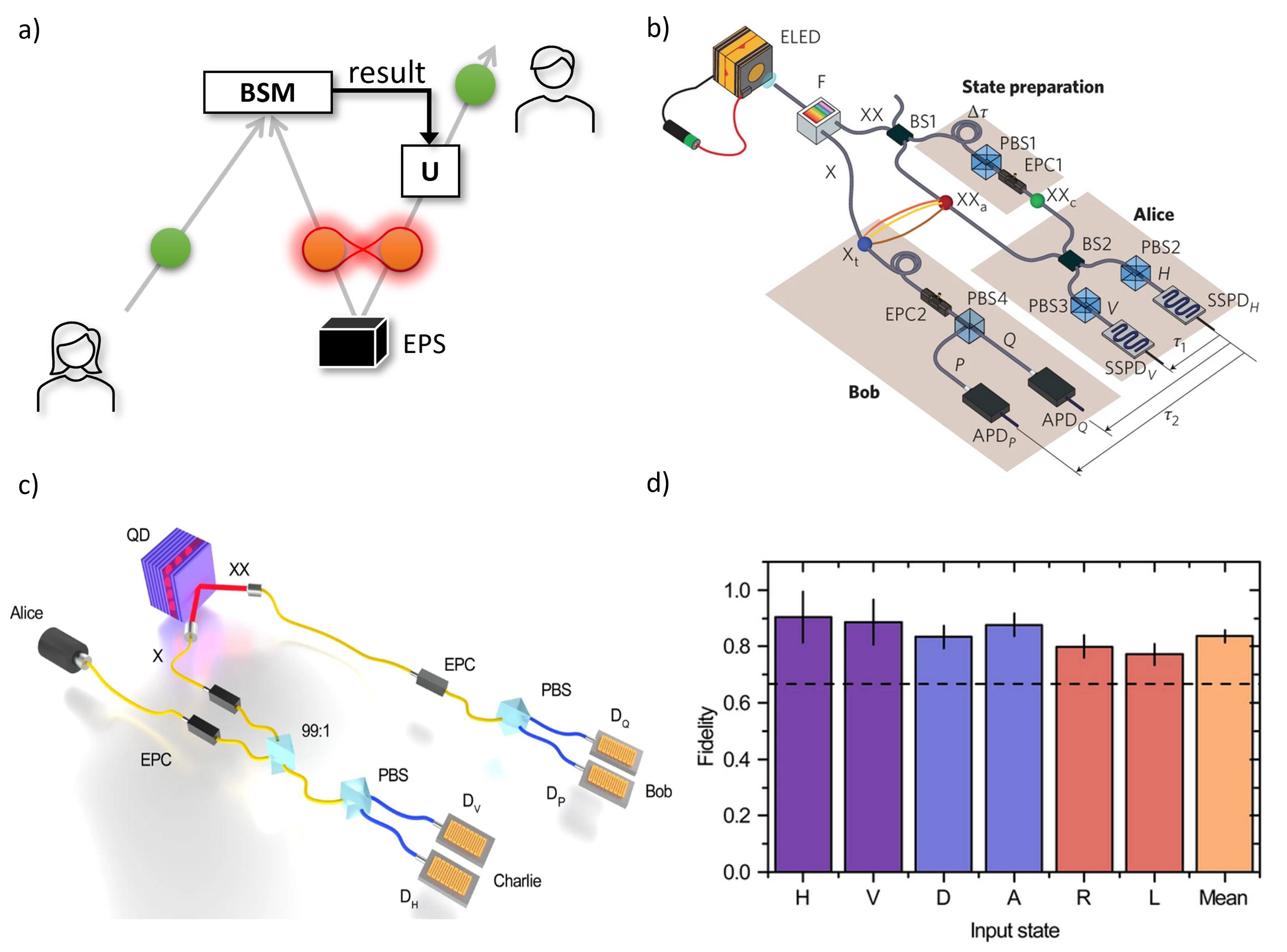}
  \caption{The quantum teleportation scheme (a) was implemented with target photon and entangled photons both emitted by the same QD by Nilsson et al. (b) \cite{Nilsson2013} and with a laser photon at telecom wavelengths as the target photon by Anderson et al. (c) achieving teleportation fidelities above the classical limit (d) \cite{Anderson2020}. (b) reprinted by permission from Springer Nature: Nature Photonics \href{http://dx.doi.org/10.1038/nphoton.2013.10}{\textit{Nilsson et al. 2013}} \cite{Nilsson2013} Copyright 2013, (c,d) reprinted from \href{https://doi.org/10.1038/s41534-020-0249-5}{\textit{Anderson et al. 2020}} \cite{Anderson2020} under Creative Commons license CC BY.}
  \label{fig:fig19}
\end{figure}
In recent years, research efforts focused on improving QD-based quantum light sources, reducing the FSS, and making the emitted photons or photon pairs more indistinguishable. However, already with imperfect QDs relevant applications are already possible. Basset et al. recently demonstrated teleportation experiments with photons from QDs, for which they deliberately chose a QD of below-average quality \cite{Basset2020a}. The authors used one photon from every second entangled photon pair as the to-be-teleported input state. They showed, that projecting into two Bell states, instead of only one (cf. \cite{Reindl2018}), in the partial BSM does not only increase the efficiency, but also reduces the impact of  an imperfect indistinguishability (cf. \textbf{Figure \ref{fig:fig20}} a and b for one-state and two-state BSM, respectively). A part of the photons that previously led to false coincidence counts since they were not indistinguishable enough, will now be identified from their polarization. The authors further enhanced the teleportation fidelities via spectral filtering, which improved the photon indistinguishability (Figure \ref{fig:fig20}c) and also proposed a model from which the process fidelity of the teleportation protocol can be predicted for given QD characteristics (FSS and photon-indistinguishability) (Figure \ref{fig:fig20}d).
\\
\begin{figure}
  \includegraphics[width= \linewidth]{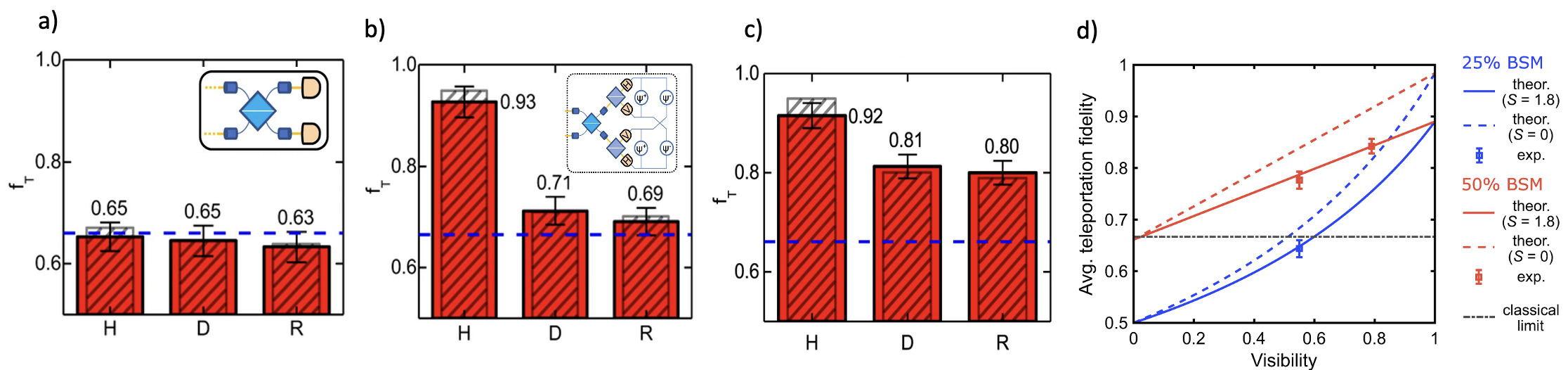}
  \caption{Teleportation with entangled photons from QDs: Comparison of teleportation fidelities with a $25\%$ BSM (a) and a $50\%$ BSM (b), both with low visibility, and one with higher visibility due to spectral filtering (c). Dependence of teleportation fidelity on visibility, FSS S and BSM type (d). Figures reprinted from \href{http://dx.doi.org/10.1038/s41534-020-00356-0}{\textit{Basset et al. 2020}} \cite{Basset2020a} under Creative Commons CC BY license.}
  \label{fig:fig20}
\end{figure}
Noteworthy, also other teleportation protocols exist, such as single mode teleportation which does not require entangled photon pairs and is related to the proposal of linear optics quantum computation \cite{Knill2001}. This protocol was experimentally demonstrated with QD single photons by Fattal el al. \cite{Fattal2004}.
\\
As teleportation relies on the same ingredients as quantum repeaters for quantum networks, namely large TPI visibilities and entangled photon pairs,  it will also profit  by further advances in the development of  QD-based quantum light sources. Due to the previous limits set to the TPI visibility, teleportation of a photon from a QD using an entangled photon pair from another QD has not been demonstrated so far. But the recent advances discussed in \textbf{section 4.2} show prospects that this can be achieved in the near future.
\subsection{Quantum Random Number Generation} \label{section4.5}
Among the building blocks required for quantum networks, quantum random number generators (QRNG) were the first to be realized using for instance a radioactive decay \cite{Schmidt1970}, for a review of random number generation see \cite{Herrero-Collantes2017}. As a result, the technology readiness level is the highest among all building blocks, as confirmed by its widespread commercial availability – recently even in consumer smartphones \cite{idquantique} - and standardization efforts are also underway \cite{Hart2017}.
\\
All existing QKD protocols require reliable sources of randomness. In addition true random numbers are of course also vital in many other domains ranging from simulations and computing to online casinos. It is not enough to have a sequence of uniformly distributed random bits, but they must also fulfill the requirements for forward and backward security. The final key will only be as secure, as the initial random numbers are actually random. Be it in the basis choices in the BB84 scheme or in the choice of 2-universal-hash functions during the privacy amplification post processing step, an initial random seed is needed to make quantum communication work.
\\
It has been shown that ideal QRNGs are the only perfect source of random numbers, as opposed to pseudo-random number generators \cite{VonNeumann1963,Peres1992}. While computers typically rely on pseudo randomness, where a random seed and an algorithm is used to generate the numbers, or on classical physical randomness, based on the complexity of classical systems such as thermodynamic systems, only quantum processes can provide true sources of randomness \cite{Ma2016}. This boils down to the proof of having no hidden variables, so that the outcome of a single projective measurement cannot be predicted by any means, which is why randomness can be certified via the violation of Bell-like inequalities \cite{Gallego2013}. The simplest optical QRNG can be described as a photon impinging on a BS and its wave function collapsing on a single-photon detector. These were in fact the first optical QRNGs that were implemented \cite{Jennewein2000,Stefanov2000}. In such a setup, it was also demonstrated that a true SPS can provide more randomness than a bright laser \cite{Oberreiter2016}. 
\\
By now, several other QRNG schemes have emerged, which provide even faster random bit rates using for instance photon arrival times \cite{Furst2010,Wayne2010,Wahl2011}. The quantum phase fluctuation and vacuum state schemes achieve Gbps bandwidths \cite{Haylock2019,Lei2020,Bai2021} and do not rely on SPSs. Even faster QRNGs are available \cite{Liu2017}, but these lack the electronics to handle this amount of data in real time – a similar problem to the classical key reconciliation in practical QKD applications. Parallel to the developments in QKD, also in QRNG researchers now shift the focus from trusted devices to the development of device independent (DI) approaches \cite{Gomez2019}. Like in QKD, full DI was not achieved yet and the intermediate approaches like Semi-Device-Independent \cite{Avesani2021} or self-certifying approaches suffer from lower bit generation rates. For an overview of different QRNG platforms and the state-of-the-art we refer to \cite{Herrero-Collantes2017} and \cite{Hart2017}.
\\
Concerning QRNG based on semiconducting QDs one must note that QKD protocols are agnostic towards where the random numbers come from, as long as their rate is sufficiently high to keep pace with QD emission rates. Nevertheless, it is interesting to note that specific ways of creating QRNG with QDs exist, which could in the future be combined with their property as a SPS to generate random numbers and true single photons on the same platform \cite{McCabe2020,Purkayastha2016}. While QDs can improve many ingredients of future quantum networks, they appear to be not the optimal choice for the generation of random numbers in terms of practicality and possible advantages.

\subsection{Towards Quantum Networks - Practical Challenges} \label{section4.6}
Finally, there are a few hands-on issues to be addressed when attempting to build a stable and functional quantum network that is immune against environmental fluctuations. Here, we review progress in maintaining the stability of potential quantum networks via active stabilization, monitoring, and optimization of the building blocks making up the network.
\\
One important issue is how one can guarantee the stability of a QKD channel for long times so that the security is not compromised by an increase in QBER \cite{Zhu2002}. Therefore, active stabilisation schemes will be necessary, e.g. using bright beacon lasers that are multiplexed in the quantum FSO channel, as demonstrated in the seminal work by Ursin et al. in 2007 using a 144 km link between the Canary Island of La Palma and Tenereiffe and entangled photon-pairs generated via an SPDC source \cite{Ursin2007}. Such stabilization approaches have been used to improve QKD experiments employing local free-space channels \cite{carrasco2014} and also to transmit quantum states from satellite to ground \cite{Yin2017}. Using QD-generated entangled photon pairs at telecom wavelengths, an advanced stabilization scheme has recently been demonstrated by Xiang et al. \cite{Xiang2019} by continuously exchanging entangled photon pairs over 18 km of optical fiber for more than a week (\textbf{Figure \ref{fig:fig21}a}). The current state-of-the-art for mechanical stabilization of free-space links was recently achieved by Liu et al. exchanging SPDC-generated qubits between flying drones \cite{Liu2021entanglement}. Moreover, it is important to define a joint reference frame for the measurement of the qubits (unless reference-frame independent schemes are employed \cite{Laing2010,Wabnig2013} ), which can also be done via auxiliary, multiplexed lasers \cite{Nauerth2013}. In addition to stabilizing the quantum channel, it should also be characterized well, to optimize the used QKD scheme for maximum performance. This can be done experimentally via optimization routines, numerically via key rate predictions or even using machine learning approach \cite{Ismail2019}.
\\
Another practical issue refers to the synchronisation between senders and receivers which can either be done by synchronisation pulse sequences or a multiplexed synchronisation signal, but also via GPS clocks \cite{Bienfang2004,Pljonkin2017}. For fast stabilization and high key rates it will also be important to achieve fast modulation of the phase and polarization of photons \cite{Grunenfelder2018,Li2019}, for which integrated photonics systems promise the best modulation rates \cite{Sibson2017,Bunandar2018}, as well as high speed detectors with low dead times, especially at Telecom wavelengths, for an overview see \cite{Zadeh2021}.
\\
Not only the stability, but also the security has to be continuously monitored. As mentioned before, multi-photon pulses enable the photon number splitting attack, which is why sources have to be well characterized in order to compensate for their non-ideal photon statistics. The amount of multi-photon contributions could however change (or be changed by an adversary) which is why it is necessary to monitor the multi-photon contributions during key exchange by using a subset of the photons for purity checks, as demonstrated by Kupko et al. in a real-time security monitoring approach (Figure \ref{fig:fig21}b) \cite{Kupko2020}. In their work, the authors also demonstrated how temporal filtering on the receiver side can enhance the signal to noise ratio to optimize the secure key rate for a given channel loss. Implementing such approaches, special care must be taken not to open the door for other side-channel attacks, as an adversary could otherwise unnoticedly steal photons outside the temporal acceptance window. Hence, an implementation which wants not only to benefit from the improved signal-to-noise ratio but also by estimating a lower $g^{(2)}(0)$, temporal filtering has to be applied on sender side alike (or $g^{(2)}(0)$ needs at least to be monitored also inside Alice). Another issue is that because of finite-key size effects, a minimum block size must be reached before the post processing step - as implemented nowadays in commercial systems \cite{Chaiwongkhot2017}.
\\
\begin{figure}[h]
  \includegraphics[width= \linewidth]{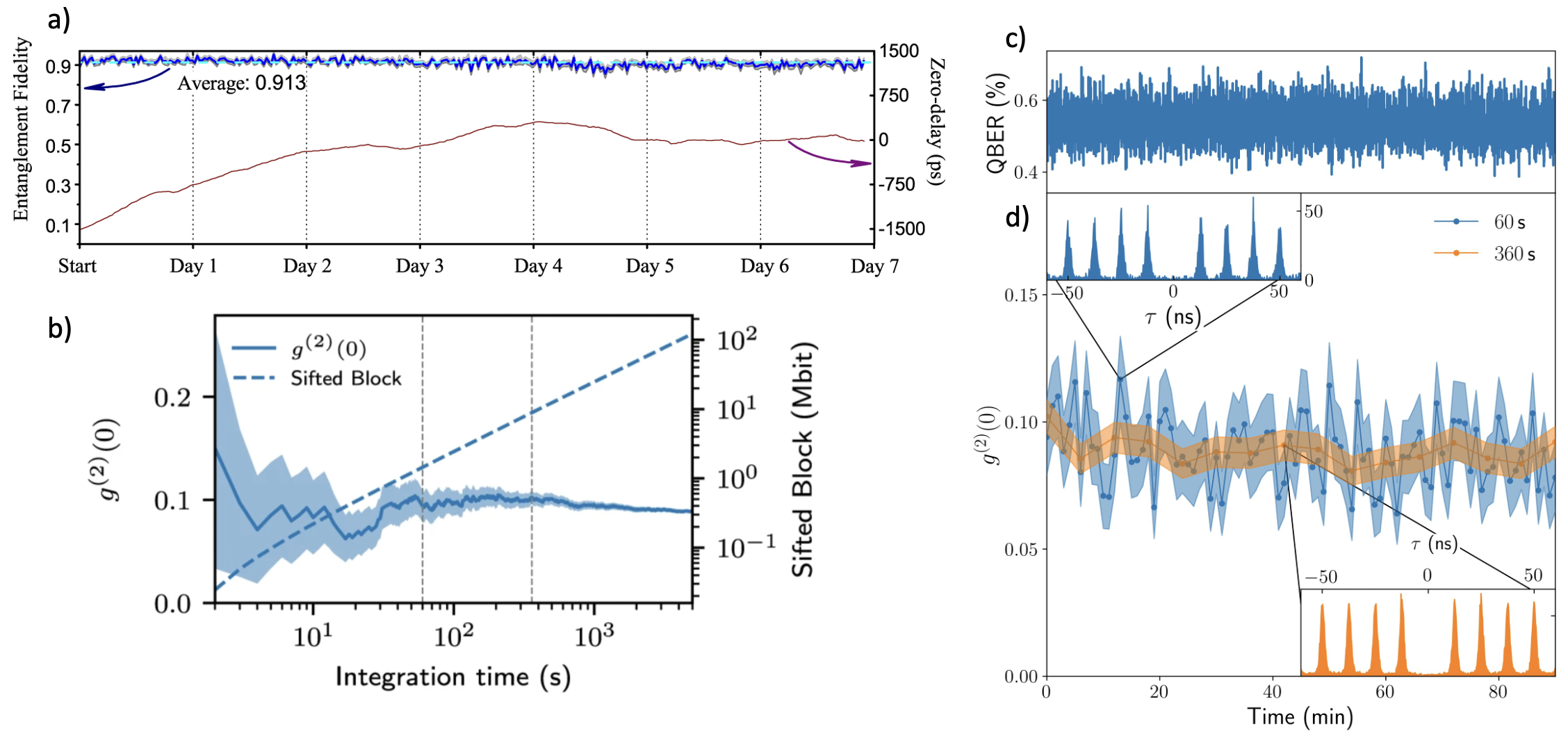}
  \caption{Stable exchange of entangled photon pairs via optical fiber for an entire week (a) \cite{Xiang2019}, continuous measurements of relevant QKD parameters such as QBER (c) and photon autocorrelation values (d) enable real-time security monitoring, as long as a minimum integration time for conclusive $g^{(2)}$ results is set (b) \cite{Kupko2020}. (a) reprinted from \href{http://www.nature.com/articles/s41598-019-40912-z}{\textit{Xiang et al. 2019}} \cite{Xiang2019} under Creative Commons Attribution 4.0 International License, (b) reprinted from \href{http://dx.doi.org/10.1038/s41534-020-0262-8}{\textit{Kupko et al. 2020}} \cite{Kupko2020} under Creative Commons Attribution 4.0 International License.}
  \label{fig:fig21}
\end{figure}
The abundance of a classical authenticated channel is always assumed implicitly in QKD protocols. Even if an initial secret is shared, authentication has to be repeated regularly which uses up a fraction of the key, which is why protocols for efficient authentication are necessary \cite{Ljunggren2000,kiktenko2020}. Alternatively, QKD can be combined with other post-quantum cryptography schemes to support the authentication \cite{Wang2021}.
\\
A challenge that increases in importance with the growing maturity of the developed QKD systems refers to the security certification. One could either rely on device-independent schemes or use special protocols to confirm that a QKD system is working properly \cite{Tomita2019}. That includes the certification of the initial randomness, the integrity of the quantum channel, the purity of the source, as well as the correctness of the post-processing steps.
\\
Finally, in order to benchmark different QKD protocols and different technology platforms, a general framework beyond stating only the maximum secure key rate or the maximum achieved distance must be developed, since these parameters are highly dependent on the specific laboratory setup used with a specific source and under certain conditions. Therefore, testing standards are envisioned, to reliably compare different approaches \cite{Alleaume2004,Langer2009,alleaume2014}. That could be done for instance by ensuring that they certify the same amount of overall $\epsilon$-security (see the discussion of security definitions in \cite{Renner2008}). Another useful figure-of-merit could be the “security-per-dollar-spent”, considering the fact that different QKD architectures, that in principle promise different levels of security, also have different levels of implementation difficulty and hence costs. And last but not least, while there is of course the ultimate aim to achieve unconditional security, ruling out even the most unlikely attacks (that are practically impossible but allowed by the laws of quantum mechanics), in practice one might be content with a more relaxed, applied form of security. This could either be a deliberate trade-off between security-gain and implementation-costs, or an intermediate step towards ultimate security. 
\\
The approach of assuming realistic restrictions on an adversary are known and even required from the field of cryptographic primitives beyond QKD in untrusted settings, e.g. quantum oblivious transfer in the so-called noisy storage model \cite{Wehner2008}, representing crucial building blocks for modern communication \cite{Broadbent2016} networks.

\section{Conclusion}
In this review we summarized the progress made in recent years in the field of quantum communication using quantum light sources based on semiconductor quantum dots. After revisiting the foundations of QKD and introducing semiconductor QDs as one of the most promising candidates for photonic implementations of quantum information, we comparatively discussed implementations of QKD using single photons as well as entangled photon pairs generated via engineered QD-devices. Next, we discussed recent progress in the development of key building blocks of future quantum networks and how they can benefit from / or become compatible with such semiconductor QDs. Considering the tremendous progress achieved in the field, functional quantum networks and real-world application appear to be within reach in the not too distant future. However, some important ingredients are still missing or require additional research efforts. For example,  it is necessary to combine the superior properties QD sources proved to be able to deliver, in terms of high efficiency, high brightness, high single-photon purity, large photon indistinguishability, and large entanglement fidelities, with practical and durable source modules operable outside shielded lab environments. Another important challenge concerns the development of efficient quantum memories with on-demand retrieval. Furthermore, protocols for multi-node architectures and schemes as well as standards for the security certification are to be developed. A major challenge, that has not yet been tackled at all using QD-devices and which was not discussed here, concerns the implementation of cryptographic primitives beyond QKD and in untrustful settings. Such primitives, however, are representing important building blocks for sensitive tasks in modern communication networks, such as the secure authentication at a bank's ATM. This highlights the rich field of quantum cryptography and new areas of research to discover.
\\
Finally, we want to emphasize that future quantum networks will certainly not be constituted of a single technology or a specific protocol. On the contrary, many different platforms and schemes will most probably be combined and coexist, including deterministic quantum light sources as well as WCP- and SPDC-based sources, two-party quantum cryptographic primitives like QKD and beyond, multi-party primitives, classical and post-quantum cryptography, different encoding schemes and various network architectures and topologies, each of which being used and optimized for its special purpose. Reviewing the achievements and success since the advent of the field of quantum cryptography, driven by ideas of S. Wiesner in the late 1960s, it seems reasonable to expect major steps towards the quantum internet within this decade.
\\
\end{justify}
\medskip

\medskip
\textbf{Acknowledgements} \par 
We gratefully acknowledge financial support from the German Federal Ministry of Education and Research (BMBF) via the project ‘QuSecure’ (Grant No. 13N14876) within the funding program Photonic Research Germany.

\medskip

\bibliographystyle{MSP}
\bibliography{references.bib}

\begin{thebibliography}{100}
\providecommand{\url}[1]{\texttt{#1}}
\providecommand{\urlprefix}{URL }

\bibitem{Singh1999}
S.~Singh,
\newblock \emph{{The code book: the evolution of secrecy from Mary, Queen of
  Scots, to quantum cryptography}},
\newblock \textbf{1999}.

\bibitem{Impagliazzo1989}
R.~Impagliazzo, M.~Luby,
\newblock In \emph{30th Annual Symposium on Foundations of Computer Science},
  January. IEEE,
\newblock ISBN 0-8186-1982-1,
\newblock ISSN 02725428, \textbf{1989} 230--235,
\newblock \urlprefix\url{http://ieeexplore.ieee.org/document/63483/}.

\bibitem{Shor1997}
P.~W. Shor,
\newblock \emph{SIAM Journal on Computing} \textbf{1997}, \emph{26}, 5 1484.

\bibitem{Bernstein2017}
D.~J. Bernstein, T.~Lange,
\newblock \emph{Nature} \textbf{2017}, \emph{549}, 7671 188.

\bibitem{Shor2000}
P.~W. Shor, J.~Preskill,
\newblock \emph{Physical Review Letters} \textbf{2000}, \emph{85}, 2 441.

\bibitem{Gisin2002}
N.~Gisin, G.~Ribordy, W.~Tittel, H.~Zbinden,
\newblock \emph{Reviews of Modern Physics} \textbf{2002}, \emph{74}, 1 145.

\bibitem{Wootters1982}
W.~K. Wootters, W.~H. Zurek,
\newblock \emph{Nature} \textbf{1973}, \emph{246}, 5429 170.

\bibitem{DiVincenzo2000}
D.~P. DiVincenzo,
\newblock \emph{Fortschritte der Physik} \textbf{2000}, \emph{48}, 9-11 771.

\bibitem{Kimble2008}
H.~J. Kimble,
\newblock \emph{Nature} \textbf{2008}, \emph{453}, 7198 1023.

\bibitem{Wiesner}
S.~Wiesner,
\newblock \emph{ACM Sigact News} \textbf{1983}, \emph{15}, 1 78.

\bibitem{Bennett1992a}
C.~H. Bennett, F.~Bessette, G.~Brassard, L.~Salvail, J.~Smolin,
\newblock \emph{Journal of Cryptology} \textbf{1992}, \emph{5}, 1 3.

\bibitem{Bennett1984}
C.~H. Bennett, G.~Brassard,
\newblock \emph{Proceedings of IEEE International Conference on Computers,
  Systems and Signal Processing, Bangalore, India} \textbf{1984}, 175--179.

\bibitem{Ekert1991}
A.~K. Ekert,
\newblock \emph{Physical Review Letters} \textbf{1991}, \emph{67}, 6 661.

\bibitem{Clauser1969}
J.~F. Clauser, M.~A. Horne, A.~Shimony, R.~A. Holt,
\newblock \emph{Physical Review Letters} \textbf{1969}, \emph{23}, 15 880.

\bibitem{Bell1964}
J.~S. Bell,
\newblock In \emph{John S Bell on the Foundations of Quantum Mechanics},
  volume~1, 7--12. WORLD SCIENTIFIC, \textbf{2001},
\newblock
  \urlprefix\url{http://www.worldscientific.com/doi/abs/10.1142/9789812386540_0002}.

\bibitem{Bennett1992}
C.~H. Bennett, G.~Brassard, N.~D. Mermin,
\newblock \emph{Physical Review Letters} \textbf{1992}, \emph{68}, 5 557.

\bibitem{Pirandola2019}
S.~Pirandola, U.~L. Andersen, L.~Banchi, M.~Berta, D.~Bunandar, R.~Colbeck,
  D.~Englund, T.~Gehring, C.~Lupo, C.~Ottaviani, J.~L. Pereira, M.~Razavi,
  J.~{Shamsul Shaari}, M.~Tomamichel, V.~C. Usenko, G.~Vallone, P.~Villoresi,
  P.~Wallden,
\newblock \emph{Advances in Optics and Photonics} \textbf{2020}, \emph{12}, 4
  1012.

\bibitem{shannon1949}
C.~E. Shannon,
\newblock \emph{The Bell system technical journal} \textbf{1949}, \emph{28}, 4
  656.

\bibitem{Vernam1926}
G.~S. Vernam,
\newblock \emph{Journal of the A.I.E.E.} \textbf{1926}, \emph{45}, 2 109.

\bibitem{Broadbent2016}
A.~Broadbent, C.~Schaffner,
\newblock \emph{Designs, Codes and Cryptography} \textbf{2016}, \emph{78}, 1
  351.

\bibitem{OBrien2009}
J.~L. O'Brien, A.~Furusawa, J.~Vu{\v{c}}kovi{\'{c}},
\newblock \emph{Nature Photonics} \textbf{2009}, \emph{3}, 12 687.

\bibitem{Hermelin2011}
S.~Hermelin, S.~Takada, M.~Yamamoto, S.~Tarucha, A.~D. Wieck, L.~Saminadayar,
  C.~B{\"{a}}uerle, T.~Meunier,
\newblock \emph{Nature} \textbf{2011}, \emph{477}, 7365 435.

\bibitem{McNeil2011}
R.~P.~G. McNeil, M.~Kataoka, C.~J.~B. Ford, C.~H.~W. Barnes, D.~Anderson,
  G.~A.~C. Jones, I.~Farrer, D.~A. Ritchie,
\newblock \emph{Nature} \textbf{2011}, \emph{477}, 7365 439.

\bibitem{flamini2018}
F.~Flamini, N.~Spagnolo, F.~Sciarrino,
\newblock \emph{Reports on Progress in Physics} \textbf{2018}, \emph{82}, 1
  016001.

\bibitem{Muller1993}
A.~Muller, J.~Breguet, N.~Gisin,
\newblock \emph{Europhysics Letters (EPL)} \textbf{1993}, \emph{23}, 6 383.

\bibitem{Yuan2008}
Z.-S. Yuan, Y.-A. Chen, B.~Zhao, S.~Chen, J.~Schmiedmayer, J.-W. Pan,
\newblock \emph{Nature} \textbf{2008}, \emph{454}, 7208 1098.

\bibitem{Martinelli1992}
M.~Martinelli,
\newblock \emph{Journal of Modern Optics} \textbf{1992}, \emph{39}, 3 451.

\bibitem{Sun1995}
P.~C. Sun, Y.~Mazurenko, Y.~Fainman,
\newblock \emph{Optics Letters} \textbf{1995}, \emph{20}, 9 1062.

\bibitem{Monteiro2015}
F.~Monteiro, V.~C. Vivoli, T.~Guerreiro, A.~Martin, J.-D. Bancal, H.~Zbinden,
  R.~T. Thew, N.~Sangouard,
\newblock \emph{Physical Review Letters} \textbf{2015}, \emph{114}, 17 170504.

\bibitem{Zhong2015}
T.~Zhong, H.~Zhou, R.~D. Horansky, C.~Lee, V.~B. Verma, A.~E. Lita,
  A.~Restelli, J.~C. Bienfang, R.~P. Mirin, T.~Gerrits, S.~W. Nam, F.~Marsili,
  M.~D. Shaw, Z.~Zhang, L.~Wang, D.~Englund, G.~W. Wornell, J.~H. Shapiro,
  F.~N.~C. Wong,
\newblock \emph{New Journal of Physics} \textbf{2015}, \emph{17}, 2 022002.

\bibitem{Kupchak2017}
C.~Kupchak, P.~J. Bustard, K.~Heshami, J.~Erskine, M.~Spanner, D.~G. England,
  B.~J. Sussman,
\newblock \emph{Physical Review A} \textbf{2017}, \emph{96}, 5 053812.

\bibitem{Lounis2005}
B.~Lounis, M.~Orrit,
\newblock \emph{Reports on Progress in Physics} \textbf{2005}, \emph{68}, 5
  1129.

\bibitem{shields2010}
A.~J. Shields,
\newblock \emph{Nanoscience And Technology: A Collection of Reviews from Nature
  Journals} \textbf{2010}, 221--229.

\bibitem{Eisaman2011}
M.~D. Eisaman, J.~Fan, A.~Migdall, S.~V. Polyakov,
\newblock \emph{Review of Scientific Instruments} \textbf{2011}, \emph{82}, 7
  071101.

\bibitem{Aharonovich2016}
I.~Aharonovich, D.~Englund, M.~Toth,
\newblock \emph{Nature Photonics} \textbf{2016}, \emph{10}, 10 631.

\bibitem{Chen2009}
T.-Y. Chen, H.~Liang, Y.~Liu, W.-Q. Cai, L.~Ju, W.-Y. Liu, J.~Wang, H.~Yin,
  K.~Chen, Z.-B. Chen, C.-Z. Peng, J.-W. Pan,
\newblock \emph{Optics Express} \textbf{2009}, \emph{17}, 8 6540.

\bibitem{Zhao2006}
Y.~Zhao, B.~Qi, X.~Ma, H.-K. Lo, L.~Qian,
\newblock \emph{Physical Review Letters} \textbf{2006}, \emph{96}, 7 070502.

\bibitem{Jennewein2000a}
T.~Jennewein, C.~Simon, G.~Weihs, H.~Weinfurter, A.~Zeilinger,
\newblock \emph{Physical Review Letters} \textbf{2000}, \emph{84}, 20 4729.

\bibitem{Poppe2004}
A.~Poppe, A.~Fedrizzi, R.~Ursin, H.~R. B{\"{o}}hm, T.~Lor{\"{u}}nser,
  O.~Maurhardt, M.~Peev, M.~Suda, C.~Kurtsiefer, H.~Weinfurter, T.~Jennewein,
  A.~Zeilinger,
\newblock \emph{Optics Express} \textbf{2004}, \emph{12}, 16 3865.

\bibitem{Boaron2018a}
A.~Boaron, G.~Boso, D.~Rusca, C.~Vulliez, C.~Autebert, M.~Caloz, M.~Perrenoud,
  G.~Gras, F.~Bussi{\`{e}}res, M.-J. Li, D.~Nolan, A.~Martin, H.~Zbinden,
\newblock \emph{Physical Review Letters} \textbf{2018}, \emph{121}, 19 190502.

\bibitem{Lutkenhaus2002}
N.~L{\"{u}}tkenhaus, M.~Jahma,
\newblock \emph{New Journal of Physics} \textbf{2002}, \emph{4} 344.

\bibitem{Wang2005}
X.-B. Wang,
\newblock \emph{Physical Review Letters} \textbf{2005}, \emph{94}, 23 230503.

\bibitem{Xu2020}
F.~Xu, X.~Ma, Q.~Zhang, H.-K. Lo, J.-W. Pan,
\newblock \emph{Reviews of Modern Physics} \textbf{2020}, \emph{92}, 2 025002.

\bibitem{Kwiat1995}
P.~G. Kwiat, K.~Mattle, H.~Weinfurter, A.~Zeilinger, A.~V. Sergienko, Y.~Shih,
\newblock \emph{Physical Review Letters} \textbf{1995}, \emph{75}, 24 4337.

\bibitem{Couteau2018}
C.~Couteau,
\newblock \emph{Contemporary Physics} \textbf{2018}, \emph{59}, 3 291.

\bibitem{Tapster1998}
P.~R. Tapster, J.~G. Rarity,
\newblock \emph{Journal of Modern Optics} \textbf{1998}, \emph{45}, 3 595.

\bibitem{Avenhaus2008}
M.~Avenhaus, H.~B. Coldenstrodt-Ronge, K.~Laiho, W.~Mauerer, I.~A. Walmsley,
  C.~Silberhorn,
\newblock \emph{Physical Review Letters} \textbf{2008}, \emph{101}, 5 053601.

\bibitem{Hosak2021}
R.~Ho{\v{s}}{\'{a}}k, I.~Straka, A.~Predojevi{\'{c}}, R.~Filip, M.~Je{\v{z}}ek,
\newblock \emph{Physical Review A} \textbf{2021}, \emph{103}, 4 042411.

\bibitem{Chen2018}
Y.~Chen, M.~Zopf, R.~Keil, F.~Ding, O.~G. Schmidt,
\newblock \emph{Nature Communications} \textbf{2018}, \emph{9}, 1 2994.

\bibitem{Liu2019}
J.~Liu, R.~Su, Y.~Wei, B.~Yao, S.~F.~C. da~Silva, Y.~Yu, J.~Iles-Smith,
  K.~Srinivasan, A.~Rastelli, J.~Li, X.~Wang,
\newblock \emph{Nature Nanotechnology} \textbf{2019}, \emph{14}, 6 586.

\bibitem{Wang2019}
H.~Wang, H.~Hu, T.-H. Chung, J.~Qin, X.~Yang, J.-P. Li, R.-Z. Liu, H.-S. Zhong,
  Y.-M. He, X.~Ding, Y.-H. Deng, Q.~Dai, Y.-H. Huo, S.~H{\"{o}}fling, C.-Y. Lu,
  J.-W. Pan,
\newblock \emph{Physical Review Letters} \textbf{2019}, \emph{122}, 11 113602.

\bibitem{Scarani2009}
V.~Scarani, H.~Bechmann-Pasquinucci, N.~J. Cerf, M.~Du{\v{s}}ek,
  N.~L{\"{u}}tkenhaus, M.~Peev,
\newblock \emph{Reviews of Modern Physics} \textbf{2009}, \emph{81}, 3 1301.

\bibitem{cai2009}
R.~Y. Cai, V.~Scarani,
\newblock \emph{New Journal of Physics} \textbf{2009}, \emph{11}, 4 045024.

\bibitem{Nauerth2013}
S.~Nauerth, F.~Moll, M.~Rau, C.~Fuchs, J.~Horwath, S.~Frick, H.~Weinfurter,
\newblock \emph{Nature Photonics} \textbf{2013}, \emph{7}, 5 382.

\bibitem{Liao2017}
S.-K. Liao, W.-Q. Cai, W.-Y. Liu, L.~Zhang, Y.~Li, J.-G. Ren, J.~Yin, Q.~Shen,
  Y.~Cao, Z.-P. Li, F.-Z. Li, X.-W. Chen, L.-H. Sun, J.-J. Jia, J.-C. Wu, X.-J.
  Jiang, J.-F. Wang, Y.-M. Huang, Q.~Wang, Y.-L. Zhou, L.~Deng, T.~Xi, L.~Ma,
  T.~Hu, Q.~Zhang, Y.-A. Chen, N.-L. Liu, X.-B. Wang, Z.-C. Zhu, C.-Y. Lu,
  R.~Shu, C.-Z. Peng, J.-Y. Wang, J.-W. Pan,
\newblock \emph{Nature} \textbf{2017}, \emph{549}, 7670 43.

\bibitem{Mandel1983}
L.~Mandel,
\newblock \emph{Physical Review A} \textbf{1983}, \emph{28}, 2 929.

\bibitem{SingleQuantum}
{Single Quantum} superconducting nanowire single photon detector,
\newblock \url{https://singlequantum.com/products/single-quantum-eos},
\newblock Accessed: 2021-08-16.

\bibitem{Gordon2005}
K.~J. Gordon, V.~Fernandez, G.~S. Buller, I.~Rech, S.~D. Cova, P.~D. Townsend,
\newblock \emph{Optics Express} \textbf{2005}, \emph{13}, 8 3015.

\bibitem{Dixon2009}
A.~R. Dixon, Z.~L. Yuan, J.~F. Dynes, A.~W. Sharpe, A.~J. Shields,
\newblock In \emph{Optical Fiber Communication Conference and National Fiber
  Optic Engineers Conference}. OSA, Washington, D.C.,
\newblock ISBN 978-1-55752-865-0, \textbf{2009} OThL3,
\newblock
  \urlprefix\url{https://www.osapublishing.org/abstract.cfm?URI=OFC-2009-OThL3}.

\bibitem{Gottesman2004}
D.~Gottesman, {Hoi-Kwong Lo}, N.~Lutkenhaus, J.~Preskill,
\newblock In \emph{International Symposium onInformation Theory, 2004. ISIT
  2004. Proceedings.}, volume~4. IEEE,
\newblock ISBN 0-7803-8280-3,
\newblock ISSN 15337146, \textbf{2004} 135--135,
\newblock \urlprefix\url{http://ieeexplore.ieee.org/document/1365172/}.

\bibitem{eraerds2010quantum}
P.~Eraerds, N.~Walenta, M.~Legr{\'e}, N.~Gisin, H.~Zbinden,
\newblock \emph{New Journal of Physics} \textbf{2010}, \emph{12}, 6 063027.

\bibitem{Scarani2008}
V.~Scarani, R.~Renner,
\newblock In \emph{Lecture Notes in Computer Science (including subseries
  Lecture Notes in Artificial Intelligence and Lecture Notes in
  Bioinformatics)}, volume 5106 LNCS, 83--95,
\newblock ISBN 3540893032, \textbf{2008},
\newblock \urlprefix\url{http://link.springer.com/10.1007/978-3-540-89304-2_8}.

\bibitem{Renner2008}
R.~RENNER,
\newblock \emph{International Journal of Quantum Information} \textbf{2008},
  \emph{06}, 01 1.

\bibitem{Chaiwongkhot2017}
P.~Chaiwongkhot, S.~Sajeed, L.~Lydersen, V.~Makarov,
\newblock \emph{Quantum Science and Technology} \textbf{2017}, \emph{2}, 4
  044003.

\bibitem{Coles2016}
P.~J. Coles, E.~M. Metodiev, N.~L{\"{u}}tkenhaus,
\newblock \emph{Nature Communications} \textbf{2016}, \emph{7}, 1 11712.

\bibitem{Winick2018}
A.~Winick, N.~L{\"{u}}tkenhaus, P.~J. Coles,
\newblock \emph{Quantum} \textbf{2018}, \emph{2} 77.

\bibitem{George2021}
I.~George, J.~Lin, N.~L{\"{u}}tkenhaus,
\newblock \emph{Physical Review Research} \textbf{2021}, \emph{3}, 1 013274.

\bibitem{Einstein1905}
A.~Einstein,
\newblock \emph{Annalen der Physik} \textbf{1905}, \emph{322}, 6 132.

\bibitem{LEWIS1926}
G.~N. LEWIS,
\newblock \emph{Nature} \textbf{1926}, \emph{118}, 2981 874.

\bibitem{kimble1977}
H.~J. Kimble, M.~Dagenais, L.~Mandel,
\newblock \emph{Physical Review Letters} \textbf{1977}, \emph{39}, 11 691.

\bibitem{ripka2018}
F.~Ripka, H.~K{\"u}bler, R.~L{\"o}w, T.~Pfau,
\newblock \emph{Science} \textbf{2018}, \emph{362}, 6413 446.

\bibitem{Leonard1993}
D.~Leonard, M.~Krishnamurthy, C.~M. Reaves, S.~P. Denbaars, P.~M. Petroff,
\newblock \emph{Applied Physics Letters} \textbf{1993}, \emph{63}, 23 3203.

\bibitem{Grundmann1995}
M.~Grundmann, J.~Christen, N.~N. Ledentsov, J.~B{\"{o}}hrer, D.~Bimberg, S.~S.
  Ruvimov, ‡, P.~Werner, U.~Richter, U.~G{\"{o}}sele, J.~Heydenreich, V.~M.
  Ustinov, A.~Y. Egorov, A.~E. Zhukov, P.~S. Kop'ev, Z.~I. Alferov,
\newblock \emph{Physical Review Letters} \textbf{1995}, \emph{74}, 20 4043.

\bibitem{Stranski1939}
I.~N. Stranski, L.~Krastanow,
\newblock \emph{Monatshefte f{\"{u}}r Chemie} \textbf{1939}, \emph{72}, 1 76.

\bibitem{Lay1978}
G.~{Le Lay}, R.~Kern,
\newblock \emph{Journal of Crystal Growth} \textbf{1978}, \emph{44}, 2 197.

\bibitem{Arakawa1982}
Y.~Arakawa, H.~Sakaki,
\newblock \emph{Applied Physics Letters} \textbf{1982}, \emph{40}, 11 939.

\bibitem{Imamoglu1999}
A.~Imamoglu, D.~D. Awschalom, G.~Burkard, D.~P. DiVincenzo, D.~Loss,
  M.~Sherwin, A.~Small,
\newblock \emph{Physical Review Letters} \textbf{1999}, \emph{83}, 20 4204.

\bibitem{Michler2009a}
P.~Michler,
\newblock \emph{{Single Semiconductor Quantum Dots}},
\newblock Springer, \textbf{2009}.

\bibitem{Shields2007}
A.~J. Shields,
\newblock \emph{Nature Photonics} \textbf{2007}, \emph{1} 215.

\bibitem{Buckley2012}
S.~Buckley, K.~Rivoire, J.~Vu{\v{c}}kovi{\'{c}},
\newblock \emph{Reports on Progress in Physics} \textbf{2012}, \emph{75}, 12
  126503.

\bibitem{Senellart2017}
P.~Senellart, G.~Solomon, A.~White,
\newblock \emph{Nature Nanotechnology} \textbf{2017}, \emph{12}, 11 1026.

\bibitem{Trivedi2020}
R.~Trivedi, K.~A. Fischer, J.~Vu{\v{c}}kovi{\'{c}}, K.~M{\"{u}}ller,
\newblock \emph{Advanced Quantum Technologies} \textbf{2020}, \emph{3}, 1
  1900007.

\bibitem{Rodt2020}
S.~Rodt, S.~Reitzenstein, T.~Heindel,
\newblock \emph{Journal of Physics: Condensed Matter} \textbf{2020}, \emph{32},
  15 153003.

\bibitem{Michler2000}
P.~Michler, A.~Kiraz, C.~Becher, W.~V. Schoenfeld, P.~M. Petroff, L.~Zhang,
  E.~Hu, A.~Imamoglu,
\newblock \emph{Science} \textbf{2000}, \emph{290}, 5500 2282.

\bibitem{Michler2000a}
P.~Michler, A.~Imamoğlu, M.~D. Mason, P.~J. Carson, G.~F. Strouse, S.~K.
  Buratto,
\newblock \emph{Nature} \textbf{2000}, \emph{406}, 6799 968.

\bibitem{Kako2006}
S.~Kako, C.~Santori, K.~Hoshino, S.~G{\"{o}}tzinger, Y.~Yamamoto, Y.~Arakawa,
\newblock \emph{Nature Materials} \textbf{2006}, \emph{5}, 11 887.

\bibitem{Arians2008}
R.~Arians, A.~Gust, T.~K{\"{u}}mmell, C.~Kruse, S.~Zaitsev, G.~Bacher,
  D.~Hommel,
\newblock \emph{Applied Physics Letters} \textbf{2008}, \emph{93}, 17 173506.

\bibitem{Deshpande2013}
S.~Deshpande, A.~Das, P.~Bhattacharya,
\newblock \emph{Applied Physics Letters} \textbf{2013}, \emph{102}, 16 161114.

\bibitem{Cui2019}
J.~Cui, Y.~E. Panfil, S.~Koley, D.~Shamalia, N.~Waiskopf, S.~Remennik,
  I.~Popov, M.~Oded, U.~Banin,
\newblock \emph{Nature Communications} \textbf{2019}, \emph{10}, 1 5401.

\bibitem{Schlehahn2018}
A.~Schlehahn, S.~Fischbach, R.~Schmidt, A.~Kaganskiy, A.~Strittmatter, S.~Rodt,
  T.~Heindel, S.~Reitzenstein,
\newblock \emph{Scientific Reports} \textbf{2018}, \emph{8}, 1 1340.

\bibitem{Schweickert2018}
L.~Schweickert, K.~D. J{\"{o}}ns, K.~D. Zeuner, S.~F. {Covre da Silva},
  H.~Huang, T.~Lettner, M.~Reindl, J.~Zichi, R.~Trotta, A.~Rastelli,
  V.~Zwiller,
\newblock \emph{Applied Physics Letters} \textbf{2018}, \emph{112}, 9 093106.

\bibitem{Rakher2010}
M.~T. Rakher, L.~Ma, O.~Slattery, X.~Tang, K.~Srinivasan,
\newblock \emph{Nature Photonics} \textbf{2010}, \emph{4}, 11 786.

\bibitem{Morrison2021}
C.~L. Morrison, M.~Rambach, Z.~X. Koong, F.~Graffitti, F.~Thorburn, A.~K. Kar,
  Y.~Ma, S.-I. Park, J.~D. Song, N.~G. Stoltz, D.~Bouwmeester, A.~Fedrizzi,
  B.~D. Gerardot,
\newblock \emph{Applied Physics Letters} \textbf{2021}, \emph{118}, 17 174003.

\bibitem{Liu2021}
S.~Liu, K.~Srinivasan, J.~Liu,
\newblock \emph{Laser \& Photonics Reviews} \textbf{2021}, 2100223.

\bibitem{keizer2012}
J.~Keizer, A.~Henriques, A.~Maia, A.~Quivy, P.~Koenraad,
\newblock \emph{Applied Physics Letters} \textbf{2012}, \emph{101}, 24 243113.

\bibitem{reithmaier2004}
J.~P. Reithmaier, G.~Sek, A.~L{\"o}ffler, C.~Hofmann, S.~Kuhn, S.~Reitzenstein,
  L.~Keldysh, V.~Kulakovskii, T.~Reinecke, A.~Forchel,
\newblock \emph{Nature} \textbf{2004}, \emph{432}, 7014 197.

\bibitem{santori2001}
C.~Santori, M.~Pelton, G.~Solomon, Y.~Dale, Y.~Yamamoto,
\newblock \emph{Physical Review Letters} \textbf{2001}, \emph{86}, 8 1502.

\bibitem{Muller2007}
A.~Muller, E.~B. Flagg, P.~Bianucci, X.~Y. Wang, D.~G. Deppe, W.~Ma, J.~Zhang,
  G.~J. Salamo, M.~Xiao, C.~K. Shih,
\newblock \emph{Physical Review Letters} \textbf{2007}, \emph{99}, 18 187402.

\bibitem{Kuhlmann2013}
A.~V. Kuhlmann, J.~Houel, D.~Brunner, A.~Ludwig, D.~Reuter, A.~D. Wieck, R.~J.
  Warburton,
\newblock \emph{Review of Scientific Instruments} \textbf{2013}, \emph{84}, 7
  073905.

\bibitem{Ates2009a}
S.~Ates, S.~M. Ulrich, S.~Reitzenstein, A.~L{\"{o}}ffler, A.~Forchel,
  P.~Michler,
\newblock \emph{Physical Review Letters} \textbf{2009}, \emph{103}, 16 167402.

\bibitem{Makhonin2014}
M.~N. Makhonin, J.~E. Dixon, R.~J. Coles, B.~Royall, I.~J. Luxmoore, E.~Clarke,
  M.~Hugues, M.~S. Skolnick, A.~M. Fox,
\newblock \emph{Nano Letters} \textbf{2014}, \emph{14}, 12 6997.

\bibitem{He2019}
Y.-M. He, H.~Wang, C.~Wang, M.-C. Chen, X.~Ding, J.~Qin, Z.-C. Duan, S.~Chen,
  J.-P. Li, R.-Z. Liu, C.~Schneider, M.~Atat{\"{u}}re, S.~H{\"{o}}fling, C.-Y.
  Lu, J.-W. Pan,
\newblock \emph{Nature Physics} \textbf{2019}, \emph{15}, 9 941.

\bibitem{holewa2017}
P.~Holewa, A.~Mary{\'n}ski, A.~Musia{\l}, T.~Heuser, N.~Srocka, D.~Quandt,
  A.~Strittmatter, S.~Rodt, J.~Misiewicz, S.~Reitzenstein, et~al.,
\newblock \emph{Optics express} \textbf{2017}, \emph{25}, 25 31122.

\bibitem{Vural2020}
H.~Vural, S.~L. Portalupi, P.~Michler,
\newblock \emph{Applied Physics Letters} \textbf{2020}, \emph{117}, 3 030501.

\bibitem{Hu1990}
Y.~Z. Hu, S.~W. Koch, M.~Lindberg, N.~Peyghambarian, E.~L. Pollock, F.~F.
  Abraham,
\newblock \emph{Physical Review Letters} \textbf{1990}, \emph{64}, 15 1805.

\bibitem{Kulakovskii1999}
V.~D. Kulakovskii, G.~Bacher, R.~Weigand, T.~K{\"{u}}mmell, A.~Forchel,
  E.~Borovitskaya, K.~Leonardi, D.~Hommel,
\newblock \emph{Physical Review Letters} \textbf{1999}, \emph{82}, 8 1780.

\bibitem{Moreau2001}
E.~Moreau, I.~Robert, L.~Manin, V.~Thierry-Mieg, J.~M. G{\'{e}}rard, I.~Abram,
\newblock \emph{Physical Review Letters} \textbf{2001}, \emph{87}, 18 183601.

\bibitem{Rodt2003}
S.~Rodt, R.~Heitz, A.~Schliwa, R.~L. Sellin, F.~Guffarth, D.~Bimberg,
\newblock \emph{Physical Review B} \textbf{2003}, \emph{68}, 3 035331.

\bibitem{Seguin2006}
R.~Seguin, A.~Schliwa, T.~D. Germann, S.~Rodt, K.~P{\"{o}}tschke,
  A.~Strittmatter, U.~W. Pohl, D.~Bimberg, M.~Winkelnkemper, T.~Hammerschmidt,
  P.~Kratzer,
\newblock \emph{Applied Physics Letters} \textbf{2006}, \emph{89}, 26 263109.

\bibitem{Sarkar2006}
D.~Sarkar, H.~P. van~der Meulen, J.~M. Calleja, J.~M. Becker, R.~J. Haug,
  K.~Pierz,
\newblock \emph{Journal of Applied Physics} \textbf{2006}, \emph{100}, 2
  023109.

\bibitem{Benson2000}
O.~Benson, C.~Santori, M.~Pelton, Y.~Yamamoto,
\newblock \emph{Phys. Rev. Lett.} \textbf{2000}, \emph{84} 2513.

\bibitem{Akopian2006}
N.~Akopian, N.~H. Lindner, E.~Poem, Y.~Berlatzky, J.~Avron, D.~Gershoni, B.~D.
  Gerardot, P.~M. Petroff,
\newblock \emph{Physical Review Letters} \textbf{2006}, \emph{96}, 13 130501.

\bibitem{Young2006}
R.~J. Young, R.~M. Stevenson, P.~Atkinson, K.~Cooper, D.~A. Ritchie, A.~J.
  Shields,
\newblock \emph{New Journal of Physics} \textbf{2006}, \emph{8}, 2 29.

\bibitem{Avron2008}
J.~E. Avron, G.~Bisker, D.~Gershoni, N.~H. Lindner, E.~A. Meirom, R.~J.
  Warburton,
\newblock \emph{Phys. Rev. Lett.} \textbf{2008}, \emph{100} 120501.

\bibitem{Heindel2017}
T.~Heindel, A.~Thoma, M.~von Helversen, M.~Schmidt, A.~Schlehahn, M.~Gschrey,
  P.~Schnauber, J.~H. Schulze, A.~Strittmatter, J.~Beyer, S.~Rodt, A.~Carmele,
  A.~Knorr, S.~Reitzenstein,
\newblock \emph{Nature Communications} \textbf{2017}, \emph{8}, 1 14870.

\bibitem{Moroni2019}
S.~Moroni, S.~Varo, G.~Juska, T.~Chung, A.~Gocalinska, E.~Pelucchi,
\newblock \emph{Journal of Crystal Growth} \textbf{2019}, \emph{506} 36.

\bibitem{Stevenson2006}
R.~M. Stevenson, R.~J. Young, P.~Atkinson, K.~Cooper, D.~A. Ritchie, A.~J.
  Shields,
\newblock \emph{Nature} \textbf{2006}, \emph{439}, 7073 179.

\bibitem{huber2018semiconductor}
D.~Huber, M.~Reindl, J.~Aberl, A.~Rastelli, R.~Trotta,
\newblock \emph{Journal of Optics} \textbf{2018}, \emph{20}, 7 073002.

\bibitem{Schimpf2021}
C.~Schimpf, M.~Reindl, F.~{Basso Basset}, K.~D. J{\"{o}}ns, R.~Trotta,
  A.~Rastelli,
\newblock \emph{Applied Physics Letters} \textbf{2021}, \emph{118}, 10 100502.

\bibitem{Zhou2018}
Y.~Zhou, Z.~Wang, A.~Rasmita, S.~Kim, A.~Berhane, Z.~Bodrog, G.~Adamo, A.~Gali,
  I.~Aharonovich, W.-b. Gao,
\newblock \emph{Science Advances} \textbf{2018}, \emph{4}, 3 eaar3580.

\bibitem{Barnes2002}
W.~Barnes, G.~Björk, J.~Gérard, P.~Jonsson, J.~Wasey, P.~Worthing,
  V.~Zwiller,
\newblock \emph{The European Physical Journal D - Atomic, Molecular, Optical
  and Plasma Physics} \textbf{2002}, \emph{18} 197.

\bibitem{Santori2002}
C.~Santori, D.~Fattal, J.~Vu{\v{c}}kovi{\'{c}}, G.~S. Solomon, Y.~Yamamoto,
\newblock \emph{Nature} \textbf{2002}, \emph{419}, 6907 594.

\bibitem{Heindel2010}
T.~Heindel, C.~Schneider, M.~Lermer, S.~H. Kwon, T.~Braun, S.~Reitzenstein,
  S.~H{\"{o}}fling, M.~Kamp, A.~Forchel,
\newblock \emph{Applied Physics Letters} \textbf{2010}, \emph{96}, 1 011107.

\bibitem{Reischle2010}
M.~Reischle, C.~Kessler, W.-M. Schulz, M.~Eichfelder, R.~Ro{\ss}bach,
  M.~Jetter, P.~Michler,
\newblock \emph{Applied Physics Letters} \textbf{2010}, \emph{97}, 14 143513.

\bibitem{Muller2018}
T.~M{\"{u}}ller, J.~Skiba-Szymanska, A.~B. Krysa, J.~Huwer, M.~Felle,
  M.~Anderson, R.~M. Stevenson, J.~Heffernan, D.~A. Ritchie, A.~J. Shields,
\newblock \emph{Nature Communications} \textbf{2018}, \emph{9}, 1 862.

\bibitem{Takemoto2007}
K.~Takemoto, M.~Takatsu, S.~Hirose, N.~Yokoyama, Y.~Sakuma, T.~Usuki,
  T.~Miyazawa, Y.~Arakawa,
\newblock \emph{Journal of Applied Physics} \textbf{2007}, \emph{101}, 8
  081720.

\bibitem{huber2017highly}
D.~Huber, M.~Reindl, Y.~Huo, H.~Huang, J.~S. Wildmann, O.~G. Schmidt,
  A.~Rastelli, R.~Trotta,
\newblock \emph{Nature communications} \textbf{2017}, \emph{8}, 1 1.

\bibitem{Gerard1998}
J.~M. G\'erard, B.~Sermage, B.~Gayral, B.~Legrand, E.~Costard, V.~Thierry-Mieg,
\newblock \emph{Phys. Rev. Lett.} \textbf{1998}, \emph{81} 1110.

\bibitem{Solomon2001}
G.~S. Solomon, M.~Pelton, Y.~Yamamoto,
\newblock \emph{Physical Review Letters} \textbf{2001}, \emph{86}, 17 3903.

\bibitem{Waks2002a}
E.~Waks, K.~Inoue, C.~Santori, D.~Fattal, J.~Vuckovic, G.~S. Solomon,
  Y.~Yamamoto,
\newblock \emph{Nature} \textbf{2002}, \emph{420}, 6917 762.

\bibitem{Intallura2009}
P.~M. Intallura, M.~B. Ward, O.~Z. Karimov, Z.~L. Yuan, P.~See, P.~Atkinson,
  D.~A. Ritchie, A.~J. Shields,
\newblock \emph{Journal of Optics A: Pure and Applied Optics} \textbf{2009},
  \emph{11}, 5 054005.

\bibitem{Gazzano2013}
O.~Gazzano, S.~{Michaelis de Vasconcellos}, C.~Arnold, A.~Nowak, E.~Galopin,
  I.~Sagnes, L.~Lanco, A.~Lema{\^{i}}tre, P.~Senellart,
\newblock \emph{Nature Communications} \textbf{2013}, \emph{4}, 1 1425.

\bibitem{Somaschi2016}
N.~Somaschi, V.~Giesz, L.~{De Santis}, J.~C. Loredo, M.~P. Almeida,
  G.~Hornecker, S.~L. Portalupi, T.~Grange, C.~Ant{\'{o}}n, J.~Demory,
  C.~G{\'{o}}mez, I.~Sagnes, N.~D. Lanzillotti-Kimura, A.~Lema{\'{i}}tre,
  A.~Auffeves, A.~G. White, L.~Lanco, P.~Senellart,
\newblock \emph{Nature Photonics} \textbf{2016}, \emph{10}, 5 340.

\bibitem{Ding2016}
X.~Ding, Y.~He, Z.-C. Duan, N.~Gregersen, M.-C. Chen, S.~Unsleber, S.~Maier,
  C.~Schneider, M.~Kamp, S.~H{\"{o}}fling, C.-Y. Lu, J.-W. Pan,
\newblock \emph{Physical Review Letters} \textbf{2016}, \emph{116}, 2 020401.

\bibitem{Heindel2012}
T.~Heindel, C.~A. Kessler, M.~Rau, C.~Schneider, M.~F{\"{u}}rst, F.~Hargart,
  W.-M. Schulz, M.~Eichfelder, R.~Ro{\ss}bach, S.~Nauerth, M.~Lermer, H.~Weier,
  M.~Jetter, M.~Kamp, S.~Reitzenstein, S.~H{\"{o}}fling, P.~Michler,
  H.~Weinfurter, A.~Forchel,
\newblock \emph{New Journal of Physics} \textbf{2012}, \emph{14}, 8 083001.

\bibitem{Boeckler2008}
C.~Böckler, S.~Reitzenstein, C.~Kistner, R.~Debusmann, A.~Löffler, T.~Kida,
  S.~Höfling, A.~Forchel, L.~Grenouillet, J.~Claudon, J.~M. Gérard,
\newblock \emph{Applied Physics Letters} \textbf{2008}, \emph{92}, 9 091107.

\bibitem{Schlehahn2016a}
A.~Schlehahn, A.~Thoma, P.~Munnelly, M.~Kamp, S.~H{\"{o}}fling, T.~Heindel,
  C.~Schneider, S.~Reitzenstein,
\newblock \emph{APL Photonics} \textbf{2016}, \emph{1}, 1 011301.

\bibitem{Hargart2013}
F.~Hargart, C.~A. Kessler, T.~Schwarzbäck, E.~Koroknay, S.~Weidenfeld,
  M.~Jetter, P.~Michler,
\newblock \emph{Applied Physics Letters} \textbf{2013}, \emph{102}, 1 011126.

\bibitem{Salter2010}
C.~L. Salter, R.~M. Stevenson, I.~Farrer, C.~A. Nicoll, D.~A. Ritchie, A.~J.
  Shields,
\newblock \emph{Nature} \textbf{2010}, \emph{465}, 7298 594.

\bibitem{Dzurnak2015}
B.~Dzurnak, R.~M. Stevenson, J.~Nilsson, J.~F. Dynes, Z.~L. Yuan,
  J.~Skiba-Szymanska, I.~Farrer, D.~A. Ritchie, A.~J. Shields,
\newblock \emph{Applied Physics Letters} \textbf{2015}, \emph{107}, 26 261101.

\bibitem{Takemoto2010}
K.~Takemoto, Y.~Nambu, T.~Miyazawa, K.~Wakui, S.~Hirose, T.~Usuki, M.~Takatsu,
  N.~Yokoyama, K.~Yoshino, A.~Tomita, S.~Yorozu, Y.~Sakuma, Y.~Arakawa,
\newblock \emph{Applied Physics Express} \textbf{2010}, \emph{3}, 9 092802.

\bibitem{Basset2021}
F.~{Basso Basset}, M.~Valeri, E.~Roccia, V.~Muredda, D.~Poderini, J.~Neuwirth,
  N.~Spagnolo, M.~B. Rota, G.~Carvacho, F.~Sciarrino, R.~Trotta,
\newblock \emph{Science Advances} \textbf{2021}, \emph{7}, 12 eabe6379.

\bibitem{Schimpf2021a}
C.~Schimpf, M.~Reindl, D.~Huber, B.~Lehner, S.~F. {Covre Da Silva}, S.~Manna,
  M.~Vyvlecka, P.~Walther, A.~Rastelli,
\newblock \emph{Science Advances} \textbf{2021}, \emph{7}, 16 eabe8905.

\bibitem{Claudon2010}
J.~Claudon, J.~Bleuse, N.~S. Malik, M.~Bazin, P.~Jaffrennou, N.~Gregersen,
  C.~Sauvan, P.~Lalanne, J.-M. G{\'{e}}rard,
\newblock \emph{Nature Photonics} \textbf{2010}, \emph{4}, 3 174.

\bibitem{Madsen2014}
K.~H. Madsen, S.~Ates, J.~Liu, A.~Javadi, S.~M. Albrecht, I.~Yeo, S.~Stobbe,
  P.~Lodahl,
\newblock \emph{Physical Review B} \textbf{2014}, \emph{90}, 15 155303.

\bibitem{Kim2016}
J.-H. Kim, T.~Cai, C.~J.~K. Richardson, R.~P. Leavitt, E.~Waks,
\newblock \emph{Optica} \textbf{2016}, \emph{3}, 6 577.

\bibitem{Davanco2011}
M.~Davan{\c{c}}o, M.~T. Rakher, D.~Schuh, A.~Badolato, K.~Srinivasan,
\newblock \emph{Applied Physics Letters} \textbf{2011}, \emph{99}, 4 041102.

\bibitem{Tomm2020}
N.~Tomm, A.~Javadi, N.~O. Antoniadis, D.~Najer, M.~C. L{\"{o}}bl, A.~R. Korsch,
  R.~Schott, S.~R. Valentin, A.~D. Wieck, A.~Ludwig, R.~J. Warburton,
\newblock \emph{Nature Nanotechnology} \textbf{2021}, \emph{16}, 4 399.

\bibitem{Uppu2020}
R.~Uppu, F.~T. Pedersen, Y.~Wang, C.~T. Olesen, C.~Papon, X.~Zhou, L.~Midolo,
  S.~Scholz, A.~D. Wieck, A.~Ludwig, P.~Lodahl,
\newblock \emph{Science Advances} \textbf{2020}, \emph{6}, 50 eabc8268.

\bibitem{Gschrey2015}
M.~Gschrey, A.~Thoma, P.~Schnauber, M.~Seifried, R.~Schmidt, B.~Wohlfeil,
  L.~Krüger, J.~H. Schulze, T.~Heindel, S.~Burger, F.~Schmidt,
  A.~Strittmatter, S.~Rodt, S.~Reitzenstein,
\newblock \emph{Nature Communications} \textbf{2015}, \emph{6} 7662.

\bibitem{Musial2020}
A.~Musia{\l}, K.~{\.{Z}}o{\l}nacz, N.~Srocka, O.~Kravets, J.~Gro{\ss}e,
  J.~Olszewski, K.~Poturaj, G.~W{\'{o}}jcik, P.~Mergo, K.~Dybka, M.~Dyrkacz,
  M.~D{\l}ubek, K.~Lauritsen, A.~B{\"{u}}lter, P.~Schneider, L.~Zschiedrich,
  S.~Burger, S.~Rodt, W.~Urba{\'{n}}czyk, G.~S{\c{e}}k, S.~Reitzenstein,
\newblock \emph{Advanced Quantum Technologies} \textbf{2020}, \emph{3}, 6
  2000018.

\bibitem{Kupko2020}
T.~Kupko, M.~von Helversen, L.~Rickert, J.-H. Schulze, A.~Strittmatter,
  M.~Gschrey, S.~Rodt, S.~Reitzenstein, T.~Heindel,
\newblock \emph{npj Quantum Information} \textbf{2020}, \emph{6}, 1 29.

\bibitem{Kupko2021}
T.~Kupko, L.~Rickert, F.~Urban, J.~Gro{\ss}e, N.~Srocka, S.~Rodt, A.~Musia{\l},
  K.~{\.{Z}}o{\l}nacz, P.~Mergo, K.~Dybka, W.~Urba{\'{n}}czyk, G.~S{\c{e}}k,
  S.~Burger, S.~Reitzenstein, T.~Heindel,
\newblock \emph{arxiv ID: 2105.03473} \textbf{2021}, 1--8.

\bibitem{Beveratos2002}
A.~Beveratos, R.~Brouri, T.~Gacoin, A.~Villing, J.-P. Poizat, P.~Grangier,
\newblock \emph{Physical Review Letters} \textbf{2002}, \emph{89}, 18 187901.

\bibitem{Alleaume2004}
R.~All{\'{e}}aume, F.~Treussart, G.~Messin, Y.~Dumeige, J.-F. Roch,
  A.~Beveratos, R.~Brouri-Tualle, J.-P. Poizat, P.~Grangier,
\newblock \emph{New Journal of Physics} \textbf{2004}, \emph{6}, 1 92.

\bibitem{Leifgen2014}
M.~Leifgen, T.~Schr{\"{o}}der, F.~G{\"{a}}deke, R.~Riemann, V.~M{\'{e}}tillon,
  E.~Neu, C.~Hepp, C.~Arend, C.~Becher, K.~Lauritsen, O.~Benson,
\newblock \emph{New Journal of Physics} \textbf{2014}, \emph{16}, 2 023021.

\bibitem{Brown1956}
R.~H. Brown, R.~Q. Twiss,
\newblock \emph{Nature} \textbf{1956}, \emph{177}, 4497 27.

\bibitem{Waks2002}
E.~Waks, C.~Santori, Y.~Yamamoto,
\newblock \emph{Physical Review A} \textbf{2002}, \emph{66}, 4 042315.

\bibitem{Lutkenhaus2000}
N.~L{\"{u}}tkenhaus,
\newblock \emph{Physical Review A} \textbf{2000}, \emph{61}, 5 052304.

\bibitem{Grunwald2019}
P.~Gr{\"{u}}nwald,
\newblock \emph{New Journal of Physics} \textbf{2019}, \emph{21}, 9 093003.

\bibitem{Aichele2004}
T.~Aichele, G.~Reinaudi, O.~Benson,
\newblock \emph{Physical Review B} \textbf{2004}, \emph{70}, 23 235329.

\bibitem{Bedington2017}
R.~Bedington, J.~M. Arrazola, A.~Ling,
\newblock \emph{npj Quantum Information} \textbf{2017}, \emph{3}, 1 30.

\bibitem{Yin2020}
J.~Yin, Y.-H. Li, S.-K. Liao, M.~Yang, Y.~Cao, L.~Zhang, J.-G. Ren, W.-Q. Cai,
  W.-Y. Liu, S.-L. Li, R.~Shu, Y.-M. Huang, L.~Deng, L.~Li, Q.~Zhang, N.-L.
  Liu, Y.-A. Chen, C.-Y. Lu, X.-B. Wang, F.~Xu, J.-Y. Wang, C.-Z. Peng, A.~K.
  Ekert, J.-W. Pan,
\newblock \emph{Nature} \textbf{2020}, \emph{582}, 7813 501.

\bibitem{Sidhu2021}
J.~S. Sidhu, S.~K. Joshi, M.~Gundogan, T.~Brougham, D.~Lowndes, L.~Mazzarella,
  M.~Krutzik, S.~Mohapatra, D.~Dequal, G.~Vallone, P.~Villoresi, A.~Ling,
  T.~Jennewein, M.~Mohageg, J.~Rarity, I.~Fuentes, S.~Pirandola, D.~K.~L. Oi,
\newblock \emph{arxiv ID: 2103.12749} \textbf{2021}, 1--26.

\bibitem{Collins2010}
R.~J. Collins, P.~J. Clarke, V.~Fern{\'{a}}ndez, K.~J. Gordon, M.~N. Makhonin,
  J.~A. Timpson, A.~Tahraoui, M.~Hopkinson, A.~M. Fox, M.~S. Skolnick, G.~S.
  Buller,
\newblock \emph{Journal of Applied Physics} \textbf{2010}, \emph{107}, 7
  073102.

\bibitem{Ward2005}
M.~B. Ward, O.~Z. Karimov, D.~C. Unitt, Z.~L. Yuan, P.~See, D.~G. Gevaux, A.~J.
  Shields, P.~Atkinson, D.~A. Ritchie,
\newblock \emph{Applied Physics Letters} \textbf{2005}, \emph{86}, 20 201111.

\bibitem{Hadfield2005}
R.~H. Hadfield, M.~J. Stevens, S.~S. Gruber, A.~J. Miller, R.~E. Schwall, R.~P.
  Mirin, S.~W. Nam,
\newblock \emph{Optics Express} \textbf{2005}, \emph{13}, 26 10846.

\bibitem{Takemoto2015}
K.~Takemoto, Y.~Nambu, T.~Miyazawa, Y.~Sakuma, T.~Yamamoto, S.~Yorozu,
  Y.~Arakawa,
\newblock \emph{Scientific Reports} \textbf{2015}, \emph{5}, 1 14383.

\bibitem{Yuan2002}
Z.~Yuan,
\newblock \emph{Science} \textbf{2002}, \emph{295}, 5552 102.

\bibitem{Rau2014}
M.~Rau, T.~Heindel, S.~Unsleber, T.~Braun, J.~Fischer, S.~Frick, S.~Nauerth,
  C.~Schneider, G.~Vest, S.~Reitzenstein, M.~Kamp, A.~Forchel,
  S.~H{\"{o}}fling, H.~Weinfurter,
\newblock \emph{New Journal of Physics} \textbf{2014}, \emph{16}, 4 043003.

\bibitem{Olbrich2017}
F.~Olbrich, J.~H{\"{o}}schele, M.~M{\"{u}}ller, J.~Kettler, S.~{Luca
  Portalupi}, M.~Paul, M.~Jetter, P.~Michler,
\newblock \emph{Applied Physics Letters} \textbf{2017}, \emph{111}, 13 133106.

\bibitem{Shooter2020}
G.~Shooter, Z.-H. Xiang, J.~R.~A. M{\"{u}}ller, J.~Skiba-Szymanska, J.~Huwer,
  J.~Griffiths, T.~Mitchell, M.~Anderson, T.~M{\"{u}}ller, A.~B. Krysa,
  R.~{Mark Stevenson}, J.~Heffernan, D.~A. Ritchie, A.~J. Shields,
\newblock \emph{Optics Express} \textbf{2020}, \emph{28}, 24 36838.

\bibitem{Schlehahn2015}
A.~Schlehahn, L.~Kr{\"{u}}ger, M.~Gschrey, J.-H. Schulze, S.~Rodt,
  A.~Strittmatter, T.~Heindel, S.~Reitzenstein,
\newblock \emph{Review of Scientific Instruments} \textbf{2015}, \emph{86}, 1
  013113.

\bibitem{Xiang2019}
Z.-H. Xiang, J.~Huwer, R.~M. Stevenson, J.~Skiba-Szymanska, M.~B. Ward,
  I.~Farrer, D.~A. Ritchie, A.~J. Shields,
\newblock \emph{Scientific Reports} \textbf{2019}, \emph{9}, 1 4111.

\bibitem{Wengerowsky2019}
S.~Wengerowsky, S.~K. Joshi, F.~Steinlechner, J.~R. Zichi, S.~M. Dobrovolskiy,
  R.~van~der Molen, J.~W.~N. Los, V.~Zwiller, M.~A.~M. Versteegh, A.~Mura,
  D.~Calonico, M.~Inguscio, H.~H{\"{u}}bel, L.~Bo, T.~Scheidl, A.~Zeilinger,
  A.~Xuereb, R.~Ursin,
\newblock \emph{Proceedings of the National Academy of Sciences} \textbf{2019},
  \emph{116}, 14 6684.

\bibitem{Yin2017}
J.~Yin, Y.~Cao, Y.-H. Li, J.-G. Ren, S.-K. Liao, L.~Zhang, W.-Q. Cai, W.-Y.
  Liu, B.~Li, H.~Dai, M.~Li, Y.-M. Huang, L.~Deng, L.~Li, Q.~Zhang, N.-L. Liu,
  Y.-A. Chen, C.-Y. Lu, R.~Shu, C.-Z. Peng, J.-Y. Wang, J.-W. Pan,
\newblock \emph{Physical Review Letters} \textbf{2017}, \emph{119}, 20 200501.

\bibitem{gobby2004}
C.~Gobby, a.~Yuan, A.~Shields,
\newblock \emph{Applied Physics Letters} \textbf{2004}, \emph{84}, 19 3762.

\bibitem{rusca2018finite}
D.~Rusca, A.~Boaron, F.~Gr{\"u}nenfelder, A.~Martin, H.~Zbinden,
\newblock \emph{Applied Physics Letters} \textbf{2018}, \emph{112}, 17 171104.

\bibitem{elkouss2009}
D.~Elkouss, A.~Leverrier, R.~All{\'e}aume, J.~J. Boutros,
\newblock In \emph{2009 IEEE International Symposium on Information Theory}.
  IEEE, \textbf{2009} 1879--1883.

\bibitem{Schlehahn2016}
A.~Schlehahn, R.~Schmidt, C.~Hopfmann, J.-H. Schulze, A.~Strittmatter,
  T.~Heindel, L.~Gantz, E.~R. Schmidgall, D.~Gershoni, S.~Reitzenstein,
\newblock \emph{Applied Physics Letters} \textbf{2016}, \emph{108}, 2 021104.

\bibitem{Acin2007}
A.~Ac{\'{i}}n, N.~Brunner, N.~Gisin, S.~Massar, S.~Pironio, V.~Scarani,
\newblock \emph{Physical Review Letters} \textbf{2007}, \emph{98}, 23 230501.

\bibitem{Masanes2011}
L.~Masanes, S.~Pironio, A.~Ac{\'{i}}n,
\newblock \emph{Nature Communications} \textbf{2011}, \emph{2}, 1 238.

\bibitem{Zapatero2019}
V.~Zapatero, M.~Curty,
\newblock \emph{Scientific Reports} \textbf{2019}, \emph{9}, 1 17749.

\bibitem{Vazirani2019}
U.~Vazirani, T.~Vidick,
\newblock \emph{Communications of the ACM} \textbf{2019}, \emph{62}, 4 133.

\bibitem{Schwonnek2021}
R.~Schwonnek, K.~T. Goh, I.~W. Primaatmaja, E.~Y.-Z. Tan, R.~Wolf, V.~Scarani,
  C.~C.-W. Lim,
\newblock \emph{Nature Communications} \textbf{2021}, \emph{12}, 1 2880.

\bibitem{Koashi2003}
M.~Koashi, J.~Preskill,
\newblock \emph{Physical Review Letters} \textbf{2003}, \emph{90}, 5 057902.

\bibitem{Lo2012}
H.-K. Lo, M.~Curty, B.~Qi,
\newblock \emph{Physical Review Letters} \textbf{2012}, \emph{108}, 13 130503.

\bibitem{Rubenok2013}
A.~Rubenok, J.~A. Slater, P.~Chan, I.~Lucio-Martinez, W.~Tittel,
\newblock \emph{Physical Review Letters} \textbf{2013}, \emph{111}, 13 130501.

\bibitem{Liu2013}
Y.~Liu, T.-Y. Chen, L.-J. Wang, H.~Liang, G.-L. Shentu, J.~Wang, K.~Cui, H.-L.
  Yin, N.-L. Liu, L.~Li, X.~Ma, J.~S. Pelc, M.~M. Fejer, C.-Z. Peng, Q.~Zhang,
  J.-W. Pan,
\newblock \emph{Physical Review Letters} \textbf{2013}, \emph{111}, 13 130502.

\bibitem{DaSilva2013}
T.~{Ferreira da Silva}, D.~Vitoreti, G.~B. Xavier, G.~C. do~Amaral, G.~P.
  Tempor{\~{a}}o, J.~P. von~der Weid,
\newblock \emph{Physical Review A} \textbf{2013}, \emph{88}, 5 052303.

\bibitem{Cao2020}
Y.~Cao, Y.-H. Li, K.-X. Yang, Y.-F. Jiang, S.-L. Li, X.-L. Hu, M.~Abulizi,
  C.-L. Li, W.~Zhang, Q.-C. Sun, W.-Y. Liu, X.~Jiang, S.-K. Liao, J.-G. Ren,
  H.~Li, L.~You, Z.~Wang, J.~Yin, C.-Y. Lu, X.-B. Wang, Q.~Zhang, C.-Z. Peng,
  J.-W. Pan,
\newblock \emph{Physical Review Letters} \textbf{2020}, \emph{125}, 26 260503.

\bibitem{Semenenko2020}
H.~Semenenko, P.~Sibson, A.~Hart, M.~G. Thompson, J.~G. Rarity, C.~Erven,
\newblock \emph{Optica} \textbf{2020}, \emph{7}, 3 238.

\bibitem{Wei2020}
K.~Wei, W.~Li, H.~Tan, Y.~Li, H.~Min, W.-J. Zhang, H.~Li, L.~You, Z.~Wang,
  X.~Jiang, T.-Y. Chen, S.-K. Liao, C.-Z. Peng, F.~Xu, J.-W. Pan,
\newblock \emph{Physical Review X} \textbf{2020}, \emph{10}, 3 031030.

\bibitem{Kaneda2017}
F.~Kaneda, F.~Xu, J.~Chapman, P.~G. Kwiat,
\newblock \emph{Optica} \textbf{2017}, \emph{4}, 9 1034.

\bibitem{Zhai2021}
L.~Zhai, G.~N. Nguyen, C.~Spinnler, J.~Ritzmann, M.~C. L{\"{o}}bl, A.~D. Wieck,
  A.~Ludwig, A.~Javadi, R.~J. Warburton,
\newblock \emph{arxiv Id: 2106.03871} \textbf{2021}, 1--7.

\bibitem{Lee2021}
D.~Lee, Y.-W. Cho, H.-T. Lim, S.-W. Han, H.~Jung, S.~Moon, Y.-S. Kim,
\newblock \emph{Quantum Information Processing} \textbf{2021}, \emph{20}, 4 1.

\bibitem{Curty2014}
M.~Curty, F.~Xu, W.~Cui, C.~C.~W. Lim, K.~Tamaki, H.-K. Lo,
\newblock \emph{Nature Communications} \textbf{2014}, \emph{5}, 1 3732.

\bibitem{Tang2016}
Y.-L. Tang, H.-L. Yin, Q.~Zhao, H.~Liu, X.-X. Sun, M.-Q. Huang, W.-J. Zhang,
  S.-J. Chen, L.~Zhang, L.-X. You, Z.~Wang, Y.~Liu, C.-Y. Lu, X.~Jiang, X.~Ma,
  Q.~Zhang, T.-Y. Chen, J.-W. Pan,
\newblock \emph{Physical Review X} \textbf{2016}, \emph{6}, 1 011024.

\bibitem{Barnett2017}
S.~M. Barnett, A.~Beige, A.~Ekert, B.~M. Garraway, C.~H. Keitel, V.~Kendon,
  M.~Lein, G.~J. Milburn, H.~M. Moya-Cessa, M.~Murao, J.~K. Pachos, G.~M.
  Palma, E.~Paspalakis, S.~J. Phoenix, B.~Piraux, M.~B. Plenio, B.~C. Sanders,
  J.~Twamley, A.~Vidiella-Barranco, M.~Kim,
\newblock \emph{Progress in Quantum Electronics} \textbf{2017}, \emph{54} 19.

\bibitem{Uppu2021}
R.~Uppu, L.~Midolo, X.~Zhou, J.~Carolan, P.~Lodahl,
\newblock \emph{arxiv ID: 2103.01110} \textbf{2021}, 1--15.

\bibitem{Xu2007}
X.~Xu, I.~Toft, R.~T. Phillips, J.~Mar, K.~Hammura, D.~A. Williams,
\newblock \emph{Applied Physics Letters} \textbf{2007}, \emph{90}, 6 061103.

\bibitem{Muller2009}
A.~Muller, E.~B. Flagg, M.~Metcalfe, J.~Lawall, G.~S. Solomon,
\newblock \emph{Applied Physics Letters} \textbf{2009}, \emph{95}, 17 173101.

\bibitem{Cadeddu2016}
D.~Cadeddu, J.~Teissier, F.~R. Braakman, N.~Gregersen, P.~Stepanov, J.-M.
  G{\'{e}}rard, J.~Claudon, R.~J. Warburton, M.~Poggio, M.~Munsch,
\newblock \emph{Applied Physics Letters} \textbf{2016}, \emph{108}, 1 011112.

\bibitem{Haupt2010}
F.~Haupt, S.~S.~R. Oemrawsingh, S.~M. Thon, H.~Kim, D.~Kleckner, D.~Ding, D.~J.
  Suntrup, P.~M. Petroff, D.~Bouwmeester,
\newblock \emph{Applied Physics Letters} \textbf{2010}, \emph{97}, 13 131113.

\bibitem{Snijders2018}
H.~Snijders, J.~A. Frey, J.~Norman, V.~P. Post, A.~C. Gossard, J.~E. Bowers,
  M.~P. van Exter, W.~L{\"{o}}ffler, D.~Bouwmeester,
\newblock \emph{Physical Review Applied} \textbf{2018}, \emph{9}, 3 031002.

\bibitem{Rickert2021}
L.~Rickert, F.~Schr{\"{o}}der, T.~Kupko, C.~Schneider, S.~H{\"{o}}fling,
  T.~Heindel,
\newblock \emph{arxiv ID: 2102.12836} \textbf{2021}, 1--9.

\bibitem{Zonacz2019}
K.~{\.{Z}}o{\l}nacz, A.~Musia{\l}, N.~Srocka, J.~Gro{\ss}e, M.~J.
  Schl{\"{o}}singer, P.-I. Schneider, O.~Kravets, M.~Mikulicz, J.~Olszewski,
  K.~Poturaj, G.~W{\'{o}}jcik, P.~Mergo, K.~Dybka, M.~Dyrkacz, M.~D{\l}ubek,
  S.~Rodt, S.~Burger, L.~Zschiedrich, G.~S{\c{e}}k, S.~Reitzenstein,
  W.~Urba{\'{n}}czyk,
\newblock \emph{Optics Express} \textbf{2019}, \emph{27}, 19 26772.

\bibitem{Northeast2021}
D.~B. Northeast, J.~F. Weber, D.~Dalacu, J.~Phoenix, P.~J. Poole, G.~Aers,
  J.~Lapointe, R.~L. Williams,
\newblock \emph{arxiv ID: 2104.11197} \textbf{2021}, 1--16.

\bibitem{Rickert2019}
L.~Rickert, T.~Kupko, S.~Rodt, S.~Reitzenstein, T.~Heindel,
\newblock \emph{Optics Express} \textbf{2019}, \emph{27}, 25 36824.

\bibitem{sunpower}
{Sunpower} stirling cryocoolers,
\newblock
  \url{https://www.sunpowerinc.com/products/stirling-cryocoolers/cryotel-cryocoolers},
\newblock Accessed: 2021-08-16.

\bibitem{stock2013}
E.~Stock, F.~Albert, C.~Hopfmann, M.~Lermer, C.~Schneider, S.~H{\"o}fling,
  A.~Forchel, M.~Kamp, S.~Reitzenstein,
\newblock \emph{Advanced Materials} \textbf{2013}, \emph{25}, 5 707.

\bibitem{Munnelly2017}
P.~Munnelly, T.~Heindel, A.~Thoma, M.~Kamp, S.~H{\"{o}}fling, C.~Schneider,
  S.~Reitzenstein,
\newblock \emph{ACS Photonics} \textbf{2017}, \emph{4}, 4 790.

\bibitem{Lee2017}
J.~P. Lee, E.~Murray, A.~J. Bennett, D.~J.~P. Ellis, C.~Dangel, I.~Farrer,
  P.~Spencer, D.~A. Ritchie, A.~J. Shields,
\newblock \emph{Applied Physics Letters} \textbf{2017}, \emph{110}, 7 071102.

\bibitem{Xiang2020}
Z.-H. Xiang, J.~Huwer, J.~Skiba-Szymanska, R.~M. Stevenson, D.~J.~P. Ellis,
  I.~Farrer, M.~B. Ward, D.~A. Ritchie, A.~J. Shields,
\newblock \emph{Communications Physics} \textbf{2020}, \emph{3}, 1 121.

\bibitem{Briegel1998}
H.-J. Briegel, W.~D{\"{u}}r, J.~I. Cirac, P.~Zoller,
\newblock \emph{Physical Review Letters} \textbf{1998}, \emph{81}, 26 5932.

\bibitem{Dur1999}
W.~D{\"{u}}r, H.-J. Briegel, J.~I. Cirac, P.~Zoller,
\newblock \emph{Physical Review A} \textbf{1999}, \emph{59}, 1 169.

\bibitem{Einstein1935}
A.~Einstein, B.~Podolsky, N.~Rosen,
\newblock \emph{Physical Review} \textbf{1935}, \emph{47}, 10 777.

\bibitem{Duan2001}
L.-M. Duan, M.~D. Lukin, J.~I. Cirac, P.~Zoller,
\newblock \emph{Nature} \textbf{2001}, \emph{414}, 6862 413.

\bibitem{Loock2020}
P.~Loock, W.~Alt, C.~Becher, O.~Benson, H.~Boche, C.~Deppe, J.~Eschner,
  S.~H{\"{o}}fling, D.~Meschede, P.~Michler, F.~Schmidt, H.~Weinfurter,
\newblock \emph{Advanced Quantum Technologies} \textbf{2020}, \emph{3}, 11
  1900141.

\bibitem{Pan1998}
J.-W. Pan, D.~Bouwmeester, H.~Weinfurter, A.~Zeilinger,
\newblock \emph{Physical Review Letters} \textbf{1998}, \emph{80}, 18 3891.

\bibitem{Zwerger2012}
M.~Zwerger, W.~D{\"{u}}r, H.~J. Briegel,
\newblock \emph{Physical Review A} \textbf{2012}, \emph{85}, 6 062326.

\bibitem{Azuma2015}
K.~Azuma, K.~Tamaki, H.-K. Lo,
\newblock \emph{Nature Communications} \textbf{2015}, \emph{6}, 1 6787.

\bibitem{Zwerger2016}
M.~Zwerger, H.~J. Briegel, W.~D{\"{u}}r,
\newblock \emph{Applied Physics B} \textbf{2016}, \emph{122}, 3 50.

\bibitem{Li2019}
Z.-D. Li, R.~Zhang, X.-F. Yin, L.-Z. Liu, Y.~Hu, Y.-Q. Fang, Y.-Y. Fei,
  X.~Jiang, J.~Zhang, L.~Li, N.-L. Liu, F.~Xu, Y.-A. Chen, J.-W. Pan,
\newblock \emph{Nature Photonics} \textbf{2019}, \emph{13}, 9 644.

\bibitem{raussendorf2001}
R.~Raussendorf, H.~J. Briegel,
\newblock \emph{Physical Review Letters} \textbf{2001}, \emph{86}, 22 5188.

\bibitem{Schwartz2016}
I.~Schwartz, D.~Cogan, E.~R. Schmidgall, Y.~Don, L.~Gantz, O.~Kenneth, N.~H.
  Lindner, D.~Gershoni,
\newblock \emph{Science} \textbf{2016}, \emph{354}, 6311 434.

\bibitem{Istrati2020}
D.~Istrati, Y.~Pilnyak, J.~C. Loredo, C.~Ant{\'{o}}n, N.~Somaschi, P.~Hilaire,
  H.~Ollivier, M.~Esmann, L.~Cohen, L.~Vidro, C.~Millet, A.~Lema{\^{i}}tre,
  I.~Sagnes, A.~Harouri, L.~Lanco, P.~Senellart, H.~S. Eisenberg,
\newblock \emph{Nature Communications} \textbf{2020}, \emph{11}, 1 5501.

\bibitem{Braunstein1995a}
S.~L. Braunstein, A.~Mann,
\newblock \emph{Physical Review A} \textbf{1995}, \emph{51}, 3 R1727.

\bibitem{Hong1987}
L.~L. ROSENBERG, G.~LAROCHE, J.~M. EHLERT,
\newblock \emph{Endocrinology} \textbf{1966}, \emph{79}, 5 927.

\bibitem{Basset2020a}
F.~{Basso Basset}, F.~Salusti, L.~Schweickert, M.~B. Rota, D.~Tedeschi, S.~F.
  {Covre da Silva}, E.~Roccia, V.~Zwiller, K.~D. J{\"{o}}ns, A.~Rastelli,
  R.~Trotta,
\newblock \emph{npj Quantum Information} \textbf{2021}, \emph{7}, 1 7.

\bibitem{DeRiedmatten2003}
H.~de~Riedmatten, I.~Marcikic, W.~Tittel, H.~Zbinden, N.~Gisin,
\newblock \emph{Physical Review A} \textbf{2003}, \emph{67}, 2 022301.

\bibitem{Llewellyn2020}
D.~Llewellyn, Y.~Ding, I.~I. Faruque, S.~Paesani, D.~Bacco, R.~Santagati, Y.-J.
  Qian, Y.~Li, Y.-F. Xiao, M.~Huber, M.~Malik, G.~F. Sinclair, X.~Zhou,
  K.~Rottwitt, J.~L. O'Brien, J.~G. Rarity, Q.~Gong, L.~K. Oxenlowe, J.~Wang,
  M.~G. Thompson,
\newblock \emph{Nature Physics} \textbf{2020}, \emph{16}, 2 148.

\bibitem{Beugnon2006}
J.~Beugnon, M.~P.~A. Jones, J.~Dingjan, B.~Darqui{\'{e}}, G.~Messin,
  A.~Browaeys, P.~Grangier,
\newblock \emph{Nature} \textbf{2006}, \emph{440}, 7085 779.

\bibitem{Vural2018}
H.~Vural, S.~L. Portalupi, J.~Maisch, S.~Kern, J.~H. Weber, M.~Jetter,
  J.~Wrachtrup, R.~L{\"{o}}w, I.~Gerhardt, P.~Michler,
\newblock \emph{Optica} \textbf{2018}, \emph{5}, 4 367.

\bibitem{Maunz2007}
P.~Maunz, D.~L. Moehring, S.~Olmschenk, K.~C. Younge, D.~N. Matsukevich,
  C.~Monroe,
\newblock \emph{Nature Physics} \textbf{2007}, \emph{3}, 8 538.

\bibitem{Sipahigil2014}
A.~Sipahigil, K.~D. Jahnke, L.~J. Rogers, T.~Teraji, J.~Isoya, A.~S. Zibrov,
  F.~Jelezko, M.~D. Lukin,
\newblock \emph{Physical Review Letters} \textbf{2014}, \emph{113}, 11 113602.

\bibitem{Bernien2012}
H.~Bernien, L.~Childress, L.~Robledo, M.~Markham, D.~Twitchen, R.~Hanson,
\newblock \emph{Physical Review Letters} \textbf{2012}, \emph{108}, 4 043604.

\bibitem{Humphreys2018}
P.~C. Humphreys, N.~Kalb, J.~P.~J. Morits, R.~N. Schouten, R.~F.~L. Vermeulen,
  D.~J. Twitchen, M.~Markham, R.~Hanson,
\newblock \emph{Nature} \textbf{2018}, \emph{558}, 7709 268.

\bibitem{Lettow2010}
R.~Lettow, Y.~L.~A. Rezus, A.~Renn, G.~Zumofen, E.~Ikonen, S.~G{\"{o}}tzinger,
  V.~Sandoghdar,
\newblock \emph{Physical Review Letters} \textbf{2010}, \emph{104}, 12 123605.

\bibitem{Giesz2015}
V.~Giesz, S.~L. Portalupi, T.~Grange, C.~Ant{\'{o}}n, L.~{De Santis},
  J.~Demory, N.~Somaschi, I.~Sagnes, A.~Lema{\^{i}}tre, L.~Lanco,
  A.~Auff{\`{e}}ves, P.~Senellart,
\newblock \emph{Physical Review B} \textbf{2015}, \emph{92}, 16 161302.

\bibitem{Thoma2016}
A.~Thoma, P.~Schnauber, M.~Gschrey, M.~Seifried, J.~Wolters, J.-H. Schulze,
  A.~Strittmatter, S.~Rodt, A.~Carmele, A.~Knorr, T.~Heindel, S.~Reitzenstein,
\newblock \emph{Physical Review Letters} \textbf{2016}, \emph{116}, 3 033601.

\bibitem{Flagg2010}
E.~Flagg, A.~Muller, S.~Polyakov, A.~Ling, A.~Migdall, G.~Solomon,
\newblock \emph{Physical Review Letters} \textbf{2010}, \emph{104}, 13 137401.

\bibitem{Beetz2013}
J.~Beetz, T.~Braun, C.~Schneider, S.~H{\"{o}}fling, M.~Kamp,
\newblock \emph{Semiconductor Science and Technology} \textbf{2013}, \emph{28},
  12 122002.

\bibitem{Reindl2017}
M.~Reindl, K.~D. J{\"{o}}ns, D.~Huber, C.~Schimpf, Y.~Huo, V.~Zwiller,
  A.~Rastelli, R.~Trotta,
\newblock \emph{Nano Letters} \textbf{2017}, \emph{17}, 7 4090.

\bibitem{Moczaa-Dusanowska2020}
M.~Mocza{\l}a-Dusanowska, {\L}.~Dusanowski, O.~Iff, T.~Huber, S.~Kuhn,
  T.~Czyszanowski, C.~Schneider, S.~H{\"{o}}fling,
\newblock \emph{ACS Photonics} \textbf{2020}, \emph{7}, 12 3474.

\bibitem{zhai2020a}
L.~Zhai, M.~C. L{\"{o}}bl, G.~N. Nguyen, J.~Ritzmann, A.~Javadi, C.~Spinnler,
  A.~D. Wieck, A.~Ludwig, R.~J. Warburton,
\newblock \emph{Nature Communications} \textbf{2020}, \emph{11}, 1 4745.

\bibitem{Patel2010}
R.~B. Patel, A.~J. Bennett, I.~Farrer, C.~A. Nicoll, D.~A. Ritchie, A.~J.
  Shields,
\newblock \emph{Nature Photonics} \textbf{2010}, \emph{4}, 9 632.

\bibitem{zhai2020}
L.~Zhai, M.~C. L{\"{o}}bl, J.-P. Jahn, Y.~Huo, P.~Treutlein, O.~G. Schmidt,
  A.~Rastelli, R.~J. Warburton,
\newblock \emph{Applied Physics Letters} \textbf{2020}, \emph{117}, 8 083106.

\bibitem{Kambs2018}
B.~Kambs, C.~Becher,
\newblock \emph{New Journal of Physics} \textbf{2018}, \emph{20}, 11 115003.

\bibitem{Fischer2016}
K.~A. Fischer, K.~M{\"{u}}ller, K.~G. Lagoudakis, J.~Vu{\v{c}}kovi{\'{c}},
\newblock \emph{New Journal of Physics} \textbf{2016}, \emph{18}, 11 113053.

\bibitem{Wang2016}
H.~Wang, Z.-C. Duan, Y.-H. Li, S.~Chen, J.-P. Li, Y.-M. He, M.-C. Chen, Y.~He,
  X.~Ding, C.-Z. Peng, C.~Schneider, M.~Kamp, S.~H{\"{o}}fling, C.-Y. Lu, J.-W.
  Pan,
\newblock \emph{Physical Review Letters} \textbf{2016}, \emph{116}, 21 213601.

\bibitem{Schmidt2020}
M.~Schmidt, M.~V. Helversen, S.~Fischbach, A.~Kaganskiy, R.~Schmidt,
  A.~Schliwa, T.~Heindel, S.~Rodt, S.~Reitzenstein,
\newblock \emph{Optical Materials Express} \textbf{2020}, \emph{10}, 1 76.

\bibitem{Zopf2018}
M.~Zopf, T.~Macha, R.~Keil, E.~Uru{\~{n}}uela, Y.~Chen, W.~Alt,
  L.~Ratschbacher, F.~Ding, D.~Meschede, O.~G. Schmidt,
\newblock \emph{Physical Review B} \textbf{2018}, \emph{98}, 16 161302.

\bibitem{Weber2019}
J.~H. Weber, B.~Kambs, J.~Kettler, S.~Kern, J.~Maisch, H.~Vural, M.~Jetter,
  S.~L. Portalupi, C.~Becher, P.~Michler,
\newblock \emph{Nature Nanotechnology} \textbf{2019}, \emph{14}, 1 23.

\bibitem{You2021}
X.~You, M.-Y. Zheng, S.~Chen, R.-Z. Liu, J.~Qin, M.~C. Xu, Z.~X. Ge, T.~H.
  Chung, Y.~K. Qiao, Y.~F. Jiang, H.~S. Zhong, M.~C. Chen, H.~Wang, Y.~M. He,
  X.~P. Xie, H.~Li, L.~X. You, C.~Schneider, J.~Yin, T.~Y. Chen, M.~Benyoucef,
  Y.-H. Huo, S.~Hoefling, Q.~Zhang, C.-Y. Lu, J.-W. Pan,
\newblock \emph{arxiv ID: 2106.15545} \textbf{2021}.

\bibitem{Cabrillo1999}
C.~Cabrillo, J.~I. Cirac, P.~Garc{\'{i}}a-Fern{\'{a}}ndez, P.~Zoller,
\newblock \emph{Physical Review A} \textbf{1999}, \emph{59}, 2 1025.

\bibitem{Delteil2015}
A.~Delteil, Z.~Sun, W.-b. Gao, E.~Togan, S.~Faelt, A.~Imamoğlu,
\newblock \emph{Nature Physics} \textbf{2016}, \emph{12}, 3 218.

\bibitem{Stockill2017}
R.~Stockill, M.~J. Stanley, L.~Huthmacher, E.~Clarke, M.~Hugues, A.~J. Miller,
  C.~Matthiesen, C.~{Le Gall}, M.~Atat{\"{u}}re,
\newblock \emph{Physical Review Letters} \textbf{2017}, \emph{119}, 1 010503.

\bibitem{gold2014}
P.~Gold, A.~Thoma, S.~Maier, S.~Reitzenstein, C.~Schneider, S.~H{\"o}fling,
  M.~Kamp,
\newblock \emph{Physical Review B} \textbf{2014}, \emph{89}, 3 035313.

\bibitem{Kim2016a}
J.-H. Kim, C.~J.~K. Richardson, R.~P. Leavitt, E.~Waks,
\newblock \emph{Nano Letters} \textbf{2016}, \emph{16}, 11 7061.

\bibitem{Zopf2019}
M.~Zopf, R.~Keil, Y.~Chen, J.~Yang, D.~Chen, F.~Ding, O.~G. Schmidt,
\newblock \emph{Physical Review Letters} \textbf{2019}, \emph{123}, 16 160502.

\bibitem{Basset2019}
F.~{Basso Basset}, M.~B. Rota, C.~Schimpf, D.~Tedeschi, K.~D. Zeuner, S.~F.
  {Covre da Silva}, M.~Reindl, V.~Zwiller, K.~D. J{\"{o}}ns, A.~Rastelli,
  R.~Trotta,
\newblock \emph{Physical Review Letters} \textbf{2019}, \emph{123}, 16 160501.

\bibitem{Huber2018}
D.~Huber, M.~Reindl, S.~F. {Covre da Silva}, C.~Schimpf,
  J.~Mart{\'{i}}n-S{\'{a}}nchez, H.~Huang, G.~Piredda, J.~Edlinger,
  A.~Rastelli, R.~Trotta,
\newblock \emph{Physical Review Letters} \textbf{2018}, \emph{121}, 3 033902.

\bibitem{Kok2005}
P.~Kok, W.~J. Munro, K.~Nemoto, T.~C. Ralph, J.~P. Dowling, G.~J. Milburn,
\newblock \emph{Reviews of Modern Physics} \textbf{2007}, \emph{79}, 1 135.

\bibitem{Giovannetti2011}
V.~Giovannetti, S.~Lloyd, L.~Maccone,
\newblock \emph{Nature Photonics} \textbf{2011}, \emph{5}, 4 222.

\bibitem{Biamonte2017}
J.~Biamonte, P.~Wittek, N.~Pancotti, P.~Rebentrost, N.~Wiebe, S.~Lloyd,
\newblock \emph{Nature} \textbf{2017}, \emph{549}, 7671 195.

\bibitem{Imamoglu2002}
A.~Imamoḡlu,
\newblock \emph{Physical Review Letters} \textbf{2002}, \emph{89}, 16 163602.

\bibitem{Bussieres2013}
F.~Bussi{\`{e}}res, N.~Sangouard, M.~Afzelius, H.~de~Riedmatten, C.~Simon,
  W.~Tittel,
\newblock \emph{Journal of Modern Optics} \textbf{2013}, \emph{60}, 18 1519.

\bibitem{Borregaard2020}
J.~Borregaard, H.~Pichler, T.~Schr{\"{o}}der, M.~D. Lukin, P.~Lodahl, A.~S.
  S{\o}rensen,
\newblock \emph{Physical Review X} \textbf{2020}, \emph{10}, 2 021071.

\bibitem{Ma2019}
L.~Ma, O.~Slattery, X.~Tang,
\newblock \emph{Journal of Research of the National Institute of Standards and
  Technology} \textbf{2019}, \emph{124}, December 124039.

\bibitem{DeRiedmatten2008}
H.~de~Riedmatten, M.~Afzelius, M.~U. Staudt, C.~Simon, N.~Gisin,
\newblock \emph{Nature} \textbf{2008}, \emph{456}, 7223 773.

\bibitem{Bao2012}
X.-H. Bao, A.~Reingruber, P.~Dietrich, J.~Rui, A.~D{\"{u}}ck, T.~Strassel,
  L.~Li, N.-L. Liu, B.~Zhao, J.-W. Pan,
\newblock \emph{Nature Physics} \textbf{2012}, \emph{8}, 7 517.

\bibitem{Eisaman2005}
M.~D. Eisaman, A.~Andr{\'{e}}, F.~Massou, M.~Fleischhauer, A.~S. Zibrov, M.~D.
  Lukin,
\newblock \emph{Nature} \textbf{2005}, \emph{438}, 7069 837.

\bibitem{Heshami2016}
K.~Heshami, D.~G. England, P.~C. Humphreys, P.~J. Bustard, V.~M. Acosta,
  J.~Nunn, B.~J. Sussman,
\newblock \emph{Journal of Modern Optics} \textbf{2016}, \emph{63}, 20 2005.

\bibitem{Neuwirth2021}
J.~Neuwirth, F.~B. Basset, M.~B. Rota, E.~Roccia, C.~Schimpf, K.~D. J{\"{o}}ns,
  A.~Rastelli, R.~Trotta,
\newblock \emph{arxiv ID: 2104.07076} \textbf{2021}.

\bibitem{hau1999}
L.~V. Hau, S.~E. Harris, Z.~Dutton, C.~H. Behroozi,
\newblock \emph{Nature} \textbf{1999}, \emph{397}, 6720 594.

\bibitem{Borregaard2016}
J.~Borregaard, M.~Zugenmaier, J.~M. Petersen, H.~Shen, G.~Vasilakis, K.~Jensen,
  E.~S. Polzik, A.~S. S{\o}rensen,
\newblock \emph{Nature Communications} \textbf{2016}, \emph{7}, 1 11356.

\bibitem{Lvovsky2009}
A.~I. Lvovsky, B.~C. Sanders, W.~Tittel,
\newblock \emph{Nature Photonics} \textbf{2009}, \emph{3}, 12 706.

\bibitem{Fleischhauer2000}
M.~Fleischhauer, M.~D. Lukin,
\newblock \emph{Physical Review Letters} \textbf{2000}, \emph{84}, 22 5094.

\bibitem{Ma2017}
L.~Ma, O.~Slattery, X.~Tang,
\newblock \emph{Journal of Optics} \textbf{2017}, \emph{19}, 4 043001.

\bibitem{Reim2011}
K.~F. Reim, P.~Michelberger, K.~C. Lee, J.~Nunn, N.~K. Langford, I.~A.
  Walmsley,
\newblock \emph{Physical Review Letters} \textbf{2011}, \emph{107}, 5 053603.

\bibitem{Rakher2013}
M.~T. Rakher, R.~J. Warburton, P.~Treutlein,
\newblock \emph{Physical Review A} \textbf{2013}, \emph{88}, 5 053834.

\bibitem{Wolters2017}
J.~Wolters, G.~Buser, A.~Horsley, L.~B{\'{e}}guin, A.~J{\"{o}}ckel, J.-P. Jahn,
  R.~J. Warburton, P.~Treutlein,
\newblock \emph{Physical Review Letters} \textbf{2017}, \emph{119}, 6 060502.

\bibitem{Hosseini2011}
M.~Hosseini, G.~Campbell, B.~M. Sparkes, P.~K. Lam, B.~C. Buchler,
\newblock \emph{Nature Physics} \textbf{2011}, \emph{7}, 10 794.

\bibitem{Guo2019}
J.~Guo, X.~Feng, P.~Yang, Z.~Yu, L.~Q. Chen, C.-h. Yuan, W.~Zhang,
\newblock \emph{Nature Communications} \textbf{2019}, \emph{10}, 1 148.

\bibitem{England2012}
D.~G. England, P.~S. Michelberger, T.~F.~M. Champion, K.~F. Reim, K.~C. Lee,
  M.~R. Sprague, X.-M. Jin, N.~K. Langford, W.~S. Kolthammer, J.~Nunn, I.~A.
  Walmsley,
\newblock \emph{Journal of Physics B: Atomic, Molecular and Optical Physics}
  \textbf{2012}, \emph{45}, 12 124008.

\bibitem{Namazi2017}
M.~Namazi, C.~Kupchak, B.~Jordaan, R.~Shahrokhshahi, E.~Figueroa,
\newblock \emph{Physical Review Applied} \textbf{2017}, \emph{8}, 3 034023.

\bibitem{namazi2017free}
M.~Namazi, G.~Vallone, B.~Jordaan, C.~Goham, R.~Shahrokhshahi, P.~Villoresi,
  E.~Figueroa,
\newblock \emph{Physical Review Applied} \textbf{2017}, \emph{8}, 6 064013.

\bibitem{Ding2018}
D.-S. Ding,
\newblock In \emph{Broad Bandwidth and High Dimensional Quantum Memory Based on
  Atomic Ensembles}, 91--107. Springer, \textbf{2018},
\newblock \urlprefix\url{http://link.springer.com/10.1007/978-981-10-7476-9_6
  http://link.springer.com/10.1007/978-981-10-7476-9}.

\bibitem{Akopian2011}
N.~Akopian, L.~Wang, A.~Rastelli, O.~G. Schmidt, V.~Zwiller,
\newblock \emph{Nature Photonics} \textbf{2011}, \emph{5}, 4 230.

\bibitem{Jahn2015}
J.-P. Jahn, M.~Munsch, L.~B{\'{e}}guin, A.~V. Kuhlmann, M.~Renggli, Y.~Huo,
  F.~Ding, R.~Trotta, M.~Reindl, O.~G. Schmidt, A.~Rastelli, P.~Treutlein,
  R.~J. Warburton,
\newblock \emph{Physical Review B} \textbf{2015}, \emph{92}, 24 245439.

\bibitem{Bremer2020}
L.~Bremer, S.~Fischbach, S.~Park, S.~Rodt, J.~Song, T.~Heindel,
  S.~Reitzenstein,
\newblock \emph{Advanced Quantum Technologies} \textbf{2020}, \emph{3}, 2
  1900071.

\bibitem{Vamivakas2009}
A.~{Nick Vamivakas}, Y.~Zhao, C.-Y. Lu, M.~Atat{\"{u}}re,
\newblock \emph{Nature Physics} \textbf{2009}, \emph{5}, 3 198.

\bibitem{Ulrich2014}
S.~M. Ulrich, S.~Weiler, M.~Oster, M.~Jetter, A.~Urvoy, R.~L{\"{o}}w,
  P.~Michler,
\newblock \emph{Physical Review B} \textbf{2014}, \emph{90}, 12 125310.

\bibitem{Usmani2010}
I.~Usmani, M.~Afzelius, H.~de~Riedmatten, N.~Gisin,
\newblock \emph{Nature Communications} \textbf{2010}, \emph{1}, 1 12.

\bibitem{Zhou2012}
{Xin Zhou}, T.~Schoepf,
\newblock In \emph{26th International Conference on Electrical Contacts (ICEC
  2012)}, 605 CP. IET,
\newblock ISBN 978-1-84919-508-9, \textbf{2012} 288--295,
\newblock
  \urlprefix\url{https://digital-library.theiet.org/content/conferences/10.1049/cp.2012.0663}.

\bibitem{Kroutvar2004}
M.~Kroutvar, Y.~Ducommun, D.~Heiss, M.~Bichler, D.~Schuh, G.~Abstreiter, J.~J.
  Finley,
\newblock \emph{Nature} \textbf{2004}, \emph{432}, 7013 81.

\bibitem{McFarlane2009}
J.~McFarlane, P.~A. Dalgarno, B.~D. Gerardot, R.~H. Hadfield, R.~J. Warburton,
  K.~Karrai, A.~Badolato, P.~M. Petroff,
\newblock \emph{Applied Physics Letters} \textbf{2009}, \emph{94}, 9 093113.

\bibitem{Taylor2003}
J.~M. Taylor, C.~M. Marcus, M.~D. Lukin,
\newblock \emph{Physical Review Letters} \textbf{2003}, \emph{90}, 20 206803.

\bibitem{Kurucz2009}
Z.~Kurucz, M.~W. S{\o}rensen, J.~M. Taylor, M.~D. Lukin, M.~Fleischhauer,
\newblock \emph{Physical Review Letters} \textbf{2009}, \emph{103}, 1 010502.

\bibitem{Shim2013}
J.~H. Shim, I.~Niemeyer, J.~Zhang, D.~Suter,
\newblock \emph{Physical Review A} \textbf{2013}, \emph{87}, 1 012301.

\bibitem{Morton2008}
J.~J.~L. Morton, A.~M. Tyryshkin, R.~M. Brown, S.~Shankar, B.~W. Lovett,
  A.~Ardavan, T.~Schenkel, E.~E. Haller, J.~W. Ager, S.~A. Lyon,
\newblock \emph{Nature} \textbf{2008}, \emph{455}, 7216 1085.

\bibitem{Bussieres2014}
F.~Bussi{\`{e}}res, C.~Clausen, A.~Tiranov, B.~Korzh, V.~B. Verma, S.~W. Nam,
  F.~Marsili, A.~Ferrier, P.~Goldner, H.~Herrmann, C.~Silberhorn, W.~Sohler,
  M.~Afzelius, N.~Gisin,
\newblock \emph{Nature Photonics} \textbf{2014}, \emph{8}, 10 775.

\bibitem{Bennett1993}
C.~H. Bennett, G.~Brassard, C.~Cr{\'{e}}peau, R.~Jozsa, A.~Peres, W.~K.
  Wootters,
\newblock \emph{Physical Review Letters} \textbf{1993}, \emph{70}, 13 1895.

\bibitem{Varnava2016}
C.~Varnava, R.~M. Stevenson, J.~Nilsson, J.~Skiba-Szymanska, B.~Dzurň{\'{a}}k,
  M.~Lucamarini, R.~V. Penty, I.~Farrer, D.~A. Ritchie, A.~J. Shields,
\newblock \emph{npj Quantum Information} \textbf{2016}, \emph{2}, 1 16006.

\bibitem{Huwer2017}
J.~Huwer, R.~M. Stevenson, J.~Skiba-Szymanska, M.~B. Ward, A.~J. Shields,
  M.~Felle, I.~Farrer, D.~A. Ritchie, R.~V. Penty,
\newblock \emph{Physical Review Applied} \textbf{2017}, \emph{8}, 2 024007.

\bibitem{Nilsson2013}
J.~Nilsson, R.~M. Stevenson, K.~H.~A. Chan, J.~Skiba-Szymanska, M.~Lucamarini,
  M.~B. Ward, A.~J. Bennett, C.~L. Salter, I.~Farrer, D.~A. Ritchie, A.~J.
  Shields,
\newblock \emph{Nature Photonics} \textbf{2013}, \emph{7}, 4 311.

\bibitem{Anderson2020}
M.~Anderson, T.~M{\"{u}}ller, J.~Huwer, J.~Skiba-Szymanska, A.~B. Krysa, R.~M.
  Stevenson, J.~Heffernan, D.~A. Ritchie, A.~J. Shields,
\newblock \emph{npj Quantum Information} \textbf{2020}, \emph{6}, 1 14.

\bibitem{Reindl2018}
M.~Reindl, D.~Huber, C.~Schimpf, S.~F.~C. da~Silva, M.~B. Rota, H.~Huang,
  V.~Zwiller, K.~D. J{\"{o}}ns, A.~Rastelli, R.~Trotta,
\newblock \emph{Science Advances} \textbf{2018}, \emph{4}, 12 eaau1255.

\bibitem{Knill2001}
E.~Knill, R.~Laflamme, G.~J. Milburn,
\newblock \emph{Nature} \textbf{2001}, \emph{409}, 6816 46.

\bibitem{Fattal2004}
D.~Fattal, E.~Diamanti, K.~Inoue, Y.~Yamamoto,
\newblock \emph{Physical Review Letters} \textbf{2004}, \emph{92}, 3 037904.

\bibitem{Schmidt1970}
H.~Schmidt,
\newblock \emph{Journal of Applied Physics} \textbf{1970}, \emph{41}, 2 462.

\bibitem{Herrero-Collantes2017}
M.~Herrero-Collantes, J.~C. Garcia-Escartin,
\newblock \emph{Reviews of Modern Physics} \textbf{2017}, \emph{89}, 1 015004.

\bibitem{idquantique}
{Press Release} idquantique,
\newblock
  \url{https://marketing.idquantique.com/acton/attachment/11868/f-b551a075-4944-4740-8563-ec5115ed08d0/1/-/-/-/-/IDQ-SKT%205G%20QRNG%20smartphone_PR.pdf},
\newblock Accessed: 2021-08-16.

\bibitem{Hart2017}
J.~D. Hart, Y.~Terashima, A.~Uchida, G.~B. Baumgartner, T.~E. Murphy, R.~Roy,
\newblock \emph{APL Photonics} \textbf{2017}, \emph{2}, 9 090901.

\bibitem{VonNeumann1963}
J.~von Neumann,
\newblock \emph{Collected Work} \textbf{1963}, \emph{512} 768.

\bibitem{Peres1992}
Y.~Peres,
\newblock \emph{The Annals of Statistics} \textbf{1992}, \emph{20}, 1 590.

\bibitem{Ma2016}
X.~Ma, X.~Yuan, Z.~Cao, B.~Qi, Z.~Zhang,
\newblock \emph{npj Quantum Information} \textbf{2016}, \emph{2}, 1 16021.

\bibitem{Gallego2013}
R.~Gallego, L.~Masanes, G.~{De La Torre}, C.~Dhara, L.~Aolita, A.~Ac{\'{i}}n,
\newblock \emph{Nature Communications} \textbf{2013}, \emph{4}, 1 2654.

\bibitem{Jennewein2000}
T.~Jennewein, U.~Achleitner, G.~Weihs, H.~Weinfurter, A.~Zeilinger,
\newblock \emph{Review of Scientific Instruments} \textbf{2000}, \emph{71}, 4
  1675.

\bibitem{Stefanov2000}
A.~Stefanov, N.~Gisin, O.~Guinnard, L.~Guinnard, H.~Zbinden,
\newblock \emph{Journal of Modern Optics} \textbf{2000}, \emph{47}, 4 595.

\bibitem{Oberreiter2016}
L.~Oberreiter, I.~Gerhardt,
\newblock \emph{Laser and Photonics Reviews} \textbf{2016}, \emph{10}, 1 108.

\bibitem{Furst2010}
M.~F{\"{u}}rst, H.~Weier, S.~Nauerth, D.~G. Marangon, C.~Kurtsiefer,
  H.~Weinfurter,
\newblock \emph{Optics Express} \textbf{2010}, \emph{18}, 12 13029.

\bibitem{Wayne2010}
M.~A. Wayne, P.~G. Kwiat,
\newblock \emph{Optics Express} \textbf{2010}, \emph{18}, 9 9351.

\bibitem{Wahl2011}
M.~Wahl, M.~Leifgen, M.~Berlin, T.~R{\"{o}}hlicke, H.-J. Rahn, O.~Benson,
\newblock \emph{Applied Physics Letters} \textbf{2011}, \emph{98}, 17 171105.

\bibitem{Haylock2019}
B.~Haylock, D.~Peace, F.~Lenzini, C.~Weedbrook, M.~Lobino,
\newblock \emph{Quantum} \textbf{2019}, \emph{3} 141.

\bibitem{Lei2020}
W.~Lei, Z.~Xie, Y.~Li, J.~Fang, W.~Shen,
\newblock \emph{Quantum Information Processing} \textbf{2020}, \emph{19}, 11
  405.

\bibitem{Bai2021}
B.~Bai, J.~Huang, G.-R. Qiao, Y.-Q. Nie, W.~Tang, T.~Chu, J.~Zhang, J.-W. Pan,
\newblock \emph{arxiv ID: 2105.13518} \textbf{2021}, 1--5.

\bibitem{Liu2017}
J.~Liu, J.~Yang, Z.~Li, Q.~Su, W.~Huang, B.~Xu, H.~Guo,
\newblock \emph{IEEE Photonics Technology Letters} \textbf{2017}, \emph{29}, 3
  283.

\bibitem{Gomez2019}
S.~G{\'{o}}mez, A.~Mattar, I.~Machuca, E.~S. G{\'{o}}mez, D.~Cavalcanti, O.~J.
  Far{\'{i}}as, A.~Ac{\'{i}}n, G.~Lima,
\newblock \emph{Physical Review A} \textbf{2019}, \emph{99}, 3 032108.

\bibitem{Avesani2021}
M.~Avesani, H.~Tebyanian, P.~Villoresi, G.~Vallone,
\newblock \emph{Physical Review Applied} \textbf{2021}, \emph{15}, 3 034034.

\bibitem{McCabe2020}
H.~McCabe, S.~M. Koziol, G.~L. Snider, E.~P. Blair,
\newblock \emph{IEEE Transactions on Nanotechnology} \textbf{2020}, \emph{19}
  292.

\bibitem{Purkayastha2016}
T.~Purkayastha, D.~De, K.~Das,
\newblock \emph{Microprocessors and Microsystems} \textbf{2016}, \emph{45} 32.

\bibitem{Zhu2002}
{Xiaoming Zhu}, J.~Kahn,
\newblock \emph{IEEE Transactions on Communications} \textbf{2002}, \emph{50},
  8 1293.

\bibitem{Ursin2007}
R.~Ursin, F.~Tiefenbacher, T.~Schmitt-Manderbach, H.~Weier, T.~Scheidl,
  M.~Lindenthal, B.~Blauensteiner, T.~Jennewein, J.~Perdigues, P.~Trojek,
  B.~{\"{O}}mer, M.~F{\"{u}}rst, M.~Meyenburg, J.~Rarity, Z.~Sodnik,
  C.~Barbieri, H.~Weinfurter, A.~Zeilinger,
\newblock \emph{Nature Physics} \textbf{2007}, \emph{3}, 7 481.

\bibitem{carrasco2014}
A.~Carrasco-Casado, N.~Denisenko, V.~Fernandez,
\newblock \emph{Optical Engineering} \textbf{2014}, \emph{53}, 8 084112.

\bibitem{Liu2021entanglement}
H.-Y. Liu, X.-H. Tian, C.~Gu, P.~Fan, X.~Ni, R.~Yang, J.-N. Zhang, M.~Hu,
  J.~Guo, X.~Cao, X.~Hu, G.~Zhao, Y.-Q. Lu, Y.-X. Gong, Z.~Xie, S.-N. Zhu,
\newblock \emph{Physical Review Letters} \textbf{2021}, \emph{126}, 2 020503.

\bibitem{Laing2010}
A.~Laing, V.~Scarani, J.~G. Rarity, J.~L. O'Brien,
\newblock \emph{Physical Review A} \textbf{2010}, \emph{82}, 1 012304.

\bibitem{Wabnig2013}
J.~Wabnig, D.~Bitauld, H.~W. Li, A.~Laing, J.~L. O'Brien, A.~O. Niskanen,
\newblock \emph{New Journal of Physics} \textbf{2013}, \emph{15}, 7 073001.

\bibitem{Ismail2019}
Y.~Ismail, I.~Sinayskiy, F.~Petruccione,
\newblock \emph{Journal of the Optical Society of America B} \textbf{2019},
  \emph{36}, 3 B116.

\bibitem{Bienfang2004}
J.~C. Bienfang, A.~J. Gross, A.~Mink, B.~J. Hershman, A.~Nakassis, X.~Tang,
  R.~Lu, D.~H. Su, C.~W. Clark, C.~J. Williams, E.~W. Hagley, J.~Wen,
\newblock \emph{Optics Express} \textbf{2004}, \emph{12}, 9 2011.

\bibitem{Pljonkin2017}
A.~Pljonkin, K.~Rumyantsev, P.~Singh,
\newblock \emph{Cryptography} \textbf{2017}, \emph{1}, 3 18.

\bibitem{Grunenfelder2018}
F.~Gr{\"{u}}nenfelder, A.~Boaron, D.~Rusca, A.~Martin, H.~Zbinden,
\newblock \emph{Applied Physics Letters} \textbf{2018}, \emph{112}, 5 051108.

\bibitem{Sibson2017}
P.~Sibson, J.~E. Kennard, S.~Stanisic, C.~Erven, J.~L. O'Brien, M.~G. Thompson,
\newblock \emph{Optica} \textbf{2017}, \emph{4}, 2 172.

\bibitem{Bunandar2018}
D.~Bunandar, A.~Lentine, C.~Lee, H.~Cai, C.~M. Long, N.~Boynton, N.~Martinez,
  C.~DeRose, C.~Chen, M.~Grein, D.~Trotter, A.~Starbuck, A.~Pomerene,
  S.~Hamilton, F.~N. Wong, R.~Camacho, P.~Davids, J.~Urayama, D.~Englund,
\newblock \emph{Physical Review X} \textbf{2018}, \emph{8}, 2 021009.

\bibitem{Zadeh2021}
I.~{Esmaeil Zadeh}, J.~Chang, J.~W.~N. Los, S.~Gyger, A.~W. Elshaari,
  S.~Steinhauer, S.~N. Dorenbos, V.~Zwiller,
\newblock \emph{Applied Physics Letters} \textbf{2021}, \emph{118}, 19 190502.

\bibitem{Ljunggren2000}
D.~Ljunggren, M.~Bourennane, A.~Karlsson,
\newblock \emph{Physical Review A} \textbf{2000}, \emph{62}, 2 022305.

\bibitem{kiktenko2020}
E.~O. Kiktenko, A.~O. Malyshev, M.~A. Gavreev, A.~A. Bozhedarov, N.~O. Pozhar,
  M.~N. Anufriev, A.~K. Fedorov,
\newblock \emph{IEEE Transactions on Information Theory} \textbf{2020},
  \emph{66}, 10 6354.

\bibitem{Wang2021}
L.-J. Wang, K.-Y. Zhang, J.-Y. Wang, J.~Cheng, Y.-H. Yang, S.-B. Tang, D.~Yan,
  Y.-L. Tang, Z.~Liu, Y.~Yu, Q.~Zhang, J.-W. Pan,
\newblock \emph{npj Quantum Information} \textbf{2021}, \emph{7}, 1 67.

\bibitem{Tomita2019}
A.~Tomita,
\newblock \emph{Advanced Quantum Technologies} \textbf{2019}, \emph{2}, 5-6
  1900005.

\bibitem{Langer2009}
T.~L{\"{a}}nger, G.~Lenhart,
\newblock \emph{New Journal of Physics} \textbf{2009}, \emph{11}, 5 055051.

\bibitem{alleaume2014}
R.~All{\'e}aume, I.~P. Degiovanni, A.~Mink, T.~E. Chapuran, N.~Lutkenhaus,
  M.~Peev, C.~J. Chunnilall, V.~Martin, M.~Lucamarini, M.~Ward, et~al.,
\newblock In \emph{2014 IEEE Globecom Workshops (GC Wkshps)}. IEEE,
  \textbf{2014} 656--661.

\bibitem{Wehner2008}
S.~Wehner, C.~Schaffner, B.~M. Terhal,
\newblock \emph{Physical Review Letters} \textbf{2008}, \emph{100}, 22 220502.

\end{thebibliography}


\end{document}